%% file: MessQR.tex
\documentclass[aps,amssymb,amsmath,twocolumn]{revtex4} 
\usepackage{color}
\usepackage{epsfig}

\newcommand{\ket}[1]{|#1\rangle}

\newcommand{\im}{{\rm i}}

\newcommand{\iden}{1 \hspace{-1.0mm}  {\bf l}}

\newcommand{\ncd}{\newcommand}
\ncd{\QC}{$\mbox{QC}_{\cal{C}}\;$}
\ncd{\QCpr}{${\mbox{QC}_{\cal{C}}}^\prime\;$}
\ncd{\QCns}{$\mbox{QC}_{\cal{C}}$}
\ncd{\QCprns}{${\mbox{QC}_{\cal{C}}}^\prime$}
\ncd{\cskN}{{|\phi_{\{\kappa\} } \rangle}_{{\cal{C}}_N}}
\ncd{\cskNpr}{{|\phi_{\{\kappa^\prime\} } \rangle}_{{\cal{C}}_N}}
\ncd{\cskNtil}{{|\phi_{\{\tilde{\kappa} \} } \rangle}_{{\cal{C}}_N}}
\ncd{\csk}{{|\phi_{\{\kappa\} } \rangle}_{\cal{C}}}
\ncd{\csktil}{{|\phi_{\{\tilde{\kappa} \} } \rangle}_{\cal{C}}}
\ncd{\cskf}{|\phi_{\{\kappa\} } \rangle_{\cal{C}}}
\ncd{\csktilf}{|\phi_{\{\tilde{\kappa} \} } \rangle_{\cal{C}}}
\ncd{\bracsk}{\mbox{}_{\cal{C}}\langle\phi_{\{\kappa\} }|}
\ncd{\bracsktil}{\mbox{}_{\cal{C}}\langle\phi_{\{\tilde{\kappa} \} }|}
\ncd{\nbracsk}{\mbox{}_{\cal{C}}\langle\phi_{\{\kappa\} }}
\ncd{\nbracsktil}{\mbox{}_{\cal{C}}\langle\phi_{\{\tilde{\kappa} \} }}
\ncd{\cs}{|\phi \rangle_{\cal{C}}\;}
\ncd{\csns}{|\phi \rangle_{\cal{C}}}
\ncd{\nbgh}{\text{nbgh}}
\ncd{\Sab}{S^{ab}}
\ncd{\Sba}{S^{ba}}
\ncd{\ds}{\displaystyle}
\ncd{\ovl}{\overline}

\begin{document}

\newtheorem{theorem}{Theorem}
\newtheorem{definition}{Definition}
\newtheorem{procedure}{Procedure}
\newtheorem{scheme}{Scheme}
\newtheorem{errmod}{Error model}

\title{Measurement-based quantum computation on cluster states}

\author{Robert Raussendorf, Daniel
  E. Browne,$\mbox{}^\ast$ and Hans J. Briegel} 
\affiliation{Theoretische Physik, 
Ludwig-Maximilians-Universit{\"a}t M\"unchen, 
Germany}

\date{\today}

\begin{abstract}
We give a detailed account of the one-way quantum computer, a
scheme of quantum computation that consists entirely of   
one-qubit measurements on a particular class of entangled states, 
the cluster states. We prove its universality, describe why its
underlying computational model is different from the network model of
quantum computation and relate quantum algorithms to mathematical
graphs. Further we 
investigate the scaling of required resources and give a number of
examples for circuits of practical interest such as the circuit for
quantum Fourier transformation and for the quantum adder. Finally,
we describe computation with clusters of finite size.
\end{abstract}

\pacs{PACS-numbers: 3.67.Lx, 3.67.-a}

\maketitle


\section{Introduction}
\label{intro}

Recently, we introduced the scheme of the one-way quantum computer (\QC)
\cite{QCmeas}. This scheme uses a given entangled state, the so-called
cluster state \cite{BR},
as its central physical resource. The entire quantum computation consists
only of a sequence of one-qubit projective measurements on this entangled
state. Thus it uses measurements as the central tool to drive a
computation \cite{Nil97} - \cite{Nil}. We called 
this scheme the ``one-way quantum computer'' since the 
entanglement in the cluster state is destroyed by the one-qubit
measurements and therefore it can only be used once. To emphasize the
importance of the cluster state for the scheme, we use the abbreviation
\QC for ``one-way quantum computer''. 

The \QC is universal since 
any unitary quantum logic network can be simulated on it
efficiently. The \QC can thus be explained as a 
simulator of quantum logic networks. However, the computational model
that emerges for the \QC \cite{QCmodel} makes no reference to the
concept of unitary evolution and it shall be pointed
out from the beginning that the network model does not provide the
most suitable description for the \QCns. Nevertheless, the
network model is the most widely used form of describing a quantum
computer and therefore the relation between the network model and
the \QC must be clarified.  

The purpose of this paper is threefold. First, it is to give the proof
for universality of the \QCns; second, to relate quantum algorithms to
graphs; and third, to
provide a number of examples for \QCns-circuits which are
characteristic and of practical interest. 
     
In Section~\ref{universal} we give the universality proof for the
described scheme of computation in a complete and detailed form. The
proof has already been presented to a large part in
\cite{QCmeas}. What was not contained in \cite{QCmeas} was the
explanation of why and how the gate simulations on the \QC work. This
omission seemed in order since the implementation of the gates discussed there
(CNOT and arbitrary rotations) require only small clusters such that
the functioning of the gates can be easily verified in a computer
simulation. For the examples of gates and sub-circuits given in
Section~\ref{examples} this is no longer the case. Generally, we want an
analytic explanation for the functioning of the
gate simulations on the \QCns.  This explanation is given in
Section~\ref{QcorrQcomp} and applied to the gates of a universal set
in Section~\ref{unifunk} as well as to more complicated examples
in Section~\ref{examples}. 

In Section~\ref{Upperbounds} we discuss the spatial, temporal and 
operational resources required in \QCns-computations in relation to
the resources needed for the corresponding quantum logic networks. 
We find that overheads are at most
polynomial. But there do not always need to be overheads. For example,
as  shown in
Section~\ref{Onestep}, all \QCns-circuits in the Clifford group have
unit logical depth.

In Section~\ref{model} we discuss non-network aspects of the \QCns.
In Section~\ref{IP} we state the reasons why the network
model is not adequate to describe the \QC in every respect. The
network model is abandoned and replaced by a more
appropriate model \cite{QCmodel}. This model is described very
briefly. 

In Section~\ref{graphs} we relate algorithms to graphs. We show
that from every 
algorithm its Clifford part can be removed. The required
algorithm-specific non-universal quantum resource to run the remainder
of the quantum algorithm on the \QC is then a graph state
\cite{Schlingel1}. All that remains of the Clifford part is a
mathematical graph specifying this graph state. 

In Section~\ref{examples} we give examples
of larger gates and sub-circuits which may be of practical
relevance, among them the \QCns-circuit for quantum Fourier
transformation and for the $n$-qubit adder.     

In Section~\ref{wenigPlatz} we discuss the \QC computations on finite
(small) clusters and in the presence of decoherence. We describe a
variant of the scheme consisting of repeated steps of (re-)entangling a
cluster via the Ising interaction, alternating with rounds of
one-qubit measurements. Using this modified scheme it is possible to
split long computations such that they fit piecewise on a small
cluster. 

\section{Universality of quantum computation via
  one-qubit-mea\-sure\-ments}
\label{universal}

In this section we prove that the \QC is a universal quantum
computer. The technique to accomplish this is to show that any quantum
logic network can be simulated efficiently on the \QCns. 
Before we go into the details, let us state the general picture. 

For the one-way quantum computer, the entire resource for the quantum  
computation is provided  initially in the form of a specific entangled
state --the cluster state \cite{BR}-- 
of a large number of qubits. Information is then written onto the 
cluster, processed, and read out from the cluster by one-particle 
measurements only. The entangled state of the cluster thereby serves as a 
universal ``substrate'' for any quantum computation. It  provides in
advance all entanglement that is involved in the subsequent quantum  
computation. Cluster states can be created  
efficiently in any system with a quantum Ising-type interaction (at very low 
temperatures) between two-state particles in a lattice
configuration. 
 
It is important to realize
here that information processing is possible even though the result of
every measurement in any direction of the Bloch sphere is completely
random. The mathematical expression for the randomness of the
measurement results is that the reduced density operator for 
each qubit in the cluster state is $\frac{1}{2}{\bf{1}}$. The
individual measurement results are random but correlated, and these
correlations enable quantum computation on the \QCns.

For clarity, let us emphasize that in the scheme of the \QC  we
distinguish between cluster qubits on 
${\cal{C}}$  which are measured in the process of computation, and the
logical qubits. The  
logical qubits constitute the quantum information being processed while
the cluster qubits in the initial cluster state form an entanglement resource.
Measurements of their individual one-qubit state drive the
computation.

\begin{figure}[tph]
\begin{center}
 \epsfig{file=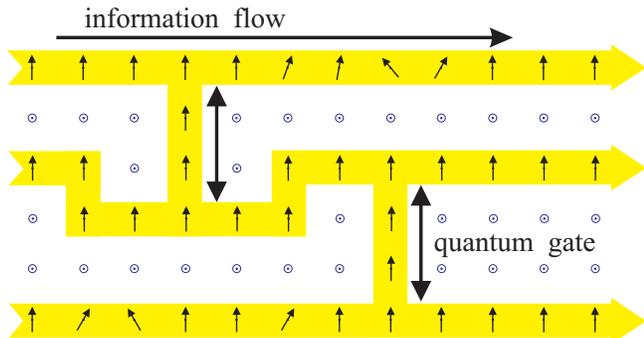,width=8.5cm}
 \caption{\label{FIGetching}Simulation of a
 quantum logic network by measuring two-state particles on a 
  lattice. Before the measurements the qubits are in the
  cluster state $|\phi\rangle_{\cal{C}}$ of (\ref{EVeqn}). 
  Circles $\odot$ symbolize measurements of $\sigma_z$, vertical arrows
  are measurements of $\sigma_x$, while tilted arrows refer to
  measurements in the x-y-plane.}
\end{center}
\end{figure}
To process quantum information with this cluster, it suffices to
measure its particles in a certain order and in a certain basis, as
depicted in Fig.~\ref{FIGetching}. Quantum 
information is thereby propagated through the cluster and
processed. Measurements of $\sigma_z$-observables effectively remove the
respective lattice qubit from the cluster. Measurements in the
$\sigma_x$- (and $\sigma_y$-) eigenbasis are
used for ``wires'', i.e. to propagate logical quantum bits through the
cluster, and for the CNOT-gate between two logical qubits. Observables
of the form $\cos(\varphi)\,\sigma_x \pm \sin(\varphi)\, \sigma_y$ are
measured to realize arbitrary rotations of logical qubits. 
For these cluster qubits, the basis in  which each of
them is  measured depends on the results of preceding 
measurements. This introduces a temporal order in which the
measurements have to be performed. The processing is finished once all qubits 
except a last one on each wire have been measured. The remaining
unmeasured qubits form the quantum register which is now ready to be
read out. At this point, the
results of previous measurements  
determine in which basis these ``output'' qubits need to be measured for the 
final readout, or if the readout measurements are in the $\sigma_x$-,
$\sigma_y$- 
or $\sigma_z$-eigenbasis, how the readout measurements have to be
interpreted. Without loss of generality, we assume in this paper that
the readout measurements are performed in the $\sigma_z$-eigenbasis. 

\subsection{Cluster states and their quantum correlations}
\label{correl}
Cluster states are pure quantum states of
two-level systems (qubits) located on a cluster ${\cal{C}}$. This
cluster is a connected subset of a simple cubic lattice $\mathbb{Z}^d$ in
$d\geq1$ dimensions. The cluster states $\csk$ obey the set of
eigenvalue equations
\begin{equation}
    \label{EVeqn}
    K^{(a)}\csk  = (-1)^{\kappa_a} \csk,
\end{equation} 
with the correlation operators
\begin{equation}
    \label{BasCorr}
    K^{(a)} =  \sigma_x^{(a)}
    \bigotimes\limits_{b \in \nbgh(a)}
    \sigma_z^{(b)}. 
\end{equation}
Therein, $\{\kappa_a \in \{0,1\}|\, a \in {\cal{C}} \}$
is a set of  
binary parameters which specify the cluster state and $\nbgh(a)$ is
the set of all neighboring lattice sites of $a$. All states
$\csk$ are equally good for computation. 
A cluster state is completely specified by the eigenvalue equations
(\ref{EVeqn}). 
To see this, first note that two states  
$\cskf$ and $\csktilf$ which obey a set of equations (\ref{EVeqn}) but
differ in at least one eigenvalue are orthogonal. This holds because
if there exists an $a \in {\cal{C}}$ such that, 
say, $K^{(a)} \cskf = \cskf$ and  $K^{(a)} \csktilf =
- \csktilf$, then $\nbracsktil \cskf= \bracsktil K^{(a)} \cskf =
- \nbracsktil \cskf = 0$. From the set of
states which obey (\ref{EVeqn}) with the
eigenvalues specified by $\{\kappa\}$ a representative $\cskf$ is
taken. There are 
$2^{|{\cal{C}}|}$ such classes of states, and hence
$2^{|{\cal{C}}|}$ mutually orthogonal representatives
$\cskf$. Therefore, the representative cluster states form a basis $\big\{
{\cskf\,| \,\{\kappa\}\in \{0,1\}^{|{\cal{C}}|}} \big\}$  of
the $|{\cal{C}}|$-qubit Hilbert space. To that end, let us now consider a state
$|\phi^\prime_{\{\kappa\}_0}\rangle_{\cal{C}}$ that obeys
(\ref{EVeqn}) with the same $\{\kappa\}_0$ as
$|\phi_{\{\kappa\}_0}\rangle_{\cal{C}}$, and expand it into the
above basis. One finds $|\phi^\prime_{\{\kappa\}_0}\rangle_{\cal{C}} =
\sum_{\{ \kappa \} } \nbracsk
|\phi^\prime_{\{\kappa\}_0}\rangle_{\cal{C}} \, \cskf =
\mbox{}_{\cal{C}} \langle \phi_{
  \{\kappa\}_0}|\phi^\prime_{\{\kappa\}_0}\rangle_{\cal{C}} \,\,
|\phi_{\{\kappa\}_0}\rangle_{\cal{C}}$. Hence, two states
$|\phi^\prime_{\{\kappa\}_0}\rangle_{\cal{C}}$ and
$|\phi_{\{\kappa\}_0}\rangle_{\cal{C}}$ which obey
(\ref{EVeqn}) with the same set ${\{\kappa\}}_0$ are the same modulo
a possible global phase. Consequently, any method that creates a
state obeying equations (\ref{EVeqn}) with a specific set
$\{\kappa_a | a \in {\cal{C}} \}$ creates the same state.

The eigenvalue equations (\ref{EVeqn}) and the quantum
correlations they imply are central for the
described scheme of computation. Also, they represent a very compact way of
characterizing the cluster states. To reflect this in the presentation,
the discussion in this paper will be based entirely on these eigenvalue
equations and we will never need to work out some cluster state 
in any specific basis. In fact, to write down a cluster state in its explicit
form would be quite space-consuming since the minimum
number of required terms scales exponentially with the number of
qubits \cite{BR}, and for computation we will be going to consider rather large
cluster states. Nevertheless, for illustration we give
a few examples of cluster states with a small
number of qubits. The cluster states on a chain of 2, 3 and 4 qubits,
fulfilling the eigenvalue equations (\ref{EVeqn}) with all
$\kappa_a = 0$, are
\begin{equation}
    \label{clusterstates}
    \begin{array}{rcl}
        {|\phi\rangle}_{{\cal{C}}_2} & = & \displaystyle{\frac{1}{\sqrt{2}}
        \left({|0\rangle_1 |+\rangle_2 + 
         |1\rangle_1 |-\rangle_2} \right),}\\
        {|\phi\rangle}_{{\cal{C}}_3} & = & \displaystyle{\frac{1}{\sqrt{2}}
        \left({|+\rangle_1 |0\rangle_2 |+\rangle_3 + 
         |-\rangle_1 |1\rangle_2 |-\rangle_3} \right),}\\ 
        {|\phi\rangle}_{{\cal{C}}_4} & = & \displaystyle{\frac{1}{2}
        |+\rangle_1 |0\rangle_2 |+\rangle_3 |0\rangle_4 + 
         \frac{1}{2} |+\rangle_1 |0\rangle_2 |-\rangle_3 |1\rangle_4
        ,}\\
        & & \displaystyle{+\frac{1}{2}
        |-\rangle_1 |1\rangle_2 |-\rangle_3 |0\rangle_4 + 
         \frac{1}{2} |-\rangle_1 |1\rangle_2 |+\rangle_3 |1\rangle_4,}
    \end{array}
\end{equation}
with the notations
\begin{equation}
    \label{01defs}
    \begin{array}{rcl}
        |0\rangle_a &:=& |0\rangle_{a,z} = \sigma_z^{(a)} |0\rangle_{a,z},\\
        |1\rangle_a &:=& |1\rangle_{a,z} = -\sigma_z^{(a)} |1\rangle_{a,z},\\
        |\pm\rangle_a &:=& \frac{1}{\sqrt{2}} (|0\rangle_a
        \pm |1\rangle_a). 
    \end{array}
\end{equation}
The state $|\phi\rangle_{{\cal{C}}_2}$ is local unitary equivalent to
a Bell state and $|\phi\rangle_{{\cal{C}}_3}$ to the
Greenberger-Horne-Zeilinger (GHZ)
state. $|\phi\rangle_{{\cal{C}}_4}$ is not equivalent to a 4-particle
GHZ state. In particular, the entanglement in
$|\phi\rangle_{{\cal{C}}_4}$ cannot be destroyed by a single local
operation \cite{BR}.

Ways to create a cluster state in principle are to measure all the correlation
operators $K^{(a)},\; a \in {\cal{C}}$ of (\ref{BasCorr})
on an arbitrary $|{\cal{C}}|$-qubit state or to cool into the
ground state of a Hamiltonian $H_K = -\hbar g \sum_{a \in {\cal{C}}}
  \kappa_a K^{(a)}$. 

Another way --likely to be more suitable for realization in the lab--
is as follows. First, a product state 
$|+\rangle_{\cal{C}} = \bigotimes_{a \in {\cal{C}}} 
|+\rangle_{a}$ is prepared. Second, the unitary
transformation $S^{({\cal{C}})}$,
\begin{equation}
    \label{Sdef}
    S^{({\cal{C}})} = \prod\limits_{a,b \in {\cal{C}}| b-a \in
    \gamma_d} \Sab,
\end{equation}
is applied to the state $|+\rangle$. Often
we will write $S$ in short for $S^{({\cal{C}})}$. In
(\ref{Sdef}), for the cases of dimension 
$d=1,2,3$, we have $\gamma_1=\{1\}$, $\gamma_2=\{(1,0)^T,(0,1)^T\}$ and
$\gamma_3=\{(1,0,0)^T,(0,1,0)^T, (0,0,1)^T \}$, and the two-qubit
transformation $\Sab$ is such that the state $|1\rangle_a
\otimes |1\rangle_b$ acquires a phase of $\pi$ under
its action while the remaining states $|0\rangle_a
\otimes |0\rangle_b$, $|0\rangle_a
\otimes |1\rangle_b$ and $|1\rangle_a
\otimes |0\rangle_b$ acquire no phase.
Thus, $\Sab$ has the form
\begin{equation}
    \label{Sabdef}
    \Sab=\frac{1}{2} \left( {\bf{1}} +\sigma_z^{(a)}
    + \sigma_z^{(b)} -
    \sigma_z^{(a)} \otimes \sigma_z^{(b)} \right). 
\end{equation}
The state $|+\rangle_{\cal{C}}$ obviously obeys the eigenvalue
equations $\sigma_x^{(a)}|+\rangle_{\cal{C}} =
|+\rangle_{\cal{C}}\;\forall a \in {\cal{C}}$ and thus the
cluster state $\cs$ generated via 
$S$ obeys
\begin{equation}
    \cs = S \sigma_x^{(a)} S^\dagger \, \cs, \;\; \forall
    a \in {\cal{C}}.
\end{equation} 
To obtain $S \sigma_x^{(a)} S^\dagger$, observe that
\begin{equation}
    \label{conjx1}
    \begin{array}{rcl}
        \Sab \sigma_x^{(a)} {\Sab}^\dagger &=&
        \sigma_x^{(a)} \otimes  \sigma_z^{(b)},\\
        \Sab \sigma_x^{(b)} {\Sab}^\dagger &=&
        \sigma_z^{(a)} \otimes  \sigma_x^{(b)},
    \end{array}
\end{equation} 
and
\begin{equation}
    \label{conjx2}
    \Sab \sigma_x^{(c)} {\Sab}^\dagger =
        \sigma_x^{(c)},\;\; \forall c \in {\cal{C}} \backslash \{ a,b \}. 
\end{equation}
Further, the Pauli phase flip operators $\sigma_z^{(d)}$
commute with all $\Sab$, i.e.
\begin{equation} 
    \label{conjz}
    \Sab \sigma_z^{(d)} {\Sab}^\dagger =
        \sigma_z^{(d)},\;\; \forall d \in {\cal{C}}. 
\end{equation}
Now, from (\ref{conjx1}), (\ref{conjx2}) and (\ref{conjz}) it follows that
\begin{equation}
    \label{SonX}
    S \sigma_x^{(a)} S^\dagger = 
    \sigma_x^{(a)} \!\!\!\!\! \bigotimes\limits_{b \in
    \nbgh(a)} \!\!\!\!\! \sigma_z^{(b)}.
\end{equation}
Thus, the state $\cs$ generated from $|+\rangle_{\cal{C}}$ via the
transformation $S$ as defined
in (\ref{Sdef}) does indeed obey eigenvalue equations of form
(\ref{EVeqn}), with
\begin{equation}
    \label{kappaS}
    \kappa_a = 0,\;\,\, \forall \, a \in {\cal{C}}.
\end{equation}

Note that all operations $\Sab$ in $S$ mutually commute and
that they can therefore be carried out at the same
time. Initial individual preparation of the cluster qubits in
$|+\rangle_{a \in {\cal{C}}}$ can also be done in
parallel. Thus, the creation of the cluster state is a two step
process. {\em{The temporal resources to create the cluster state are
constant in the size of the cluster.}} 

If a cluster state is created as described above this leads to the
specific set of eigenvalues in (\ref{EVeqn})
specified by the parameters $\kappa_a$ in (\ref{kappaS}). As
the eigenvalues are fixed in this case, we drop them in the
notation for the cluster state $\cs$. Cluster states specified
by different sets $\{ \kappa_a \}$ can be obtained by applying
Pauli phase flip operators $\sigma_z^{(a)}$. To see this,
note that
\begin{equation}
    \label{zconjK}
    \sigma_z^{(a)} K^{(b)} {\sigma_z^{(a)}}^\dagger
    = (-1)^{\delta_{a,b}}\,K^{(b)}.
\end{equation}
Therefore,
\begin{equation}
    \label{Eigenvalmod}
    \bigotimes\limits_{a \in {\cal{C}}}
    \left( \sigma_z^{(a)}\right)^{\Delta \kappa_a} \,\,
    |\phi_{ \{ 
    \kappa_a\} }  \rangle_{\cal{C}}    = 
    |\phi_{ \{ \kappa_a + \Delta
\kappa_a \} } \rangle_{\cal{C}}, 
\end{equation} 
where the addition for the $\kappa_a$ is modulo 2.

The transformation $S$ defined in (\ref{Sdef}) is generated by
the Hamiltonian 
\begin{equation}
    \label{SHamil}
    H = \hbar g\sum\limits_{a,b \in {\cal{C}}| 
          b - a \in \gamma_d}
          \frac{1-\sigma_z^{(a)}}{2}
          \frac{1-\sigma_z^{(b)}}{2},
\end{equation}
and $S$ is of the form
\begin{equation}
          \label{SexpH}
          S = \exp\left(-i
          \pi\sum\limits_{a,b \in {\cal{C}}| 
          b - a \in \gamma_d}
          \frac{1-\sigma_z^{(a)}}{2}
          \frac{1-\sigma_z^{(b)}}{2} \right).
\end{equation}
Expanding the exponent in (\ref{SexpH}), one obtains
\begin{equation}
    \label{Isingeq}
    \begin{array}{rcl}
          S &=& \displaystyle{ \left[ \prod\limits_{a,b \in
          {\cal{C}}| 
          b - a \in \gamma_d} \!\!\!\!\!\!\! e^{-i\frac{\pi}{4}}
          \exp\left(i\frac{\pi}{4} 
          \sigma_z^{(a)}\right) \exp\left(i\frac{\pi}{4}
          \sigma_z^{(b)}\right) \right] } \\ 
          & &  \displaystyle{\times \exp\left(-i\frac{\pi}{4}
          \sum\limits_{a,b 
          \in {\cal{C}}| b - a \in \gamma_d}
          \sigma_z^{(a)} \sigma_z^{(b)}  \right)}.
    \end{array}
\end{equation}
We find that the interaction part $H_I$ of the Hamiltonian
$H$  generating $S$ is of Ising form,
\begin{equation}
    \label{IsingHamil}
    H_I = \hbar \frac{g}{4} \sum_{a,b \in {\cal{C}}| b - a \in
  \gamma_d} \sigma_z^{(a)} \sigma_z^{(b)}, 
\end{equation}
and, since the local part $H_{\mbox{\footnotesize{local}}}$ of the
Hamiltonian commutes with the Ising Hamiltonian $H_I$, the interaction
$S$ generated by $H$ is local unitary equivalent to
the unitary transformation generated by a Ising Hamiltonian. 

For matter of presentation, the interaction $\Sab$ in (\ref{Sabdef}) and,
correspondingly, the local part of the Hamiltonian $H$ in
(\ref{SHamil}) has been chosen in such a way that the eigenvalue equations
(\ref{EVeqn}) take the particularly simple form with 
$\kappa_a=0$ for all $a \in {\cal{C}}$, irrespective of
the shape of the cluster.   

Concerning the creation of states that are useful as a resource for
the \QCns, i.e. cluster- or local unitary equivalent states, all 
systems with a tunable Ising interaction and a local $\sigma_z$-type
Hamiltonian, i.e. with a Hamiltonian
\begin{equation}
    \label{SuitHamil}
    H^\prime = \sum_{a \in {\cal{C}}}
    \Delta E_a \, \sigma_z^{(a)} + \hbar \frac{g(t)}{4}
    \sum_{a,b \in {\cal{C}}| b - a \in 
      \gamma_d} \sigma_z^{(a)} \sigma_z^{(b)} 
\end{equation}
are suitable, provided the coupling $g(t)$ can be switched
between zero and at least one nonzero value.

Even this condition can be relaxed. A permanent Ising interaction instead
of a globally tunable one is sufficient, if the measurement process is
much faster than the characteristic time scale for the Ising
interaction, i.e. if the measurements are stroboscopic. If it takes
the Ising interaction a time $T_\text{Ising}$ to create a cluster
state $\cs$ from a product state $|+\rangle_{\cal{C}}$, then the Ising
interaction acting for a time $2 T_\text{Ising}$ performs the identity
operation, $S^{({\cal{C}})} S^{({\cal{C}})} =
\iden^{({\cal{C}})}$. Therefore, starting with a product state
$|+\rangle_{\cal{C}}$ at time $t=0$ evolving under permanent Ising
interaction, stroboscopic measurements may be performed at times
$\left(2k+1\right) T_\text{Ising}, \; k \in \mathbb{N}$. 

Some basic notions of graph theory will later, in the universality
proof, simplify the formulation of our specifications. Therefore let
us, at this point,  establish a connection between quantum states such
as the cluster state of (\ref{EVeqn}) and graphs. The treatment here
follows that of \cite{Schlingel1}, adapted to our notation.  

First let us recall the definition of a graph. A graph  
$G(V,E)$ is a set $V$ of vertices connected via edges $e$ from the set
$E$. The information of which vertex $a \in V$ is connected to which
other vertex $b \in V$ is contained in a symmetric $|V| \times
|V|$ matrix $\Gamma$, the adjacency matrix. The matrix $\Gamma$ is
such that $\Gamma_{ab}=1$ if two vertices $a$ and $b$ are connected
via an edge $e \in E$, and $\Gamma_{ab}=0$ otherwise.  We
identify the cluster ${\cal{C}}$ with the vertices $V_{\cal{C}}$ of a graph,
${\cal{C}}=V_{\cal{C}}$, and in this way establish a connection to the
notion introduced earlier.

To relate graphs to quantum mechanics, the vertices of a graph can be
identified with local quantum systems, in this case qubits, and the 
edges with two-particle interactions \cite{Schlingel1},\cite{Rudolph},
in the present case $\sigma_z\sigma_z$-interactions. If one initially
prepares each individual qubit $a$ in the state
${(\sigma_z^{(a)})}^{\kappa_a}\,|+\rangle_a$ and subsequently
switches on, 
for an appropriately chosen finite time span, the interaction
\begin{equation}
    \label{HamilG}
    H_{G(V,E)}= \hbar g \sum_{(a,b)\in E} \frac{1-\sigma_z^{(a)}}{2}
    \frac{1-\sigma_z^{(b)}}{2},
\end{equation} 
with $(a,b) \in E$ denoting an edge between qubits $a$ and $b$, then
one obtains quantum states that are graph code words as introduced in
\cite{Schlingel1}. Henceforth we will refer to these graph code words
as graph states and use them in a context different from coding. The
graph states $|\phi \{\kappa\} \rangle_G$ 
are defined by a set of eigenvalue equations which read 
\begin{equation}
    \label{EVeqnG}
    \sigma_x^{(a)} \bigotimes_{b \in V} \left( \sigma_z^{(b)} 
    \right)^{\Gamma_{ab}} |\phi \{\kappa\} \rangle_G =
    {(-1)}^{\kappa_a} |\phi \{\kappa\} \rangle_G,
\end{equation}   
with $\kappa_a \in \{0,1\}\; \forall \, a \in V$. Here we use $G$
instead of $V$ as an index for the state $|\phi\rangle$ as the set $E
\subset V \times V$ of edges is now independent and no longer
implicitly specified by $V$ as was the case in (\ref{EVeqn}).   

Note that cluster states (\ref{EVeqn}) are a particular
case of graph states (\ref{EVeqnG}). The graph $G({\cal{C}},
E_{\cal{C}})$ which describes a cluster state is that of a square
lattice in 2D and that of a simple cubic lattice in 3D, i.e. the set
$E_{\cal{C}}$ of edges is given by  
\begin{equation}
    \label{clusteredges}
    E_{\cal{C}}=\big\{ (a,b)|\,a,b \in {\cal{C}}, b\in \text{nbgh}(a) \big\}.
\end{equation}

Let us at the end of this section mention how cluster states may be
created in practice. One possibility is via cold controlled collisions
in optical lattices, as described in \cite{BR}. Cold atoms representing
the qubits can be arranged on a two- or three dimensional lattice and
state-dependent interaction phases may be acquired via cold collisions
between neighboring atoms \cite{Jaksch} or via tunneling
\cite{Duan}. For a suitable choice of the 
collision phases $\varphi$, $\varphi = \pi \; \mbox{mod} \,\, 2\pi$,
the state resulting from a product state $|+\rangle_{\cal{C}}$ after
interaction is a cluster state obeying the eigenvalue
equations (\ref{EVeqn}), with the set $\{\kappa_a,
a \in {\cal{C}} \}$ specified by the filling pattern of the
lattice. 

\subsection{A universal set of quantum gates}
\label{unigate}

To provide something definite to discuss right
from the beginning, we now give the procedures of how to
realize a CNOT-gate and a general one-qubit rotation via one-qubit
measurements on a cluster state. The explanation of why and how these
gates work will be given in Section~\ref{unifunk}. 

\begin{figure}[htb]
    \begin{center}
        \begin{tabular}{cc}
            \multicolumn{2}{c}{{\normalsize{(a)}} \hspace{0.3cm}
            \parbox{4.7cm}{ 
            \epsfig{file=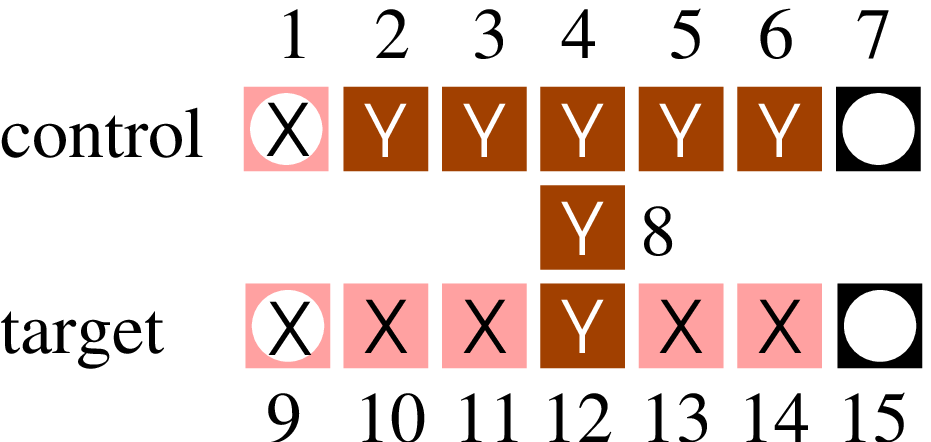, 
            width=4.7cm}}} \vspace{2mm} \\
            \multicolumn{2}{c}{CNOT-gate} \\ & \\ 
            \parbox{2.8cm}{{\normalsize{(b)}} \vspace{0.3cm}\\
            \epsfig{file=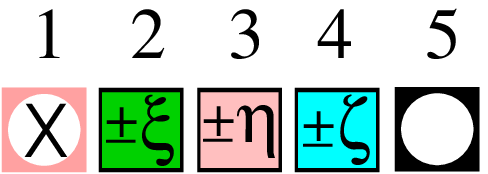,width=2.8cm}} & 
            \hspace*{0.4cm} \parbox{2.8cm}{{\normalsize{(c)}} \vspace{0.3cm}\\
            \epsfig{file=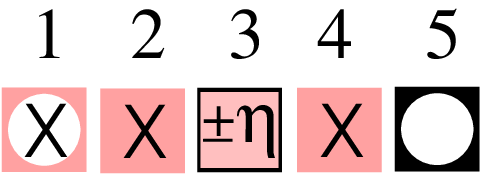,width=2.8cm}} \vspace{2mm} \\ 
            general rotation & $z$-rotation \\ & \\
            \parbox{2.8cm}{{\normalsize{(d)}} \vspace{0.3cm}\\
            \epsfig{file=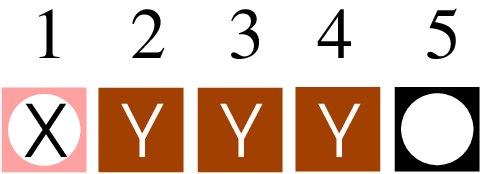,width=2.8cm}} & 
            \hspace*{0.4cm}\parbox{2.8cm}{{\normalsize{(e)}} \vspace{0.3cm}\\
            \epsfig{file=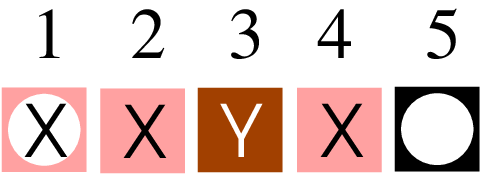,width=2.8cm}} \vspace{2mm} \\ 
            Hadamard-gate & $\pi/2$-phase gate \
        \end{tabular} \vspace*{0.3cm}

        \caption{\label{Gates}\small{Realization of 
              elementary quantum gates on the \QCns. Each square
              represents a lattice qubit. The 
    squares in the extreme left column marked with white circles
    denote the input qubits, those in the right-most column denote the
    output qubits.}}
    \end{center}
\end{figure}
A CNOT-gate can be realized on a cluster state of 15 qubits, as shown
in Fig.~\ref{Gates}. All measurements can be performed
simultaneously. The procedure to realize a CNOT-gate on a cluster with
15 qubits as displayed in Fig.~\ref{Gates} is
\begin{procedure}
    \label{CNTtproc}
    \em{
    Realization of a CNOT-gate acting on a two-qubit state
    $|\psi_\text{in}\rangle$. 
    \begin{enumerate}
        \item{Prepare the state\\
              $|\Psi_\text{in}\rangle_{{\cal{C}}_{15}} 
              = |\psi_\text{in}\rangle_{1,9} \otimes
              \left( \bigotimes_{i\in {{\cal{C}}_{15}}\backslash
                    \{1,9\} } |+\rangle_i \right)$.}
        \item{Entangle the 15 qubits of the cluster ${\cal{C}}_{15}$
              via the unitary operation $S^{({\cal{C}}_{15})}$.}
        \item{Measure all qubits of ${\cal{C}}_{15}$ except for the
              output qubits 7, 15 (following the labeling in
              Fig.~\ref{Gates}). The measurements can be performed 
              simultaneously. Qubits 1, 9, 10, 11, 13, 14 are measured in the
              $\sigma_x$-eigenbasis and qubits 2-6, 8, 12 in the
              $\sigma_y$-eigenbasis.}
    \end{enumerate}
    }    
\end{procedure}
Dependent on the measurement results, the following
gate is thereby realized:
\begin{equation}
    \label{CNOT}
    U^\prime_{CNOT} = U_{\Sigma,CNOT}\, CNOT(c,t).
\end{equation} 
Therein the byproduct operator $U_{\Sigma,CNOT}$ has the form
\begin{equation}
    \label{CNOTbyprop}
    \begin{array}{rcl}
    U_{\Sigma,CNOT}&=&{\sigma_x^{(c)}}^{\gamma_x^{(c)}}
       {\sigma_x^{(t)}}^{\gamma_x^{(t)}}
       {\sigma_z^{(c)}}^{\gamma_z^{(c)}}
       {\sigma_z^{(t)}}^{\gamma_z^{(t)}},\;\mbox{with}\\ & \\
       {\gamma_x^{(c)}}&=&{s}_2+{s}_3+{s}_5+{s}_6\\
       {\gamma_x^{(t)}}&=&{s}_2+{s}_3+{s}_8+{s}_{10}+{s}_{12}+{s}_{14}\\
       {\gamma_z^{(c)}}&=&{s}_1+{s}_3+{s}_4+{s}_5+{s}_8+{s}_9+{s}_{11}+1  \\
       {\gamma_z^{(t)}}&=&{s}_9+{s}_{11}+{s}_{13}.
    \end{array}
\end{equation}
Therein, the ${s}_i$ represent the measurement outcomes
$s_i$ on the 
qubits $i$. The expression (\ref{CNOTbyprop}) is modified if 
redundant cluster qubits are present  and/or if the cluster state on
which the CNOT gate is realized 
is specified by a set $\{\kappa_a\}$ different from (\ref{kappaS}), see
Section~\ref{redrem}. This concludes the 
presentation of the CNOT gate, the proof of its functioning is given
in Section~\ref{unifunk}. 

An arbitrary rotation $U_{Rot} \in SU(2)$ can be realized  on a
chain of 5 qubits. Consider a
rotation in its Euler representation 
\begin{equation}
    \label{Euler}
    U_{Rot}[\xi,\eta,\zeta] = U_x[\zeta]U_z[\eta]U_x[\xi],
\end{equation}
where the rotations about the $x$- and $z$-axis are 
\begin{equation}
\label{XZrots}
\begin{array}{rcl}
U_x[\alpha] &=&
\displaystyle{\mbox{exp}\left(-i\alpha\frac{\sigma_x}{2}\right)}
\vspace{1.5mm}\\ 
U_z[\alpha] &=& \displaystyle{\mbox{exp}\left(-i\alpha 
\frac{\sigma_z}{2}\right)}.
\end{array}
\end{equation}
Initially, the first qubit is prepared in some state
$|\psi_{\mbox{\footnotesize{in}}}\rangle$, which is to be rotated, and
the other qubits are prepared in 
$|+\rangle$. After the 5 qubits are entangled by the unitary transformation
$S$, the state $|\psi_{\mbox{\footnotesize{in}}}\rangle$ can be rotated by
measuring qubits 1 to 
4. At the same time, the state is also swapped to site 5. The qubits $1
\, .. \,4$ are measured in appropriately  chosen bases
\begin{equation}
    \label{Measbas}
    {\cal{B}}_j(\varphi_{j}) = \left\{
        \frac{|0\rangle_j+e^{i \varphi_{j}}
        |1\rangle_j}{\sqrt{2}} ,\, 
        \frac{|0\rangle_j-e^{i \varphi_{j}}
        |1\rangle_j}{\sqrt{2}}
    \right\},
\end{equation} 
whereby the measurement outcomes $s_{j}  \in \{ 0,1 \}$ for
$j=1\, .. \,  4$ are obtained. Here, $s_{j}=0$ means that
qubit $j$ is projected into the first state of
${\cal{B}}_j(\varphi_{j})$. In
 (\ref{Measbas}) the basis states of all possible measurement bases
lie on the equator of the Bloch sphere, i.e. on the intersection of
the Bloch sphere with the $x$-$y$-plane. Therefore, the measurement
basis for qubit $j$ can be specified by a single parameter, the
measurement angle $\varphi_{j}$. The
measurement direction of qubit $j$ is the vector on the Bloch sphere
which corresponds to the first state in the measurement basis
${\cal{B}}_j(\varphi_{j})$. Thus, the
measurement angle $\varphi_{j}$ is the angle between the measurement
direction at qubit $j$ and the positive $x$-axis. In summary,
the procedure to realize an arbitrary rotation $U_{Rot}[\xi,\eta,\zeta]$, 
specified by its Euler angles $\xi,\eta,\zeta$, is this:
\begin{procedure}
    \label{Rotproc}
    \em{
    Realization of general one-qubit rotations $U_{Rot} \in SU(2)$.
    \begin{enumerate}
        \item{Prepare the state
              $|\Psi_{\mbox{\footnotesize{in}}}\rangle_{{\cal{C}}_5} 
              = |\psi_{\mbox{\footnotesize{in}}}\rangle_1 \otimes
              \left( \bigotimes_{i=2}^5 |+\rangle_i \right)$.}
        \item{Entangle the five qubits of the cluster ${\cal{C}}_5$
              via the unitary operation $S^{({\cal{C}}_5)}$.}
        \item{\label{Mea}Measure qubits 1 - 4 in the following order and basis
              \begin{equation}
                  \label{genrot}
                  \begin{array}{rl}
                      \ref{Mea}.1 & \mbox{measure qubit 1 in} \;
                      {\cal{B}}_1(0)\\  
                      \ref{Mea}.2 & \mbox{measure qubit 2 in} \;
                      {\cal{B}}_2\left( -\xi\,(-1)^{{s}_1}
                      \right)\\
                      \ref{Mea}.3 & \mbox{measure qubit 3 in} \;
                      {\cal{B}}_3\left( 
                          -\eta\,(-1)^{{s}_2}  \right) \\  
                      \ref{Mea}.4 & \mbox{measure qubit 4 in} \;
                      {\cal{B}}_4\left( 
                          -\zeta\,(-1)^{{s}_1+{s}_3} \right)
                  \end{array}
            \end{equation}}
    \end{enumerate}
    }    
\end{procedure}
Via Procedure~\ref{Rotproc} the
rotation $U_{Rot}^\prime$ is realized:
\begin{equation}
    \label{Rotprime}
    U_{Rot}^\prime[\xi,\eta,\zeta] =  U_{\Sigma,Rot} \,U_{Rot}[\xi,\eta,\zeta].
\end{equation}
Therein, the random byproduct operator has the form
\begin{equation}
    \label{Byprod1}
    U_{\Sigma,Rot}=\sigma_x^{{s}_2+{s}_4}
    \sigma_z^{{s}_1+{s}_3}.   
\end{equation} 
It can be corrected for at the end of the computation, as will be explained in
Section~\ref{rand}.  

There is a subgroup of rotations for which the realization
procedure is somewhat simpler than Procedure~\ref{Rotproc}.  These
rotations form the subgroup of local operations in the Clifford group. The
Clifford group is the normalizer of the Pauli group. 

Among these rotations are, for example, 
the Hadamard gate and the $\pi/2$-phase gate. These gates can be
realized on a chain of 5 qubits in the following way:
\begin{procedure}
\label{Hadaproc}
    \em{
    Realization of a Hadamard- and $\pi/2$-phase gate.
    \begin{enumerate}
        \item{Prepare the state
              $|\Psi_{\mbox{\footnotesize{in}}}\rangle_{{\cal{C}}_5} 
              = |\psi_{\mbox{\footnotesize{in}}}\rangle_1 \otimes
              \left( \bigotimes_{i=2}^5 |+\rangle_i \right)$.}
        \item{Entangle the five qubits of the cluster ${\cal{C}}_5$
              via the unitary operation $S^{({\cal{C}}_5)}$.}
        \item{Measure qubits 1 - 4. This can be done
              simultaneously. For the Hadamard 
              gate, measure individually the observables $\sigma_x^{(1)}$,
              $\sigma_y^{(2)}$, $\sigma_y^{(3)}$,
              $\sigma_y^{(4)}$. For the $\pi/2$-phase gate measure 
              $\sigma_x^{(1)}$, 
              $\sigma_x^{(2)}$, $\sigma_y^{(3)}$, $\sigma_x^{(4)}$.}
    \end{enumerate}
    }    
\end{procedure}  
The difference with respect to Procedure~\ref{Rotproc} for general
rotations is that in Procedure~\ref{Hadaproc} no measurement bases
need to be adjusted according to previous measurement
results and therefore the measurements can all be performed at the
same time.

As in the cases before, the Hadamard- and the $\pi/2$-phase gate are
performed only modulo a subsequent byproduct operator which is
determined by the random measurement outcomes ${s}_k$
\begin{equation}
    \label{Hadabyprop}
    \begin{array}{rcl}
          U_{\Sigma,H} &=&
          \sigma_x^{{s}_1+{s}_3+{s}_4}\,
          \sigma_z^{{s}_2+{s}_3}\\ 
          U_{\Sigma,U_z(\pi/2)} &=&
          \sigma_x^{{s}_2+{s}_4}\,
          \sigma_z^{{s}_1+{s}_2+{s}_3+1}.
    \end{array}      
\end{equation}

Before we explain the functioning of the above gates, we would like to address 
the following questions: First,``How does one manage to occupy only
those lattice sites with cluster qubits that are required for a
particular circuit but leaves the remaining ones empty?''. The answer
to this question is that redundant qubits will not have to be removed
physically. It is sufficient to measure each of them in the
$\sigma_z$-eigenbasis, as will be described in Section~\ref{redrem}.

Second, ``How can the described procedures for gate simulation
be concatenated such that they represent a measurement based
simulation of an entire circuit?''. It seems at first sight that the
described building 
blocks would only lead to a computational scheme consisting of
repeated steps of entangling operations and measurements. This is not
the case. As will be shown in Section~\ref{connect}, the three
procedures stated are precisely of such a form that the described
measurement-based  scheme of quantum computation can be decomposed into them. 

The third question is: ``How does one deal with the randomness of the
measurement results that leads to the byproduct operators
(\ref{CNOTbyprop}), (\ref{Byprod1}) and (\ref{Hadabyprop})?''. The
appearance of byproduct operators may suggest that there 
is a need for local correction operations to counteract these unwanted extra
operators. However, there is neither a possibility for such counter
rotations within the described model of quantum computation, nor is
there a need. The scheme works with unit efficiency despite the
randomness of the individual measurement results, as will be discussed
in Section~\ref{rand}.

\subsection{Removing the redundant cluster qubits}
\label{redrem}

A cluster state on a two-dimensional cluster of rectangular shape,
say, is a resource that allows for any computation
that fits on the cluster. If one realizes a certain quantum 
circuit on this cluster state, there will always be qubits on the
cluster which are not needed for its realization.
Such cluster qubits we call redundant for this
particular circuit. 

In the description of the \QC as a quantum logic network, the first
step of each computation will be to remove these redundant cluster
qubits. Fortunately, the situation is not such that we have to remove
the qubits (or, more precisely, the carriers of the qubits) physically
from the lattice. To make them ineffective to the realized circuit,
it suffices to measure each of them in the $\sigma_z$-eigenbasis. In
this way, one is left with an entangled quantum state on the cluster
${\cal{C}}_N$ of the unmeasured qubits and a product state on
${\cal{C}}\backslash {\cal{C}}_N$,
\begin{equation}
    \label{redmeas}
    \csk \longrightarrow  |Z\rangle_{{\cal{C}} \backslash {\cal{C}}_N}
    \otimes \cskNpr, 
\end{equation}
with $|Z\rangle_{{\cal{C}} \backslash {\cal{C}}_N} =
    \left(\bigotimes_{ i \in 
    {\cal{C}}\backslash {\cal{C}}_N} |s_i \rangle_{i,z}
    \right)$ and $s_i$ the results of the
$\sigma_z$-measurements. The resulting entangled state $\cskNpr$ on the
sub-cluster ${\cal{C}}_N$ is again a cluster state obeying the set of
equations (\ref{EVeqn}). This can be seen as follows. First, by
    definition we have 
\begin{equation}
    \label{CNcs1}
    \begin{array}{l}
        \ds{|Z\rangle_{{\cal{C}} \backslash {\cal{C}}_N}
    \!\otimes\! \cskNpr}\! = \!\left( \bigotimes_{i \in {\cal{C}}
    \backslash {\cal{C}}_N} \frac{1+(-1)^{s_i}
    \sigma_z^{(i)}} {2} \right) \! \csk.
    \end{array}
\end{equation}
Using the eigenvalue equations (\ref{EVeqn}), we now insert a
correlation operator $K^{(a)}$ with $a\in
{\cal{C}}_N$ into the r.h.s of (\ref{CNcs1}) between the projector and
the state, and obtain
\begin{equation}
    \label{CNcs2}
        \ds{|Z\rangle_{{\cal{C}} \backslash {\cal{C}}_N}
          \otimes \cskNpr} = (-1)^{\kappa_a^\prime}
        K^{\prime(a)} |Z\rangle_{{\cal{C}} \backslash
          {\cal{C}}_N} \otimes \cskNpr, 
\end{equation}  
with the correlation operators
\begin{equation}
    \label{BasCorrN}
    K^{\prime(a)} = \sigma_x^{(a)}
    \!\!\!\!\bigotimes\limits_{c \in 
    \mbox{\footnotesize{nbgh}}(a) \cap {\cal{C}}_N}
    \!\!\!\!\!\sigma_z^{(c)}, 
\end{equation}
and the set $\{ \kappa_a^\prime \}$ specifying the eigenvalues 
\begin{equation}
    \label{kappaN}
    \kappa^\prime_a = \left( \kappa_a+\sum\limits_{b \in
    \mbox{\footnotesize{nbgh}}(a) \cap ({\cal{C}}\backslash
    {\cal{C}}_N)} s_b \right) \; \mbox{mod} \,\,2 . 
\end{equation}
As the new correlation operators $K^{\prime(a)}$ in
(\ref{CNcs2}) only act on 
the cluster qubits in ${\cal{C}}_N$, the states $\cskNpr$ again obey
eigenvalue equations of 
type (\ref{EVeqn}), i.e.
\begin{equation}
    \label{EVeqnN}
    K^{\prime(a)} \cskNpr = {(-1)}^{\kappa_a^\prime}
    \cskNpr, \;\forall \,\, a \in {\cal{C}}_N.
\end{equation}
There are $|{\cal{C}}_N|$ such eigenvalue equations for a state of
$|{\cal{C}}_N|$ qubits. Thus, the state $\cskNpr$ is specified by
(\ref{EVeqnN}) up to a global phase.

From (\ref{kappaN}) we find that the redundant qubits have some
remaining influence
on the process of computation. After they have been measured,
the random measurement results enter into the eigenvalues that specify the
residual cluster state 
$\cskNpr$ on the cluster ${\cal{C}}_N$. However, any cluster state
$\cskNpr$ is equally good for computation as 
stated in Section~\ref{correl}. 
From (\ref{Eigenvalmod}) it follows that
\begin{equation}
    \label{reminfl}
    \cskNpr = \bigotimes_{a \in {\cal{C}}_N}\left(
    \sigma_z^{(a)}\right)^{\kappa_a^\prime} |\phi\rangle_{{\cal{C}}_N}.
\end{equation}
The Pauli phase flip operators that appear on the r.h.s. of equation
(\ref{reminfl}) may be absorbed into the subsequent measurements.
This allows us to adopt the following two rules in the further discussion
\begin{equation}
    \label{simpl}
    \parbox{0.85\linewidth}{
      \begin{enumerate}
      \item{\em{The redundant cluster qubits are discarded. We only consider
      the sub-cluster \mbox{${\cal{C}}_N$}.}}
      \item{\em{We assume that
      \mbox{$\kappa^\prime_a=0$} for all 
      \mbox{$a \in {\cal{C}}_N$}.}}
      \end{enumerate}}
\end{equation}
This reduction will make a number of expressions such as
those for the byproduct operators more transparent and it will also simplify
the remaining part of the universality proof.

\subsection{Concatenation of gate simulations}
\label{connect}

A quantum circuit on the \QC is a spatial and temporal pattern of
measurements on individual qubits which have previously been entangled
to form a cluster state. To better understand its functioning we would
like --as in the network model of quantum computation-- to decompose
the circuit into basic building blocks. These building blocks should
be such that out of them any circuit can be assembled. In explaining
the \QC in a network language, we can relate the building
blocks of a quantum logic network --the quantum gates-- to building
blocks of \QCns-circuits.

The fact that quantum gates can be combined to quantum logic networks
is obvious. But 
the statement that, for a \QCns-computation, measurement patterns which
simulate gates 
can simply be patched together to give the measurement pattern for
the whole circuit requires a proof. This proof is given next. 

We begin by stating the general form of the procedures to realize 
gates and sub-circuits. The reason why these procedures work is explained in 
subsequent sections.
To realize a gate $g$ on the \QC consider
a cluster ${\cal{C}}(g)$. This cluster has an input section
${\cal{C}}_I(g)$, a body ${\cal{C}}_M(g)$ and an output section
${\cal{C}}_O(g)$, with 
\begin{equation}
    \label{subsetrel}
    \begin{array}{c}
        \ds{{\cal{C}}_I(g) \cup {\cal{C}}_M(g) \cup {\cal{C}}_O(g)=
          {\cal{C}}(g)}\vspace{1mm}\\
        \begin{array}{rcl}
            \ds{{\cal{C}}_I(g)\cap{\cal{C}}_M(g)} &=& \emptyset \vspace{1mm}\\
            \ds{{\cal{C}}_I(g)\cap{\cal{C}}_O(g)} &=& \emptyset \vspace{1mm}\\
            \ds{{\cal{C}}_M(g)\cap{\cal{C}}_O(g)} &=& \emptyset. 
        \end{array}
    \end{array}
\end{equation}  
The measurement bases of the qubits in ${\cal{C}}_M(g)$, the body of
the gate $g$, encode $g$.
The general scheme for procedures to realize a gate $g$ on a cluster
${\cal{C}}(g)$ is 
\begin{scheme}
    \label{gatescheme}
    {\em{Simulation of the gate $g$ on ${\cal{C}}(g)$, acting on the
    input state $|\psi\rangle_\text{in}$.
        \begin{enumerate}
            \item{Prepare the input state $|\psi_{\text{in}} \rangle$ on
                  ${\cal{C}}_I(g)$ and the qubits in ${\cal{C}}_M(g)
                \cup {\cal{C}}_O(g)$ individually in the state
                $|+\rangle=|0\rangle_x$ such that the quantum state of
                  all qubits 
                in ${\cal{C}}(g)$ becomes
                \begin{equation}
                    |\Psi_{\text{in}}\rangle_{{\cal{C}}(g)} = |\psi_{\text{in}}
                     \rangle_{{\cal{C}}_I(g)} \otimes \!\!\!\!
                     \bigotimes_{k \in 
                     {\cal{C}}_M(g) 
                     \cup {\cal{C}}_O(g)} \!\!\!\! |+\rangle_k.
                \end{equation}}
          \item{Entangle $|\Psi_{\text{in}}\rangle_{{\cal{C}}(g)}$ by the
                  interaction
                  \begin{equation}
                      S^{\left({\cal{C}}(g)\right)} =
                      \prod_{a, b \in
                              {\cal{C}}(g)|\,
                              b - a \in \gamma_d}
                        \Sab,
                  \end{equation}
                  such that the resulting quantum state is
                      $|\Psi_\varepsilon\rangle_{{\cal{C}}_N} =
                      S^{\left({\cal{C}}(g)\right)}
                      |\Psi_{\text{in}}\rangle_{{\cal{C}}(g)}$.
                      } 
          \item{Measure the cluster qubits in ${\cal{C}}_I(g)
                \cup {\cal{C}}_M(g)$, i.e. choose measurement bases
                specified by $\vec{r}_k \in S^2, \,\,k \in
                {\cal{C}}_I(g) \cup {\cal{C}}_M(g)$ and obtain
                the random measurement results $s_k$ such that the projector  
                \begin{equation}
                    P^{({\cal{C}}_I(g) \cup {\cal{C}}_M(g))} =
                  \!\!\!\!\bigotimes_{k \in  {\cal{C}}_I(g) \cup 
                  {\cal{C}}_M(g)}
                  \!\!\!\!\!\!\!\!\frac{1+{(-1)}^{s_k} 
                  \vec{r}_k \cdot \vec{\sigma}^{(k)}}{2}
                \end{equation}
                is applied. The resulting state is
                $|\Psi_{\text{out}}\rangle_{{\cal{C}}_N} =
                P^{({\cal{C}}_I(g) \cup {\cal{C}}_M(g))} | \Psi_\varepsilon
                \rangle_{{\cal{C}}_N}$.}   
        \end{enumerate} 
        }}    
\end{scheme}    
Putting all three steps of Scheme~\ref{gatescheme} together, the
relation between $|\Psi_{\text{in}}\rangle_{{\cal{C}}_N}$ and
$|\Psi_{\text{out}}\rangle_{{\cal{C}}_N}$ is 
\begin{equation}
    |\Psi_{\text{out}}\rangle_{{\cal{C}}_N} =
    P^{({\cal{C}}_I(g) \cup {\cal{C}}_M(g))} \,S^{\left({\cal{C}}(g)\right)} 
     \,|\Psi_{\text{in}}\rangle_{{\cal{C}}_N}. 
\end{equation}
As we will show later, the state
$|\Psi_{\text{out}}\rangle_{{\cal{C}}_N}$ has the form
\begin{equation}
    \label{PSIout}
    |\Psi_{\text{out}}\rangle_{{\cal{C}}_N}= \left(
        \bigotimes_{k \in {\cal{C}}_I(g) \cup
        {\cal{C}}_M(g)} \!\!\!\!\!\!\!|s_k
        \rangle_{k,\vec{r}_k}\right) \otimes  
     |\psi_{\text{out}}\rangle_{{\cal{C}}_O(g)}, 
\end{equation}
where $|s_k \rangle_{k,\vec{r}_k}$ denotes
the state of the qubit $k$ after the observable
$\vec{r}_k\cdot \vec{\sigma}^{(k)}$ has been measured
and the measurement outcome was $s_k$, and
\begin{equation}
    \label{psiout}
    |\psi_{\text{out}}\rangle = U_{\Sigma,g} U_g |\psi_{\text{in}} \rangle.
\end{equation}
Therein, $U_g$ is the desired unitary operation and $U_{\Sigma,g}$ an
extra multi-local rotation that depends on the measurement results $\big\{
s_k\, |\,\,k \in {\cal{C}}_I(g) \cup
{\cal{C}}_M(g) \big\}$. The extra rotations $U_{\Sigma,g}$ are always
in the Pauli group, i.e. 
\begin{equation}
    \label{Pauli}
    U_{\Sigma,g} = \bigotimes_{i=1}^n \left(\sigma_x^{[i]}
    \right)^{x_i} \left(\sigma_z^{[i]}
    \right)^{z_i}  
\end{equation} 
modulo a possible global phase, and $n=|I|=|O|$. In
(\ref{Pauli}) the $\sigma^{[i]}$ 
denote Pauli operators acting on the {\em{logical}} qubit $i$, not
cluster qubit. The values $x_i,z_i \in \{0,1\}$ are computed from the
measurement outcomes $\big\{s_k\,|\,\, k \in
{\cal{C}}_I(g) \cup {\cal{C}}_M(g)\big\}$.

We now have all prerequisites at hand to explain why measurement
patterns of basic gates can be combined to form the measurement
pattern of the whole circuit which is obtained by combining the gates.
The scheme of quantum computation with the \QC consists of a single
entangling operation which creates the resource cluster state and,
subsequently, of a series of one-qubit measurements on that state. We want
to view the measurement pattern of a quantum circuit as being composed
of basic blocks from whose 
function the function of the whole circuit can be deduced. To do so,
we will explain computation on the \QC as a sequential process of
performing the circuit gate by gate. Then we have to demonstrate that the
computational scheme as it is practically carried out, i.e. entangle
once and afterwards only measure, and the sequential scheme that we
use to explain the functioning of the circuit are mathematically
equivalent. 

The sequential scheme is this. Consider a circuit
$U=\prod_{i=1}^{|{\cal{N}}|} U_{g_i}$ that consists of a succession of gates
$g_1, .. , g_{|{\cal{N}}|} \in {\cal{N}}$ applied to some input state
$|\psi_{\text{in}}\rangle$, leading to an output state
$|\psi_{\text{out}}\rangle=U|\psi_{\text{in}}\rangle$ which is then
measured. ${\cal{N}}$ denotes the network, i.e. the set of gates plus a
description of their relation. For simplicity let us first assume that
each gate $g_i$ acts on all of the logical qubits. Subsequently we
will drop this assumption. 

First, a quantum
state $|\psi_{\text{in}}\rangle_I \otimes |+\rangle_{{\cal{C}}_N
  \backslash I}$ is prepared. Then each gate, one after another, is
realized on a sub-cluster ${\cal{C}}(g) \subset {\cal{C}}_N$
according to Scheme~\ref{gatescheme}. Finally, the output is measured
as usual.

In the step of carrying out the gate $g_i$ the state of the quantum
register is, besides being processed, also teleported from
${\cal{C}}_I(g_i)={\cal{C}}_O(g_{i-1})$ to
${\cal{C}}_O(g_i)={\cal{C}}_I(g_{i+1})$. In this way, by carrying out
the gate $g_i$ the input for the successor gate $g_{i+1}$ is
provided. To proceed with the realization of gate $g_{i+1}$, in
accordance with Scheme~\ref{gatescheme}, the sub-cluster
${\cal{C}}(g_{i+1})$ is entangled via $S^{({\cal{C}}(g_{i+1}))}$and
subsequently the cluster qubits in ${\cal{C}}_I(g_{i+1}) \cup
  {\cal{C}}_M(g_{i+1})$ are measured. This completes the realization
of gate $g_{i+1}$ and at the same time writes the input for $g_{i+2}$, and
so on.

The reason why the sequential scheme just described is equivalent to
the entangle-once-and-then-only-measure scheme is the following. The
entanglement 
operations at the various stages of the sequential scheme
commute with all the measurements carried out earlier. This holds
because both operations, entangling operation and earlier measurement, act on
different particles. Thus, the operations may be reordered in such a
way that in a first step all entangling operations
$S^{({\cal{C}}(g_i))}$ act on the initial state and afterwards all the
measurements are performed.

\begin{figure}
    \begin{center}
        \epsfig{file=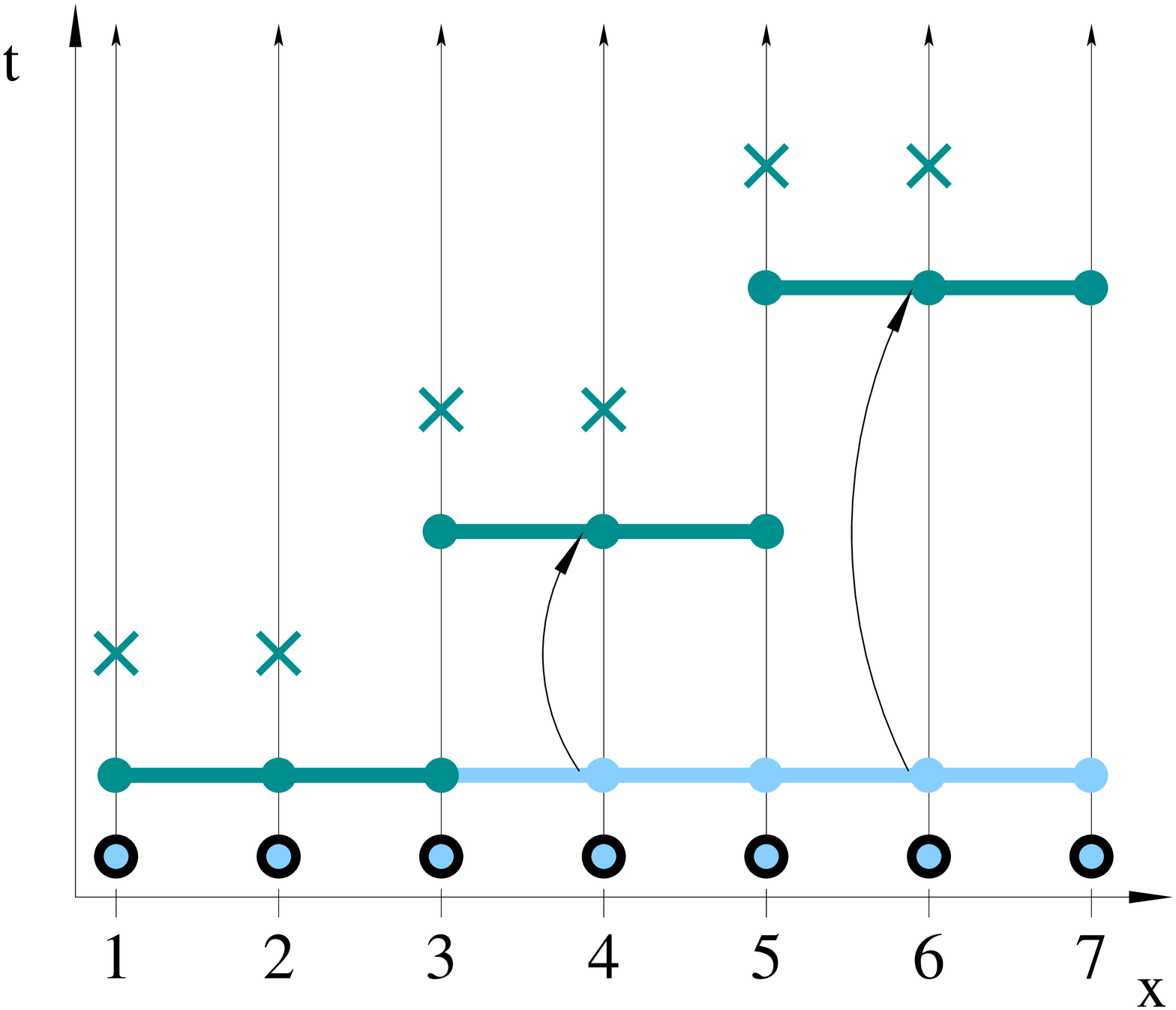, width=7.5cm}
        \caption{\label{IsingWWprop}Here the exchange of the order of
          the measurements and the entanglement operations is
          shown. The crosses ``$\times$'' denote the one-qubit
          measurements and the horizontal lines between adjacent
          cluster qubits denote the unitary transformations
          $S^{a,a+1}$.}
    \end{center}
\end{figure}

The exchange of the order of the one-particle measurements and the
two-particle Ising interactions is shown in Fig.~\ref{IsingWWprop}
for a 1D cluster. In one dimension the decomposition of a
cluster into sub-clusters, as displayed in Fig.~\ref{IsingWWprop}, is
clear. However, the interesting cases for \QCns-computations are
clusters in 2D and 3D; and there we must state more precisely what
``decomposition of a cluster into sub-clusters'' means. 
The use of basic notions from graph theory will prove helpful for this
purpose. 

To decompose a cluster into sub-clusters means in more precise terms
to decompose the associated graph $G({\cal{C}}_N,E_{{\cal{C}}_N})$
into subgraphs. That is, we have to decompose both the vertices and
the edges of the graph. Each vertex $a \in {\cal{C}}_N$ has to belong
to a subset ${\cal{C}}(g_i)$, where
\begin{equation}
    \label{vertexunion}
    {\cal{C}}_N= \bigcup_{i=1}^{|{\cal{N}}|} {\cal{C}}(g_i),
\end{equation}
and the sets ${\cal{C}}(g_i)$ of vertices corresponding to the gates
$g_i$ may overlap on their input- and output vertices.  

Correspondingly, the set $E_{{\cal{C}}_N}$ of edges, defined in the
same way as $E_{\cal{C}}$ in
(\ref{clusteredges}), is decomposed into subsets
\begin{equation}
    \label{edgeunion}
    E_{{\cal{C}}_N}= \bigcup_{i=1}^{|{\cal{N}}|} E(g_i),
\end{equation}
but the subsets $E(g_i)$ of edges are not allowed to overlap,
\begin{equation}
    \label{edgesdisjoint}
    \forall i,j=1\,..\,|{\cal{N}}|, i\neq j:\; E(g_i)\cap E(g_j)=\emptyset. 
\end{equation}
The rules for the decomposition of edges (\ref{edgeunion}) and
(\ref{edgesdisjoint}) are, as we shall see, central for the universality proof.

Further, for the decomposition to be useful, the subsets ${\cal{C}}(g_i)$
and $E(g_i)$ must fulfill a number of constraints. The first
of these is that each pair $({\cal{C}}(g_i),E(g_i))$ is again
a graph, $G({\cal{C}}(g_i),E(g_i))$. This requires, in
particular, that the endpoints of all the edges in $E(g_i)$
are in ${\cal{C}}(g_i)$,
\begin{equation}
    \label{subgraphcond}
    \forall i=1,..,|{\cal{N}}|: E(g_i) \subset {\cal{C}}(g_i) \times
    {\cal{C}}(g_i).
\end{equation}
For details on the graph decomposition, in particular for
conditions on the subgraphs imposed to guarantee
(\ref{edgeunion}) and (\ref{edgesdisjoint}) see Appendix~\ref{decomp}.

Now consider
the concatenation $g_2 \circ g_1$ of the two gates $g_1$, realized on
a cluster ${\cal{C}}(g_1)$, and $g_2$, realized on a cluster
${\cal{C}}(g_2)$, each of them by a procedure according to
scheme~\ref{gatescheme}. The composite circuit $g=g_2 \circ g_1$ is
realized on the cluster
\begin{equation}
    \label{combclust}
    {\cal{C}}(g) = {\cal{C}}(g_1) \cup {\cal{C}}(g_2),
\end{equation}
with
\begin{equation}
    \label{combclustprop}
    \begin{array}{rcl}
        \ds{{\cal{C}}_I(g)} &=& \ds{{\cal{C}}_I(g_1) \cup
            \left({\cal{C}}_I(g_2) 
            \backslash {\cal{C}}_O(g_1) \right)}\vspace{0.1cm}\\
        \ds{{\cal{C}}_O(g)} &=& \ds{{\cal{C}}_O(g_2) \cup
            \left({\cal{C}}_O(g_1) 
            \backslash {\cal{C}}_I(g_2) \right)}.
    \end{array} 
\end{equation}

Now, the procedure to perform the two gates $g_1$, $g_2$ sequentially
is
\begin{enumerate}
    \item{Prepare the state $|\Psi_{\text{in}}\rangle_{{\cal{C}}(g)}=
        |\psi_{\text{in}} \rangle_{{\cal{C}}_I(g)} \otimes
        |+\rangle_{{\cal{C}}(g)\backslash {{\cal{C}}_I(g)}}$.}
    \item{Entangle the qubits on the sub-cluster ${\cal{C}}(g_1)$ via
          \begin{equation}
              \label{S1}
              S_1:= S^{({\cal{C}}(g_1))}= \prod_{a,b \in
              {\cal{C}}(g_1)|\,b - a \in \gamma_d} \!\!\!\!\Sab.
          \end{equation}
          }
        \item{Measure the qubits in ${\cal{C}}_I(g_1) \cup
              {\cal{C}}_M(g_1)$, resulting in the projector
              $P_{{\cal{C}}_I(g_1) \cup {\cal{C}}_M(g_1)} =:P_1$,
          \begin{equation}
              \label{Proj_g1}
              P_1=
              \bigotimes_{k \in 
              {\cal{C}}_I(g_1) \cup 
              {\cal{C}}_M(g_1)} \!\!\!\!\!\!\frac{1+(-1)^{s_k}
              \vec{r}_k \cdot \vec{\sigma}^{(k)}}{2} . 
        \end{equation}
        Therein $s_k$ is the outcome of the measurement
          of qubit $k$ and $\vec{r}_k$ the respective
          measurement direction.
          }
        \item{Entangle the qubits on the sub-cluster ${\cal{C}}(g_2)$ via
          \begin{equation}
              \label{S2}
              S_2:= S^{({\cal{C}}(g_2))}= \prod_{a,b \in
              {\cal{C}}(g_2)|\,b - a \in \gamma_d} \!\!\!\!\Sab.
          \end{equation}
          }
        \item{Measure the qubits in ${\cal{C}}_I(g_2) \cup
              {\cal{C}}_M(g_2)$, resulting in the projector
              $P_{{\cal{C}}_I(g_2) \cup {\cal{C}}_M(g_2)} =:P_2$,
          \begin{equation}
              \label{Proj_g2}
              P_2=
              \bigotimes_{k \in 
              {\cal{C}}_I(g_2) \cup 
              {\cal{C}}_M(g_2)} \!\!\!\!\!\!\frac{1+(-1)^{s_k}
              \vec{r}_k \cdot \vec{\sigma}^{(k)}}{2} . 
        \end{equation}
        }
\end{enumerate}
The procedure results in an output state
\begin{equation}
    \label{Psiout}
    |\Psi_{\text{out}} \rangle_{{\cal{C}}(g)} = P_2\,S_2\, P_1 \, S_1
     |\Psi_{\text{in}} \rangle_{{\cal{C}}(g)}
\end{equation}
that has the form
\begin{equation}
    \label{Psiout2}
    |\Psi_{\text{out}} \rangle_{{\cal{C}}(g)} = \left(
              \bigotimes_{{\cal{C}}_I(g) \cup {\cal{C}}_M(g)}
              \!\!\!\!\! |
              s_k \rangle_{k,\vec{r}_k}
              \right) \otimes |\psi_{\text{out}}\rangle_{{\cal{C}}_O(g)},
\end{equation} 
with
\begin{equation}
    \label{psi_inoutrel}
    |\psi_{\text{out}} \rangle = U_{\Sigma,g_2}\, U_{g_2} U_{\Sigma,g_1}\,
     U_{g_1} |\psi_{\text{in}} \rangle, 
\end{equation}
according to (\ref{psiout}).

As will be shown next, the above procedure is equivalent to a
procedure of Scheme~\ref{gatescheme} applied to the cluster
${\cal{C}}(g) = {\cal{C}}(g_1) \cup {\cal{C}}(g_2)$, i.e. when, first,
all qubits in ${\cal{C}}(g)$ are entangled and, 
second, all but the output qubits of ${\cal{C}}_O(g)$  are measured.

The procedure according to Scheme~\ref{gatescheme} yields the state 
\begin{equation}
    \label{Psiprimeout}
    |\Psi^\prime_{\text{out}}\rangle_{{\cal{C}}(g)} = P^{({\cal{C}}_I(g)
     \cup {\cal{C}}_M(g))} \, S^{({\cal{C}}(g))} \,|\Psi_{\text{in}}
     \rangle_{{\cal{C}}(g)}, 
\end{equation}
and we now have to show that the output states
     $|\Psi_{\text{out}}\rangle_{{\cal{C}}(g)}$ in (\ref{Psiout}) and
     $|\Psi^\prime_{\text{out}}\rangle_{{\cal{C}}(g)}$ in
     (\ref{Psiprimeout}) are the same for all input states $|\Psi_{\text{in}}
     \rangle_{{\cal{C}}(g)}$. 

First note that the operations $P_1$ and $S_2$ commute since they act
on different particles. $P_1$ acts on the qubits in ${\cal{C}}_I(g_1)
\cup {\cal{C}}_M(g_1)$ while $S_2$ acts on ${\cal{C}}(g_2)$. The sub-clusters
associated with the gates 
may overlap only via their input- and output qubits. This is
intuitively clear, and also follows from the decomposition constraint
(\ref{disconnected}). As the gate $g_1$
is applied before $g_2$, of ${\cal{C}}(g_1)$ only the qubits in
${\cal{C}}_O(g_1)$ may overlap with the qubits in ${\cal{C}}_I(g_2)$.
    Thus, $( {\cal{C}}_I(g_1)
\cup {\cal{C}}_M(g_1) ) \cap {\cal{C}}(g_2) = \emptyset$. Therefore
\begin{equation}
    \label{changeorder}
    P_2 \,S_2 \,P_1 \,S_1 = P_2\, P_1 \, S_2 \, S_1. 
\end{equation} 
Now note that as a direct consequence of (\ref{combclustprop}) the union of the
input- and body section of the composite gate $g$ on the cluster
${\cal{C}}(g)$ are made up by the union of the input- and body
sections of the two individual gates $g_1$ and $g_2$, i.e. 
\begin{equation}
    \label{ccP1}
    {\cal{C}}_I(g_1) \cup {\cal{C}}_M(g_1) \cup {\cal{C}}_I(g_2) \cup
    {\cal{C}}_M(g_2) = {\cal{C}}_I(g) \cup {\cal{C}}_M(g). 
\end{equation}
Further, from the decomposition constraint (\ref{disconnected}) and
from the fact that $g_1$ is applied before $g_2$ it follows that the input- and
body sections of gates $g_1$ and $g_2$ do not intersect,
\begin{equation}
    \label{ccP2}
    \left({\cal{C}}_I(g_1) \cup {\cal{C}}_M(g_1)\right) \cap
    \left({\cal{C}}_I(g_2) \cup 
    {\cal{C}}_M(g_2)\right) = \emptyset. 
\end{equation}
Therefore,
\begin{equation}
    \label{Pmerge}
    \begin{array}{rcl}
    P_2\,P_1 &=& \ds{\bigotimes \limits_{{\cal{C}}_I(g_2) \cup
    {\cal{C}}_M(g_2)} \!\!\!\!\!\!\!
    \frac{1+ (-1)^{s_k} \vec{r}_k \cdot
    \vec{\sigma}^{(k)} }{2}} \\
    & & \mbox{ } \;\;\;\;\;\ds{\bigotimes \limits_{{\cal{C}}_I(g_1) \cup
    {\cal{C}}_M(g_1)} \!\!\!\!\!\!\!
    \frac{1+ (-1)^{s_k} \vec{r}_k \cdot
    \vec{\sigma}^{(k)} }{2}} \\
    &=& \ds{\bigotimes \limits_{{\cal{C}}_I(g) \cup
    {\cal{C}}_M(g)} \!\!\!\!\!\!\!
    \frac{1+ (-1)^{s_k} \vec{r}_k \cdot
    \vec{\sigma}^{(k)} }{2}} \\
    &=& \ds{P^{({\cal{C}}_I(g) \cup {\cal{C}}_M(g))},} 
    \end{array}
\end{equation}
where the second line holds by (\ref{ccP1}) and (\ref{ccP2}). We find
that measurement patterns corresponding to the projections $P_1$ and
$P_2$ can be patched together to form the measurement pattern on the
cluster ${\cal{C}}(g)$.

The same holds for the entangling operations. The entangling operation
$S_1$ on ${\cal{C}}(g_1)$ and $S_2$ on ${\cal{C}}(g_2)$ combined give
the entangling operation $S^{({\cal{C}}(g))}$ on ${\cal{C}}(g)$,
\begin{equation}
    \label{Smerge}
        S_2\,S_1 = S^{({\cal{C}}(g))},   
\end{equation} 
because of the central rule (\ref{edgesdisjoint}). 

Inserting (\ref{Pmerge}) and (\ref{Smerge}) into (\ref{changeorder}) yields
\begin{equation}
    \label{changeordercompr}
    P_2\, S_2 \, P_1 \, S_1 = P^{({\cal{C}}_I(g) \cup {\cal{C}}_M(g))}
    \, S^{({\cal{C}}(g))}, 
\end{equation}
and therefore, if we compare (\ref{Psiout}) and (\ref{Psiprimeout}) we
find that $|\Psi_{\text{out}}\rangle_{{\cal{C}}(g)}$, the output state
of the sequential realization of the two gates $g_1$ and $g_2$, and
$|\Psi^\prime_{\text{out}}\rangle_{{\cal{C}}(g)}$, the output state of
the standard procedure applied to the composite circuit, are indeed the
same for all inputs $|\Psi_{\text{in}}\rangle_{{\cal{C}}(g)}$. Thus
both realizations, the sequential and the non-sequential, are
equivalent. 

This composition can be iterated so that the entire circuit can be
realized via the standard procedure of Scheme~\ref{gatescheme}. The
measurement pattern of the circuit is thereby obtained by patching
together the measurement patterns of the gates the circuit is composed
of.  

From (\ref{psi_inoutrel}) it follows that the quantum input
$|\psi_{\text{in}}\rangle$ and the quantum output
$|\psi_{\text{out}}\rangle$ of the unitary evolution are related via
\begin{equation}
    \label{Gateseq}
    |\psi_{\text{out}}\rangle = \left( \prod_{i=1}^{|{\cal{N}}|}
    U_{\Sigma,g_i} \, U_{g_i} \right)
     |\psi_{\text{in}}\rangle.
\end{equation}
The random but known byproduct operators $U_{\Sigma,g_i}$ that appear
in (\ref{Gateseq}) are dealt with in Section~\ref{rand}. The gates
$g_i \in {\cal{N}}$ are labeled corresponding to the order of their
action. 

Now, we want to specify to the case where the quantum input is
{\em{known}} and where the quantum output is measured. This is the
situation which interests us most in this paper. Examples of such a
situation are Shor's factoring algorithm and Grover's
search algorithm. In both cases, the quantum input is
$|\psi_{\text{in}}\rangle = \bigotimes_{i=1}^n |+\rangle_i$.

Let us denote the input section of the whole
cluster ${\cal{C}}$, comprising the input qubits of the network simulation,
as $I$; and the output section, comprising the qubits of the readout
quantum register, as $O$. As long as the quantum input is known it is
sufficient to consider the 
state $|+\rangle_I = \bigotimes_{i \in I} |+\rangle_i$. For different
but known input states $|\psi_{\text{in}}\rangle_I$ 
one can always find a transformation $U_{\text{in}}$ such that
$|\psi_{\text{in}}\rangle_I =  U_{\text{in}} |+\rangle_I$ and instead
of realizing some unitary transformation $U$ on
$|\psi_{\text{in}}\rangle_I$ one realizes $U\, U_{\text{in}}$ on
$|+\rangle_I$. 

Preparing an input state $|+\rangle_I$ and entangling it via
$S^{({\cal{C}})}$ is the same as creating a cluster state
$|\phi\rangle_{\cal{C}}$, 
$S^{({\cal{C}})} \,|+\rangle_I \otimes |+\rangle_{{\cal{C}} \backslash
  I} = S^{({\cal{C}})} \,|+\rangle_{\cal{C}} = \cs$. This holds
because the state $ S^{({\cal{C}})}
\,|+\rangle_{\cal{C}}$ obeys the 
eigenvalue equations (\ref{EVeqn}) and, as we have stated earlier,
these eigenvalue equations determine the state completely. Thus the
created state is a cluster state $\cs$ that could as well have been
prepared by any other means. 

Once the quantum output is read then all
cluster qubits have been measured. Therefore, the entire procedure of
realizing a quantum computation on the \QC amounts to
\begin{scheme}
    \label{QCC-comp}
    {\em{Performing a computation on the \QCns.
        \begin{enumerate}
            \item{Prepare a cluster state $\csk$ of sufficient size.}
            \item{Perform a sequence of measurements on $\csk$ and
                  obtain the result of the computation from all the
                  measurement outcomes.}
        \end{enumerate}

    }}
\end{scheme} 

The link between the network model and the
\QC is established by Scheme~\ref{gatescheme}. The elementary
constituents of the quantum logic network are mapped onto the
corresponding basic blocks 
of the \QCns.  In this way, Scheme~\ref{gatescheme} helps to
understand why the \QC  
works. 

However, a \QCns-computation is more appropriately 
described by Scheme~\ref{QCC-comp} than by Scheme~\ref{gatescheme}.
Scheme~\ref{QCC-comp} does not use the notion of quantum gates, but
only of
a spatial and temporal measurement pattern. Once
universality of the \QC is established, to demonstrate  the
functioning of specific \QCns-algorithms one
would prefer decomposing a measurement
pattern directly into sub-patterns rather than decomposing a network
simulation into 
simulations of 
gates. A tool for the direct approach is provided by
Theorem~\ref{gatecond} in Section~\ref{QcorrQcomp}. 

\subsection{Randomness of the measurement results}
\label{rand}

We will now show that the described scheme of quantum computation with
the \QC works with unit efficiency despite the randomness of the
individual measurement results.

First note that a byproduct operator $U_\Sigma$ that acts after the
final unitary gate $U_{g_{|{\cal{N}}|}}$ does not jeopardize the scheme. Its
only effect is that the results of the readout measurements have to be
reinterpreted. The byproduct operator $U_\Sigma$ that acts upon the logical
output qubits $1\,..\, n$ has the form
\begin{equation}
    \label{byprod}
    U_\Sigma = \prod \limits_{i=1}^n
    \,\,{\left(\sigma_x^{[i]}\right)}^{x_i}
    {\left(\sigma_z^{[i]}\right)}^{z_i}, 
\end{equation}  
where $x_i,z_i \in \{0,1\}$ for $1\leq i \leq n$.
Let the qubits on the cluster which are left unmeasured be labeled in
the same way 
as the readout qubits of the quantum logic network. 

The qubits on the cluster which take the role of the readout qubits
are, at this point, in a state $U_\Sigma 
|\mbox{out}\rangle$,  where $|\mbox{out}\rangle$ is the output state
of the corresponding quantum logic network. The computation is
completed by measuring each qubit in 
the $\sigma_z$-eigenbasis, thereby obtaining the measurement results
$\{s_i^\prime \}$, say. In the \QCns-scheme, one measures the state
$U_\Sigma |\mbox{out}\rangle$ directly, whereby  outcomes
$\{ s_i \}$ are obtained and the readout qubits are projected
into the state $|{\cal{M}}\rangle =
\prod_{i=1}^n \frac{1+{(-1)}^{s_i} \sigma_z^{(i)}}{2}\,
\, U_\Sigma |\mbox{out}\rangle $. Depending on the byproduct
operator $U_\Sigma$, the set of measurement results $\{s\}$ in
general has a different interpretation from what the network readout
$\{s_i^\prime\}$ would have. The measurement basis is the same. From
(\ref{byprod}) one obtains
\begin{equation}
    \label{identification}
    \begin{array}{rcl}
    \ds{|{\cal{M}}\rangle} &=&
  \ds{\prod_{i=1}^n \frac{1+{(-1)}^{s_i} \sigma_z^{(i)}}{2}\,
    U_\Sigma |\mbox{out}\rangle} \vspace*{0.1cm}\\
    &=& \ds{U_\Sigma \left(U_\Sigma^\dagger \,
  \prod_{i=1}^n \frac{1+ 
  {(-1)}^{s_i} \sigma_z^{(i)}}{2}  U_\Sigma \right)
  |\mbox{out}\rangle} \vspace{0.1cm}\\
  &=& \ds{U_\Sigma \, \prod_{i=1}^n
  \frac{1+{(-1)}^{s_i+x_i} \sigma_z^{(i)}}{2}  |\mbox{out}\rangle}
  \end{array}
\end{equation}
From (\ref{identification}) we see that a $\sigma_z$-measurement on the
state $U_\Sigma \, |\mbox{out}\rangle$ with results $\{s\}$
represents the same algorithmic output as a $\sigma_z$-measurement of
the state $|\mbox{out} \rangle$ with the results $\{s_i^\prime\}$, where the
sets $\{s\}$ and $\{s_i^\prime\}$ are related by
\begin{equation}
    \label{join}
    s_i^\prime \equiv s_i+x_i\; \mbox{mod} \,\, 2.
\end{equation}
The set $\{s_i^\prime\}$ represents the result of the computation. It can be
calculated from the results $\{s_i \}$ of the $\sigma_z$-measurements
on the ``readout'' cluster qubits, and the values $\{x_i\}$ which are
determined by the byproduct operator $U_\Sigma$.

Thus we find that one can cope with the randomness of the measurement
results provided the byproduct operators $U_{\Sigma,g_i}$ in
(\ref{Gateseq}) can be propagated 
forward through the subsequent gates such that they act on the cluster
qubits representing the output register. 
This can be done. To propagate the byproduct operators we use the
propagation relations
\begin{equation}
    \label{CNOTproprel}
    \begin{array}{rcr}
        \mbox{CNOT}(c,t) \sigma_x^{(t)} &=& \sigma_x^{(t)}
        \mbox{CNOT}(c,t)\\
        \mbox{CNOT}(c,t) \sigma_x^{(c)} &=& \sigma_x^{(c)}
        \sigma_x^{(t)} \mbox{CNOT}(c,t)\\
        \mbox{CNOT}(c,t) \sigma_z^{(t)} &=& \sigma_z^{(c)}
        \sigma_z^{(t)} \mbox{CNOT}(c,t)\\
        \mbox{CNOT}(c,t) \sigma_z^{(c)} &=& \sigma_z^{(c)}
        \mbox{CNOT}(c,t)\\        
    \end{array}
\end{equation}   
for the CNOT gate, 
\begin{equation}
    \label{Rotproprel}
    \begin{array}{rcl}
        U_{Rot}[\xi,\eta,\zeta] \sigma_x &=& \sigma_x
        U_{Rot}[\xi,-\eta,\zeta]\\ 
        U_{Rot}[\xi,\eta,\zeta] \sigma_z &=& \sigma_z
        U_{Rot}[-\xi,\eta,-\zeta]
    \end{array}
\end{equation}
for general rotations  $U_{Rot}[\xi,\eta,\zeta]$ as defined in
(\ref{Euler}), and
\begin{equation}
    \label{Hadaproprel}
    \begin{array}{rcl}
        H \sigma_x &=& \sigma_z H \\
        H \sigma_z &=& \sigma_x H \\
        U_z[\pi/2] \sigma_x &=& \sigma_y U_z[\pi/2] \\
        U_z[\pi/2] \sigma_z &=& \sigma_z U_z[\pi/2]
    \end{array}
\end{equation}
for the Hadamard- and $\pi/2$-phase gate. The propagation relations
(\ref{Rotproprel}) apply to general rotations realized via
Procedure~\ref{Rotproc} --including Hadamard- and $\pi/2$-phase
gates-- while the propagation relations (\ref{Hadaproprel}) apply to
Hadamard- and $\pi/2$-phase gates as realized via
Procedure~\ref{Hadaproc}. 

Note
that the propagation relations (\ref{CNOTproprel}) -
(\ref{Hadaproprel}) are such that Pauli operators are mapped
onto Pauli operators 
under propagation and thus the byproduct operators remain in the Pauli
group when being propagated. Further note that there is a difference
between the 
relations for propagation through gates which are in the Clifford
group and through those which are not. For CNOT-, Hadamard- and
$\pi/2$-phase gates 
the byproduct operator changes under propagation while the gate remains
unchanged. This holds for all gates in the Clifford group, because the
propagation relations for Clifford gates are of the form $U_g 
U_\Sigma = (U_g U_\Sigma U_g^{-1}) U_g$ as (\ref{CNOTproprel}) and
(\ref{Hadaproprel}), i.e. the byproduct
operator $U_\Sigma$ is conjugated under the gate, and the Clifford
group by its definition as the normalizer of the Pauli group 
maps Pauli operators onto Pauli operators under
conjugation. For gates which are not in the
Clifford group this would in general not work and therefore, for 
rotations which are not in the Clifford group, the propagation
relations are different. There, the gate is conjugated under
the byproduct operator; and thus the byproduct operator
remains unchanged in propagation while the gate is modified. In both
cases, the forward propagation leaves the byproduct operators in the
Pauli group. In particular, their tensor product structure is maintained.

Let us now discuss how byproduct operator propagation affects the
scheme of computation with the \QCns. In Section~\ref{connect} we
arrived at the conclusion (\ref{Gateseq}) that by
patching the measurement patterns of individual gates together and
keeping the measurement bases fixed, we can realize a composite unitary
evolution on some input state
$|\psi_{\mbox{\footnotesize{in}}}\rangle$,
\begin{eqnarray}
    |\psi_{\mbox{\footnotesize{out}}}\rangle &=& \left(
    \prod_{i=1}^{|{\cal{N}}|} 
    U_{\Sigma,g_i} \, U_{g_i} \right)
     |\psi_{\mbox{\footnotesize{in}}}\rangle, \nonumber 
\end{eqnarray}  
where $U_{g_i}$ is the $i$-th unitary gate in the circuit and
$U_{\Sigma,g_i}$ the byproduct operator resulting from the
realization of that gate. Now using the above propagation
relations, (\ref{Gateseq}) can be rewritten in the following way
\begin{equation}
    \label{Byproped}
    |\psi_{\mbox{\footnotesize{out}}}\rangle = \left( \prod_{i=1}^{|{\cal{N}}|}
    \left.U_{\Sigma,g_i}\right|_\Omega \right) \left( \prod_{i=1}^{|{\cal{N}}|}
    U^\prime_{g_i} \right)
     |\psi_{\mbox{\footnotesize{in}}}\rangle.
\end{equation} 
Therein, $\left.U_{\Sigma,g_i}\right|_\Omega$ are forward
propagated byproduct operators resulting from the byproduct operators
$U_{\Sigma,g_i}$ of the gates $g_i$. They accumulate to the total
byproduct operator $U_\Sigma$ whose effect on
the result of the computation is contained in (\ref{join}),
\begin{equation}
    U_\Sigma = \prod_{i=1}^{|{\cal{N}}|}
    \left.U_{\Sigma,g_i}\right|_\Omega.
\end{equation} 
Further, the $U^\prime_{g_i}$ are the gates modified under the
propagation of the byproduct operators. As discussed above, for gates
in the Clifford group we have
\begin{equation}
    \label{Cliffnotmod}
    U^\prime_{g} = U_{g}, \; \forall \,\, g \in \mbox{Clifford group}, 
\end{equation}
as can be seen from (\ref{CNOTproprel}) and (\ref{Hadaproprel}).

\begin{figure}[tph]
\begin{center}
 \input{cuts.tex}
 \caption{\label{cuts}Vertical
 cuts. The vertical cuts intersect each qubit line exactly once but do not
 intersect gates. Thus, ${\cal{O}}_i$,  ${\cal{O}}_j$ and $\Omega$ are
 vertical cuts, but $\not\!\!{\cal{O}}\,\,$ is not. The cut
 ${\cal{O}}_i$ intersects the rotation $U_x$ just before its
 input. For two of the 
 rotations in the displayed network, the sub-clusters on which these
 gates are realized are symbolically displayed in gray
 underlay. Via the measurement of the cluster qubits $a$ and $b$
 (displayed as black dots with white border), the
 rotation angles of the respective rotations $U_x$ and $U_z$ are
 set.}
\end{center}
\end{figure}
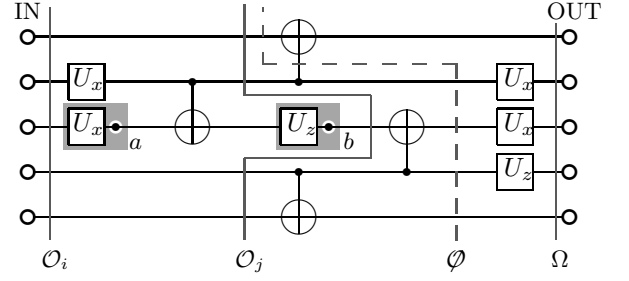

Gates which are not in the Clifford group are modified by byproduct
operator propagation. Specifically, the general rotations
(\ref{Euler}) are conjugated as can be seen from
(\ref{Rotproprel}). From the structure of (\ref{Gateseq}) we see that only the
byproduct operators  
of gates $g_k$ earlier than $g_i$ in the network may have an effect on
$U_{g_i}$, i.e. those with $k<i$. To give an
explicit expression, let 
us define $\left.U_{\Sigma,g_k}\right|_{{\cal{O}}_i}$, which are
byproduct operators $U_{\Sigma,g_k}$ propagated forward by the
propagation relations (\ref{CNOTproprel}) - (\ref{Hadaproprel}) to the
vertical cut ${\cal{O}}_i$ through the network, see Fig.~\ref{cuts}. A
vertical cut through 
a network is a cut which intersects each qubit line exactly once and
does not intersect gates. The vertical
cut ${\cal{O}}_i$ has the additional property that it intersects the
network just before the input of gate $g_i$. The relation between a
rotation $U^\prime_{g_i}$ modified by the byproduct operators and the
non-modified rotation $U_{g_i}$ is
\begin{equation}
    \label{Rotmod}
    \begin{array}{rcr}
        U^\prime_{g_i} &=& \ds{\left( \prod_{k|k<i}
            \left.U_{\Sigma,g_k}\right|_{{\cal{O}}_i} \right) U_{g_i}
        {\left( \prod_{k|k<i}
              \left.U_{\Sigma,g_k}\right|_{{\cal{O}}_i}
            \right)}^\dagger}, \\ &\\
        & & \forall \,\, U_{g_i} \in SU(2).
    \end{array}
\end{equation}
Now that we have investigated the effect of byproduct operator
propagation on the individual gates let us return to equation
(\ref{Byproped}). There, we find that the operations which act on the
input state $|\psi_{\mbox{\footnotesize{in}}}\rangle$ group into two
factors. The first is composed of the modified gate operations
$U_{g_i}^\prime$ and the second of the forward propagated byproduct
operators. The second factor gives
the accumulated byproduct operator $U_\Sigma$ and is
absorbed into the result of the computation via (\ref{join}). It does
not cause any complication. 

So what remains is the first factor, and we find that the unitary
evolution of the input state $|\psi_{\mbox{\footnotesize{in}}}\rangle$ that
is realized is composed of the modified gates $U^\prime_{g_i}$. The gates we
will realize are 
thus the $U^\prime_{g_i}$, not $U_{g_i}$. However, the standard
procedures 1 - 3 in Section~\ref{unigate} are for 
the operations $U_{g_i}$. Thus we have
to read (\ref{Rotmod}) in reverse. We need to deduce $U_{g_i}$ from
$U^\prime_{g_i}$.  Once the gates $g_k$ for all $k<i$
have been realized, this can be done for each gate $g_i$ since 
the byproduct operators $U_{\Sigma,k}$ are then
known for all $k<i$. Finally, with $U_{g_i}$ determined from
$U^\prime_{g_i}$, Procedure~\ref{Rotproc} gives the
measurement bases required for the realization of the gate $g_i$. 
Please note that it is a sufficient
criterion for the realization of the gate $g_i$ that all gates
$g_k$ with $k<i$ must have been realized before, but not a
necessary one.

Let us, at this point, address the question of temporal
ordering more explicitly. For proper discussion of the  temporal
ordering  we have to 
step out of the network frame for a moment. First note that in case of
the \QC the basic primitive are measurements. Thus, the temporal
complexity will be determined by the temporal ordering of these
measurements, unlike in quantum 
logic networks, where it depends on the ordering
of gates. The most 
efficient ordering of measurements that simulates a quantum logic network
is not pre-described by the temporal ordering of the gates in this network.  

A temporal ordering among the measurements is inferred from the
requirement to keep the computation on the \QC deterministic in spite of
of the randomness introduced by the measurements. This randomness is
accounted for by the byproduct operators. The key to obtain
the temporal ordering of measurements is eq. (\ref{Rotmod}). There, the
byproduct operators $U_{\Sigma,g_k}|_{{\cal{O}}_i}$ may modify 
Euler angles of the one-qubit rotations in the network and
consequently change measurement bases.  The temporal ordering
thus arises due to the fact that bases for one-qubit measurements must
be chosen in accordance with outcomes obtained from the measurements of
other qubits.  

For each cluster qubit $q$ that needs to be measured in a non-trivial
basis, i.e. not in the eigenbasis of $\sigma_x$, $\sigma_y$ or 
$\sigma_z$, a set of  
cluster qubits $p_i$ can be identified, whose measurement outcomes
influence the choice of the measurement basis for qubit $q$. We say 
that $q$ is in the forward cone \cite{QCmodel} of $p_i$, $q \in
\text{fc}(p_i)$. Each cluster qubit has a forward cone, and in no
forward cone there appears a qubit which is measured in a
trivial basis. 

The rule is that a cluster qubit $q$ can only be measured once all
cluster qubits $p_i$ for which $q \in \text{fc}(p_i)$ have been
measured earlier. The forward cones thereby generate an anti-reflexive partial
ordering 
among the measurements from which the most efficient measurement
strategy can be inferred, see \cite{QCmodel}. Gates in the Clifford
group do not contribute to the temporal complexity of a
\QCns-algorithm, see Section~\ref{Onestep}.
 
\subsection{Using quantum correlations for quantum
  computation} 
\label{QcorrQcomp}

In this section we give a criterion which allows do demonstrate the
functioning of the \QCns-simulations of unitary gates in a compact way.

Before we state the theorem, let us make the notion of a
measurement pattern more precise. In a \QCns-computation one can only
choose the measurement bases, while the measurement
outcomes are random. This is sufficient for deterministic computation.
Thus one can perform measurements specified by a spatial and temporal pattern
of measurement bases but one cannot control into which of the two
eigenstates the qubits are projected. 

\begin{definition}
    \label{measpattdef}
    A measurement pattern ${\cal{M}}^{({\cal{C}})}$ on a cluster
    ${\cal{C}}$ is a set of vectors
    \begin{equation}
        {\cal{M}}^{({\cal{C}})}=\left\{ \vec{r}_a \in S^2\, |
        \,\, a \in {\cal{C}} \right\},
    \end{equation}
    defining the measurement bases of the one-qubit measurements on
    ${\cal{C}}$.
\end{definition}
If this pattern ${\cal{M}}^{({\cal{C}})}$ of
measurements is applied on an initial state
$|\Psi_{\cal{E}}\rangle_{\cal{C}}$ and thereby the set of
measurement outcomes
\begin{equation}
    \{{s}\}_{\cal{C}} = \left\{{s}_a \in \{0,1\} \,|\,\, a \in
        {\cal{C}} \right\}
\end{equation}
is obtained, then the resulting state
$|\Psi_{\cal{M}}\rangle_{\cal{C}}$ is, modulo norm factor, given by
$|\Psi_{\cal{M}}\rangle_{\cal{C}} =P^{({\cal{C}})}_{{\{s\}}}({\cal{M}})\,
|\Psi_{\cal{E}}\rangle_{\cal{C}}$, where
\begin{equation}
    P^{({\cal{C}})}_{{\{{s}\}}}({\cal{M}})=
    \bigotimes_{k \in {\cal{C}}} 
    \frac{1+(-1)^{{s}_k} \vec{r}_k\cdot
      \vec{\sigma}^{(k)} }{2}.
\end{equation}

Besides, let us introduce some conventions for labeling. Be
${\cal{C}}_I(g)$ and ${\cal{C}}_O(g)$  such that $|{\cal{C}}_I(g)| =
|{\cal{C}}_O(g)|=n$ where $n$ is the number of logical qubits
processed by $g$. Operators acting on qubits $p \in {\cal{C}}_I(g)$
and $q \in {\cal{C}}_O(g)$ are 
labeled by upper indices $({\cal{C}}_I(g),i)$ and  $({\cal{C}}_O(g),i^\prime)$,
$1 \leq i,i^\prime \leq n$, respectively. The qubits $p \in {\cal{C}}_I(g)$ and
$q \in {\cal{C}}_O(g)$ are ordered
from 1 to $n$ in the same way as the logical qubits 
that they represent.

We make a distinction between the gate $g$ and the unitary
transformation $U$ it realizes. The gate $g \in {\cal{N}}$ does, besides
specifying the unitary transformation $U$, also comprise the
information about the location of the gate within the network.  
  
After these definitions and conventions we can now state
the following theorem 
\begin{theorem} 
    \label{gatecond}
    Be ${\cal{C}}(g)={\cal{C}}_I(g) \cup {\cal{C}}_M(g) \cup
    {\cal{C}}_O(g)$ with ${\cal{C}}_I(g) \cap
    {\cal{C}}_M(g)={\cal{C}}_I(g) \cap {\cal{C}}_O(g)={\cal{C}}_M(g)
    \cap {\cal{C}}_O(g)=\emptyset$ a cluster for the simulation
    of a gate $g$, realizing the unitary transformation $U$,
    and $|\phi\rangle_{{\cal{C}}(g)}$ the
    cluster state on the cluster ${\cal{C}}(g)$. 

    Suppose, the 
    state $|\psi \rangle_{{\cal{C}}(g)}=
    P_{\{{s}\}}^{({\cal{C}}_M(g))}({\cal{M}})
    \,|\phi\rangle_{{\cal{C}}(g)}$ obeys the $2n$ 
    eigenvalue equations 
    \begin{equation}
        \label{gatecheck}
        \begin{array}{rcl}
            \sigma_x^{({\cal{C}}_I(g),i)} \left( U \sigma_x^{(i)}
            U^\dagger \right)^{({\cal{C}}_O(g))}
            |\psi\rangle_{{\cal{C}}(g)} &=& {(-1)}^{\lambda_{x,i}}
            |\psi\rangle_{{\cal{C}}(g)} \\
            \sigma_z^{({\cal{C}}_I(g),i)} \left( U \sigma_z^{(i)}
            U^\dagger \right)^{({\cal{C}}_O(g))}
            |\psi\rangle_{{\cal{C}}(g)} &=& {(-1)}^{\lambda_{z,i}}
            |\psi\rangle_{{\cal{C}}(g)},
        \end{array}
    \end{equation}
with $\lambda_{x,i},\lambda_{z,i} \in \{0,1\}$ and $1 \leq i \leq n$.

Then, on the cluster ${\cal{C}}(g)$ the gate $g$ acting on an
arbitrary quantum input state $|\psi_\text{\em{in}}\rangle$ can be realized
according to Scheme~\ref{gatescheme} with the measurement directions
in ${\cal{C}}_M(g)$ described by ${\cal{M}}^{({\cal{C}}_M(g))}$ and the
measurements of the qubits in ${\cal{C}}_I(g)$ being
$\sigma_x$-measurements. Thereby, the input- and
output state in the simulation of $g$ 
are related via 
\begin{equation}
    \label{acts}
    |\psi_{\text{\em{out}}}\rangle = U U_\Sigma
     \,|\psi_{\text{\em{in}}} \rangle, 
\end{equation}  
where $U_\Sigma$ is a byproduct operator given by
\begin{equation}
    \label{ByProp}
    U_\Sigma=
    \bigotimes_{({\cal{C}}_I(g) \ni
    i)=1}^{n}(\sigma_z^{[i]})^{{s}_i+\lambda_{x,i}} 
    (\sigma_x^{[i]}) ^{\lambda_{z,i}}.
\end{equation}

\end{theorem}

The significance of the above theorem is that it provides a
comparatively simple criterion for the functioning of gate simulations
on the \QCns.  

In Scheme~\ref{gatescheme}, after read-in of the input state and the
entangling operation $S^{({\cal{C}}(g))}$, i.e. before the
measurements that realize the gate are performed, the resulting state
carries the 
quantum input in an encoded form. This state is in general not a
cluster state. It is therefore 
not clear a priori that cluster state correlations alone are
sufficient to explain the functioning of the gate. However, this is
what Theorem~\ref{gatecond} states. To prove the functioning of a gate
$g$ realized via Scheme~\ref{gatescheme} it is sufficient  to
demonstrate that a cluster state on ${\cal{C}}(g)$ exhibits certain
quantum correlations. About the variable input one does not need to
worry. 
 
This is convenient in two ways. First, we can base the
explanation of the gates directly on the eigenvalue equations
(\ref{EVeqn}) which were also used to define the cluster states in a
compact way. The quantum correlations required to explain the
functioning of the gates are derived from the basic correlations
(\ref{BasCorr}) rather easily and thus the use of
Theorem~\ref{gatecond} makes the explanation of the gates
compact. 

Second, Theorem~\ref{gatecond} is a tool to
demonstrate the functioning of \QCns-circuits without having to
repeat the whole universality proof for each particular circuit under
consideration. Scheme~\ref{QCC-comp} describes the computation as a
series of one-qubit measurements on a 
cluster state. An accordance with this, instead of decomposing a
circuit simulation into gate 
simulations as done in Scheme~\ref{gatescheme}, a measurement pattern is
decomposed into sub-patterns. The effect of these measurement
sub-patterns is tested via the criterion (\ref{gatecheck}) in
Theorem~\ref{gatecond}.

Before we turn to the proof of Theorem~\ref{gatecond} let us 
note that the measurements described by 
$P^{({\cal{C}}_M(g))}_{\{s\}}({\cal{M}}(g))$, as they have full rank,
project the initial cluster state $|\phi\rangle_{{\cal{C}}(g)}$ into a
tensor product state, $|\psi\rangle_{{\cal{C}}(g)}=
|m\rangle_{{\cal{C}}_M(g)} \otimes |\psi\rangle_{{\cal{C}}_I(g)\cup
  {\cal{C}}_O(g)}$. Thereof only the second factor,
$|\psi\rangle_{{\cal{C}}_I(g)\cup {\cal{C}}_O(g)}$, is of
interest. This state alone satisfies the eigenvalue equations
\eqref{gatecheck}, and is uniquely determined  by these
equations.  To see this, consider the 
state  $\ket{\psi'}_{{\cal{C}}_I(g)\cup
  {\cal{C}}_O(g)}=U^\dag\ket{\psi}_{{\cal{C}}_I(g)\cup {\cal{C}}_O(g)}$.  It
satisfies the $2n$ eigenvalue equations 
\begin{equation}
    \begin{array}{rcl}
        \sigma_x^{(i,{\cal{C}}_I(g))} \sigma_x^{(i,{\cal{C}}_O(g))} 
        |\psi'\rangle &=&
        {(-1)}^{\lambda_{x,i}} 
        |\psi'\rangle, \\
        \sigma_z^{(i,{\cal{C}}_I(g))}
        \sigma_z^{(i,{\cal{C}}_O(g))} 
        |\psi'\rangle &=&
        {(-1)}^{\lambda_{z,i}} 
        |\psi'\rangle,
    \end{array}
\end{equation}
where we have written in short $|\psi^\prime\rangle$ for $|\psi'\rangle_{{\cal{C}}_I(g)\cup {\cal{C}}_O(g)}$.
The state $|\psi'\rangle_{{\cal{C}}_I(g)\cup {\cal{C}}_O(g)}$ is uniquely defined by the above
set of commuting observables, it is a product of Bell
states. Therefore, $|\psi\rangle_{{\cal{C}}_I(g)\cup {\cal{C}}_O(g)}$
is uniquely defined as well.   

{\em{Proof of Theorem~\ref{gatecond}.}} We will discuss the
functioning of the gates for two 
cases of inputs. First, for all input states in the computational
basis. This leaves relative phases open which have to be
determined. To fix them, we discuss second the input state with all
qubits individually in 
$|+\rangle$. As we will see, from these two cases it can be concluded
that the gate simulation works for all input states of the
computational basis. This is sufficient because of the linearity of
the applied operations; if the gate simulations work for states of the
computational basis then they work for superpositions of such inputs
as well. 

Case 1: The input $|\psi_\text{in}\rangle$ is one of the states of the
computational  
basis, i.e. $|\psi_\text{in}\rangle=|{\bf{z}}\rangle:=\bigotimes_{i=1}^n
|z_i\rangle_{z,i}$ with $z_i \in \{0,1\}, \,i=1\,..\,n$. Then the state
$|\Psi_\text{out}({\bf{z}})\rangle_{{\cal{C}}(g)}$ of
the qubits in ${\cal{C}}$ [after performing a procedure according to
Scheme~\ref{gatescheme}, using a measurement pattern
${\cal{M}}^{({\cal{C}}_M(g))}$ on the body ${\cal{C}}_M(g)$ of the
gate $g$, and applying $\sigma_x$-measurements on ${\cal{C}}_I(g)$] is 
\begin{equation}
    \label{proc1}
    \begin{array}{l}
        \ds{n_O({\bf{z}})\,|\Psi_\text{out}({\bf{z}})\rangle_{{\cal{C}}(g)}
        =}\vspace{1.5mm}\\ 
        \ds{P^{({\cal{C}}_I(g))}_{\{s\}}\!(X)
        P^{({\cal{C}}_M(g))}_{\{s\}}\!({\cal{M}}) 
        S^{({\cal{C}}(g))} \,|{\bf{z}}\rangle_{{\cal{C}}_I(g)} \!\otimes\!
        |+\rangle_{{\cal{C}}_M(g) \cup {\cal{C}}_O(g)},}
    \end{array}
\end{equation} 
with norm factors $n_O({\bf{z}})$ that
are nonzero for all ${\bf{z}}$, as we shall show later. 

The input $|{\bf{z}}\rangle$ in (\ref{proc1}) satisfies the equation
\begin{equation}
    \label{Inter1}
    n_I({\bf{z}}) \,|{\bf{z}}\rangle = P^{({\cal{C}}_I(g))}_{Z,{\bf{z}}}
      \bigotimes_{i=1}^n |+\rangle_i,
\end{equation}
with $P^{({\cal{C}}_I(g))}_{Z,{\bf{z}}}= \bigotimes_{i=1}^n
\frac{1+{(-1)}^{z_i} \sigma_z^{[i]}}{2}$, and $n_I({\bf{z}})=1/2^{n/2}$
for all ${\bf{z}}$. Now note that $S^{({\cal{C}}(g))}$
and $P^{({\cal{C}}_I(g))}_{Z,{{\bf{z}}}}$, as well as  
$P^{({\cal{C}}_M(g))}_{\{s\}}({\cal{M}})$ and
$P^{({\cal{C}}_I(g))}_{Z,{{\bf{z}}}}$, commute. Thus, 
$|\Psi_\text{out}({\bf{z}})\rangle_{{\cal{C}}(g)}$ can be written as
\begin{equation}
    \label{Inter2}
    \begin{array}{l}
        n_O^\prime({\bf{z}})\, |\Psi_\text{out}({\bf{z}})
         \rangle_{{\cal{C}}(g)} =\vspace{1.5mm}\\ 
         \mbox{ }\hspace{2cm}=\,P^{({\cal{C}}_I(g))}_{\{s\}}(X) \,
         P^{({\cal{C}}_I(g))}_{Z,{{\bf{z}}}}\,
         P^{({\cal{C}}_M(g))}_{\{s\}}({\cal{M}})\,
         |\phi\rangle_{{\cal{C}}(g)} \vspace{1.5mm}\\ 
         \mbox{ }\hspace{2cm}=\, P^{({\cal{C}}_I(g))}_{\{s\}}(X) \,
         P^{({\cal{C}}_I(g))}_{Z,{{\bf{z}}}}\, |\psi\rangle_{{\cal{C}}(g)},
    \end{array}
\end{equation}
where $|\psi\rangle_{{\cal{C}}(g)}$ is specified by the eigenvalue
equations (\ref{gatecheck}) in Theorem~\ref{gatecond}.

Let us, at this point, emphasize that the projections
$P^{({\cal{C}}_I(g))}_{\{s\}}(X)$ and
$P^{({\cal{C}}_I(g))}_{Z,{{\bf{z}}}}$ in (\ref{Inter2}) are of very
different origin. The projector $P^{({\cal{C}}_I(g))}_{\{s\}}(X)$
describes the action of the $\sigma_x$-measurements on the qubits in
${\cal{C}}_I(g)$. These measurements are part of the procedure to
realize some gate $g$ on the cluster ${\cal{C}}(g)$. One has no
control over the
thereby obtained measurement outcomes $\{s\}$ specifying
$P^{({\cal{C}}_I(g))}_{\{s\}}(X)$. In contrast,
the projector $P^{({\cal{C}}_I(g))}_{Z,{{\bf{z}}}}$ does not correspond
to measurements that are performed in reality. Instead, it is 
introduced as an auxiliary construction that allows one to relate
the processing of quantum inputs to quantum correlations in
cluster states. The parameters ${\bf{z}}$ specifying the quantum input
$|{\bf{z}}\rangle$ and thus the projector
$P^{({\cal{C}}_I(g))}_{Z,{{\bf{z}}}}$ in (\ref{Inter1}) can be chosen
freely.   

The goal is to find for the state
$|\Psi_\text{out}({\bf{z}})\rangle_{{\cal{C}}(g)}$ an expression
involving the transformation $U$ acting on the input $|{\bf{z}}\rangle$.
To accomplish this, first observe that for the state on the r.h.s of
(\ref{Inter2}) via 
(\ref{gatecheck}) the following eigenvalue equations hold
\begin{equation}
    \label{Inter4}
    \begin{array}{l}
    {\left(U \sigma_z^{[i]} U^\dagger \right)}^{\!({\cal{C}}_O)}
    \!\left[P^{({\cal{C}}_I(g))}_{\{s\}}(X) \,
         P^{({\cal{C}}_I(g))}_{Z,{{\bf{z}}}}\,
    |\psi\rangle_{{\cal{C}}(g)}\right] = \vspace{1.5mm}\\
    \mbox{ }\hspace{1.5cm}{(-1)}^{\lambda_{z,i}+z_i} 
    \left[P^{({\cal{C}}_I(g))}_{\{s\}}(X) \,
         P^{({\cal{C}}_I(g))}_{Z,{{\bf{z}}}}\,
    |\psi\rangle_{{\cal{C}}(g)}\right]\!, 
    \end{array}
\end{equation}
with $i=1..n$.

To make use of the equations (\ref{Inter4}) we  need to prove that
$P^{({\cal{C}}_I(g))}_{\{s\}}(X) \,  
P^{({\cal{C}}_I(g))}_{Z,{{\bf{z}}}}\,
|\psi\rangle_{{\cal{C}}(g)} \neq 0$ for all ${\bf{z}}$ under the
assumptions of theorem~\ref{gatecond}. 

For this, we consider the
scalar $\mbox{}_{{\cal{C}}(g)} \langle \psi| 
P^{({\cal{C}}_I(g))}_{Z,{{\bf{z}}}} |\psi\rangle_{{\cal{C}}(g)}$ and
write $P^{({\cal{C}}_I(g))}_{Z,{{\bf{z}}}}$ in the form
\begin{equation}
    \label{Inter7}
    P^{({\cal{C}}_I(g))}_{Z,{{\bf{z}}}} =
    \frac{1}{2^n} \left(1+\sum_{k=1}^{2^n} \bigotimes_{i \in I_k}
    {(-1)}^{z_i} \sigma_z^{(i)} \right)^{({\cal{C}}_I(g))},
\end{equation}
 where $I_k \subset {\cal{C}}_I \neq \emptyset\; \forall
 k=1..2^n$. For each $I_k$ we choose an  
 $i \in I_k$ and insert the respective eigenvalue equation from the
 upper line of 
 (\ref{gatecheck}) into  
 $\mbox{}_{{\cal{C}}(g)} \langle \psi| \bigotimes_{j \in
 I_k} \sigma_z^{(j)}|\psi\rangle_{{\cal{C}}(g)}$. Since
 $ \bigotimes_{j \in I_k} \sigma_z^{(j)}$ and 
$\sigma_x^{(i,{\cal{C}}_I(g))} \left( U \sigma_x^{(i)}
     U^\dagger \right)^{({\cal{C}}_O(g))}$ anti-commute,
 $\mbox{}_{{\cal{C}}(g)} \langle \psi| \bigotimes_{j \in 
 I_k} \sigma_z^{(i)}|\psi\rangle_{{\cal{C}}(g)}=0$ for
 all $I_k$. Thus, with (\ref{Inter7}), one finds $\mbox{}_{{\cal{C}}(g)}
 \langle \psi|  
P^{({\cal{C}}_I(g))}_{Z,{{\bf{z}}}} |\psi\rangle_{{\cal{C}}(g)}=1/2^n$,
 such that $P^{({\cal{C}}_I(g))}_{Z,{{\bf{z}}}}
 |\psi\rangle_{{\cal{C}}(g)} \neq 0$ and therefore also
\begin{equation}
    \label{nonzero}
    P^{({\cal{C}}_I(g))}_{\{s\}}(X) \, P^{({\cal{C}}_I(g))}_{Z,{{\bf{z}}}}\,
    |\psi\rangle_{{\cal{C}}(g)} \neq 0,  
\end{equation}
or, in other words, $n_O^\prime({\bf{z}}) \neq 0$ for all ${\bf{z}}$.
 
Due to the fact that the projections
$P^{({\cal{C}}_I(g))}_{Z,{{\bf{z}}}}$ and
$P^{({\cal{C}}_M(g))}_{\{s\}} ({\cal{M}})$ are of full rank the
above state  has the form 
\begin{equation}
    \label{Inter3}
    \begin{array}{l}
        P^{({\cal{C}}_I(g))}_{\{s\}}(X) \,
        P^{({\cal{C}}_I(g))}_{Z,{{\bf{z}}}}\, 
        |\psi\rangle_{{\cal{C}}(g)}=\vspace{1.5mm}\\
        \mbox{ }\hspace{1.5cm} n_O^\prime({\bf{z}})
        \,|{\bf{s}}\rangle_{x,{\cal{C}}_I(g)} 
    \otimes |m\rangle_{{\cal{C}}_M(g)} \otimes
        |\psi_\text{out}({\bf{z}})\rangle_{{\cal{C}}_O(g)}, 
    \end{array}
\end{equation}
where $|{\bf{s}}\rangle_{x,{\cal{C}}_I}=\!\!\bigotimes\limits_{(
  {\cal{C}}_I \ni i)=1}^n\!\! |s_i\rangle_{x,i}$, and
$|m\rangle_{{\cal{C}}_M(g)}$ is some product state with $\|\, 
|m\rangle_{{\cal{C}}_M(g)}\|=1$. Elaborating the argument that leads 
to (\ref{nonzero}) one finds that $n_O^\prime({\bf{z}})=1/2^n$ and
$n_O({\bf{z}})=1/2^{n/2}$, but at this point the precise values of the
normalization factors are not important as long as they are nonzero.
 
In (\ref{Inter3}) only the third factor of the state on the r.h.s. is
interesting, and this factor is determined by the eigenvalue equations
(\ref{Inter4}):
\begin{equation}
    \label{psi_O}
    |\psi_\text{out}({\bf{z}})\rangle=e^{i \eta({\bf{z}})} U
        U_\Sigma |{\bf{z}}\rangle,
\end{equation}
where $U_\Sigma$ is given by (\ref{ByProp}). Now, because of
(\ref{Inter2}) with 
$n_O^\prime({\bf{z}})\neq0\; \forall\, {\bf{z}}$, a
solution (\ref{Inter3}) with (\ref{psi_O}) for the state
$P^{({\cal{C}}_I(g))}_{\{s\}}(X) \, 
P^{({\cal{C}}_I(g))}_{Z,{{\bf{z}}}}\, |\psi\rangle_{{\cal{C}}(g)}$ is
also a solution for the state
$|\Psi_\text{out}({\bf{z}})\rangle_{{\cal{C}}(g)}$, and one
finally obtains
\begin{equation}
    \label{Inter13}
    |\Psi_\text{out} ({\bf{z}})\rangle_{{\cal{C}}(g)} = e^{i
      \eta({\bf{z}})} |{\bf{s}}\rangle_{x,{\cal{C}}_I(g)}
    \otimes |m\rangle_{{\cal{C}}_M(g)} \otimes \left[ U
        U_\Sigma |{\bf{z}}\rangle \right]_{{\cal{C}}_O(g)}\!.
\end{equation} 
There appear no additional norm factors in (\ref{Inter13}) 
because the states on the l.h.s. and the r.h.s. are both normalized to
unity. 

The solution (\ref{Inter13}) still allows for one free parameter, the
phase factor  
$e^{i\eta({\bf{z}})}$. Note that, a priori, the phase factors for
different ${\bf{z}}$ can all be different.

This concludes the discussion of case 1. We have found in
(\ref{Inter13}) that the realized gate acts as 
\begin{equation}
    \label{Inter8b}
    \tilde{U}= U\,U_\Sigma\,D
\end{equation}
where the gate $D$ is diagonal in the computational basis
and contains all the phases $e^{i\eta({\bf{z}})}$. What remains is to
show that $D = {\bf{1}}$ modulo a possible global phase.  

Case 2. Now the same procedure is
applied for the input state $|\psi_\text{in}\rangle=|
+\rangle := \bigotimes_{i=1}^n
|+\rangle_i$. Then, the state $|\Psi_\text{out}
(+)\rangle_{{\cal{C}}(g)}$ that results from the gate simulation is
\begin{equation}
    \label{Inter9}
    n_O(+)\,|\Psi_\text{out} (+)\rangle_{{\cal{C}}(g)} =
    P^{({\cal{C}}_I(g))}_{\{s\}}(X)\, 
    P^{({\cal{C}}_M(g))}_{\{s\}}({\cal{M}})
    \,|\phi\rangle_{{\cal{C}}(g)},
\end{equation}
with a nonzero norm factor $n_O(+)$.
Using the upper line of eigenvalue equations (\ref{gatecheck}), the
state $|\Psi_\text{out} 
(+)\rangle_{{\cal{C}}(g)}$ is found to obey the eigenvalue equations
\begin{equation}
    \label{Inter10}
    \left(\! U \sigma_x^{[i]}
     U^\dagger \!\right)^{\!({\cal{C}}_O(g))} \! |\Psi_\text{out}
     (+)\rangle_{{\cal{C}}(g)} \! = \! {(-1)}^{\lambda_{x,i}+s_i}
     |\Psi_\text{out} (+)\rangle_{{\cal{C}}(g)}. 
\end{equation} 
The eigenvalue equations (\ref{Inter10}) in combination with
(\ref{Inter9}) imply that
\begin{equation}
    \label{Inter14}
    |\Psi_\text{out} (+)\rangle_{{\cal{C}}(g)} = e^{i
      \chi} |{\bf{s}}\rangle_{x,{\cal{C}}_I(g)}
    \otimes |m\rangle_{{\cal{C}}_M(g)} \otimes \left[ U
        U_\Sigma |+\rangle \right]_{{\cal{C}}_O(g)}\!,
\end{equation}
with $\chi$ being a free parameter.
Therefore, on the input state $|+\rangle$ the gate
simulation acts as
\begin{equation}
    \label{Inter11b}
    \tilde{U}= e^{i\chi}\,U\, U_\Sigma.
\end{equation}
This observation concludes the discussion of case 2.

The fact that (\ref{Inter13}) and (\ref{Inter14}) hold simultaneously
imposes stringent conditions on the phases $\eta({\bf{z}})$. To see
this, let us evaluate the scalar product
\begin{equation}
    \label{Inter16}
    c_\chi=\mbox{}_{{\cal{C}}(g)}\langle \Psi_\text{out}(+)| U
    U_\Sigma |{\bf{s}}\rangle_{x,{\cal{C}}_I(g)} \otimes
    |m\rangle_{{\cal{C}}_M(g)} \otimes |+
    \rangle_{{\cal{C}}_O(g)}. 
\end{equation}
From (\ref{Inter14}) it follows immediately that
\begin{equation}
    \label{Inter17}
    c_\chi=e^{-i\chi}.
\end{equation}
On the other hand, since $|+\rangle = 1/2^{n/2}
\sum_{{\bf{z}}\in {\{0,1\}}^n} |{\bf{z}}\rangle$ and, by
linearity, $|\Psi_\text{out}(+)\rangle = 1/2^{n/2}
\sum_{{\bf{z}}\in  {\{0,1\}}^n }
|\Psi_\text{out}({\bf{z}})\rangle$, from (\ref{Inter13}) it follows
that
\begin{equation}
    \label{Inter18}
    c_\chi = \frac{1}{2^n} \sum_{{\bf{z}}\in  {\{0,1\}}^n}
    e^{-i\eta({\bf{z}})}. 
\end{equation}
The sum in (\ref{Inter18}) runs over $2^n$ terms. Thus, with
$|e^{-i\eta({\bf{z}})}|=1$ for all ${\bf{z}}$, it follows from
  the triangle inequality that $|c_\chi| \leq 1$. The modulus of
  $c_\chi$ can be unity only if all $e^{-i\eta({\bf{z}})}$ are
equal. As (\ref{Inter17}) shows, $|c_\chi|$ is indeed equal to
unity. Therefore, the phase factors $e^{i\eta({\bf{z}})}$ must all be
the same, and with (\ref{Inter17}) and (\ref{Inter18}),
\begin{equation}
    \label{Inter19}
    e^{i\eta({\bf{z}})} = e^{i \chi},\; \forall\, {\bf{z}}.
\end{equation}   
If we now insert (\ref{Inter19}) into (\ref{Inter13}) we find that the
gate simulation acts upon every input state in the computational
basis, and thus upon every input state, as $\tilde{U}_g = e^{i
  \chi} U\, U_\Sigma$. Therein, the global phase factor $e^{i
  \chi}$ has no effect. Thus we find that the gate simulation
indeed acts as stated in (\ref{acts}) and (\ref{ByProp}). \hfill $\Box$
\medskip

We would like to acknowledge that a similar theorem restricted to gates in the
Clifford group has been obtained in \cite{Perdrix}.

Let us conclude this section with some comments on how to use this
    theorem. First, note that {\em{Theorem~\ref{gatecond} 
    does not imply anything about the temporal order of measurements
    within a gate simulation.}} In particular it should be understood
that a procedure according to Scheme~\ref{gatescheme} is not such that
first the measurements on the 
cluster qubits in ${\cal{C}}_M(g)$ and thereafter the measurements in
${\cal{C}}_I(g)$ are performed. 

Instead, first all those cluster
qubits $q \in {\cal{C}}_I(g) \cup {\cal{C}}_M(g)$ are measured whose
measurement basis is the eigenbasis of either $\sigma_x$ or $\sigma_y$
(remember that, after the removal of the redundant cluster qubits as
described in Section~\ref{redrem}, we are dealing with clusters
${\cal{C}}_N$ such that, 
apart from the readout, no measurements in the $\sigma_z$-eigenbasis
occur). Second, possibly in several subsequent rounds, the
remaining measurements are performed in bases which are chosen
according to previous measurement results.

Let us now discuss how to choose
the appropriate measurement bases. First note that the unitary
operations $U_{\Sigma}$ and $U$ in (\ref{acts}) both depend on
measurement results of qubits in ${\cal{C}}(g)$,
\begin{equation}
    \begin{array}{rcl}
        \ds{U_\Sigma} &=&
        \ds{U_\Sigma\left( \{ s_i| \, i  
        \in {\cal{C}}_I(g) \cup {\cal{C}}_M(g) \} \right),}\\
        \ds{U} &=& \ds{U\left({\cal{M}}^{({\cal{C}}_M(g))}, \{ s_i| \, i
        \in {\cal{C}}_M(g) \} \right).}        
    \end{array}
\end{equation}
The dependence of $U_\Sigma$ on $\{ s_i |\, i \in {\cal{C}}_M(g)
\}$ arises through the $\{ \lambda_{x,i}, \lambda_{z,i}| \, i =1\,
.. \, n\}$ of (\ref{gatecheck}). 

Now note that in (\ref{acts}) the order of the unitary gate
$U$ and the byproduct operator $U_\Sigma$ is the opposite of
what is required in (\ref{Gateseq}). Therefore, the order of these
operators has to be interchanged, which is achieved by propagating the
byproduct operator $U_\Sigma$ through the gate $U$. For gates
or sub-circuits given as a quantum logic network composed of
CNOT-gates and one-qubit rotations, this task can
be performed using the propagation relations (\ref{CNOTproprel}),
(\ref{Rotproprel}) and (\ref{Hadaproprel}).  The result is
\begin{equation}
    \label{Exchorder}
     \begin{array}{l}
        \tilde{U}= U\left({\cal{M}}^{({\cal{C}}_M(g))},
        \{s\}_{{\cal{C}}_M}\} \right) \,
        U_\Sigma\left(\{s\}_{{\cal{C}}_{I} \cup {\cal{C}}_{M}} \right)
        = \vspace{1.5mm}\\ 
        \mbox{ }\hspace{1cm}U_\Sigma^\prime\left(\{s\}_{{\cal{C}}_{I}
        \cup {\cal{C}}_{M}} 
        \right)\, 
        U^\prime\left({\cal{M}}^{({\cal{C}}_M(g))},
        \{s\}_{{\cal{C}}_{I} \cup {\cal{C}}_{M}} \right).  
     \end{array}
\end{equation}

Now, the choice of measurement bases in
${\cal{M}}^{({\cal{C}}_M(g))}$ is allowed to be adaptive, that is the
measurement bases may depend on measurement outcomes at other cluster qubits,
${\cal{M}}^{({\cal{C}}_M(g))}= {\cal{M}}^{({\cal{C}}_M(g))}(\{s\})$. 
For the realization of the gate $U_g$, 
the measurement bases must be chosen in such a way that 
\begin{equation}
    \label{baschoice}
    U^\prime\left({\cal{M}}^{({\cal{C}}_M(g))}, \{s\}\right)
    = U_g.
\end{equation}

This induces the identification,
\begin{equation}
U_\Sigma^\prime\left(\{s\}_{{\cal{C}}_{IM}}
        \right)=U_{\Sigma,g}
\end{equation}
of the byproduct operators.
Now, the order of the desired unitary operation $U_g$ and the
byproduct operator
$U_{\Sigma,g}$ is as required
    in (\ref{Gateseq}). With adaptive measurement bases the effect of
the randomness introduced by the measurements can be
counteracted. What remains is  
a random byproduct operator which does not affect the
deterministic character of a \QCns-computation and which is accounted
for in the post-processing of the measurement results.  

In subsequent sections we will illustrate in a number of examples how
Theorem~\ref{gatecond} is used to demonstrate the functioning of
quantum gate simulations on the \QCns, and how the strategies for
adapting the measurement bases are found.  

\subsection{Function of CNOT-gate and general one-qubit
  rotations}   
\label{unifunk}

In this section, we demonstrate that the measurement patterns which we
have introduced do indeed realize the desired quantum logic gates. 

The basis for all our considerations is the set~(\ref{EVeqn}) of
eigenvalue equations fulfilled by the cluster states. Therefore let
us, before we turn to the realization of the gates in the universal
set, describe how the eigenvalue equations can be
manipulated. Equations~(\ref{EVeqn}) are not the only eigenvalue
equations satisfied by the cluster state. Instead, a vast number of
other eigenvalue equations can be derived from them. 

The operators $K^{(a)}$ may for example be added,
multiplied by a scalar and multiplied with each other.  
In this way, a large number of eigenvalue equations can be generated
from equations~(\ref{EVeqn}). 
Note, however, that  not all operators generated in this way are
correlation operators.  Non-Hermitian operators can be generated which
do not represent observables, yet will prove to be useful for the
construction of new correlation operators. 

Furthermore, if quantum correlation operator $K$ for state
$\ket{\phi}$ commutes with measured observable
$\vec{r}_i\cdot\vec{\sigma}^{(i)}$, the correlation will still apply
to the measured state. More specifically, if the state $\ket{\phi}$
satisfies the eigenvalue equation $K\ket{\phi}=\lambda\ket{\phi}$ and
$[K,\vec{r}_i\cdot\vec{\sigma} ]=0$, then the state resulting from the
measurement, $P_{s_i}^{(i)}\ket{\phi}$, where
$P_{s_i}^{(i)}=\frac{1+(-1)^{s_i}\vec{r}_i\cdot\vec{\sigma}^{(i)}}{2}$,
satisfies the same eigenvalue equation since $\lambda [P_{s_i}^{(i)}
\ket{\phi} ]=[P_{s_i}^{(i)}K \ket{\phi} ]= K [P_{s_i}^{(i)} \ket{\phi}
]$. Thus the correlation $K$ is inherited to the resultant state,
$P_{s_i}^{(i)}\ket{\phi}$. 

To demonstrate and explain the measurement patterns
realizing certain quantum gates, the program is as follows. 
First, from the set of eigenvalue equations which define the cluster
state $\ket{\phi}_{{\cal C}(g)}$, we derive a set of eigenvalue
equations which is compatible with the measurement pattern on ${\cal
  C}_M$. Then, we use these to deduce the set of eigenvalue equations
which define the state $\ket{\psi}_{{\cal C}(g)}$, where the qubits in ${\cal
  C}_M$ have been measured. Thus we demonstrate that the assumptions
for Theorem~\ref{gatecond}, that is the set of
equations~\eqref{gatecheck}, are satisfied with the appropriate
unitary transformation $U$. Third,  $U_\Sigma$ is obtained from
equation~(\ref{ByProp}) as a function of the measurement results. The
order of $U$ and $U_\Sigma$ is then interchanged and, in this way, the
temporal ordering of the measurements  becomes apparent.

\subsubsection{Identity gate}
\label{sec:id}

As a simple example, let us first consider a gate which realizes the
identity operation $\iden$ on a single logical qubit. 

For the identity gate ${\cal C}_I$, ${\cal C}_M$ and ${\cal C}_O$
each consist of a single qubit, so labeling the qubits 1, 2 and 3,
$1\in{\cal C}_I$,  $2\in{\cal C}_M$ and  $3\in{\cal C}_O$. The pattern
${\cal  M}(\iden)$ corresponds to a measurement of qubit 2 in the
$\sigma_x$ basis.  

Let $\ket{\phi}_{{\cal{C}}(\iden)}$ be the cluster state on these
three qubits. The state is defined by the 
following set of eigenvalue equations. 
\begin{subequations}\label{3qubeig}
\begin{align}
\sigma_{x}^{(1)}\sigma_{z}^{(2)}\phantom{\sigma_{z}^{(3)}}\quad
    \ket{\phi}_{{\cal 
    C}(\iden)}=&\ket{\phi}_{{\cal C}(\iden)},\label{3qubeig1} \\ 
\sigma_{z}^{(1)}\sigma_{x}^{(2)}\sigma_{z}^{(3)}\quad\ket{\phi}_{{\cal 
    C}(\iden)} =&\ket{\phi}_{{\cal C}(\iden)},\label{3qubeig2}\\
\phantom{\sigma_{z}^{(1)}}\sigma_{z}^{(2)}\sigma_{x}^{(3)}\quad
    \ket{\phi}_{{\cal C}(\iden)} =& \ket{\phi}_{{\cal C}(\iden)}.
    \label{3qubeig3}  
\end{align}
\end{subequations}

After the measurement of qubit 2, the resulting state of the cluster is 
\begin{equation}\label{psiiddef}
\ket{\psi}_{{\cal C}(\iden)}=P_{x,s_2}^{(2)}\ket{\phi}_{{\cal C}(\iden)}, 
\end{equation} 
where $s_2\in\{0,1\}$, and
$P_{x,s_2}^{(2)}=\frac{1+(-1)^{s_2}\sigma_x^{(2)}}{2}$. 

$P_{x,s_2}^{(2)}$ and $\sigma_x^{(2)}$ obey the following relation, 

\begin{equation}\label{idenproj}
P_{x,s_2}^{(2)}\sigma_x^{(2)}=(-1)^{s_2}P_{x,s_2}^{(2)}.
\end{equation}
Applying $P_{x,s_2}^{(2)}$ to both sides of equation~\eqref{3qubeig2},
and using equation~\eqref{idenproj}, one obtains for
$\ket{\psi}_{{\cal C}(\iden)}$, defined in equation~\eqref{psiiddef}, 

\begin{equation}
\sigma_{z}^{(1)}\sigma_{z}^{(3)}\quad\ket{\psi}_{{\cal C}(\iden)}
= (-1)^{s_2}\ket{\psi}_{{\cal C}(\iden)}.
\end{equation}
Also from equations~\eqref{3qubeig1} and \eqref{3qubeig3} we have 
\begin{equation}
\sigma_{x}^{(1)}\sigma_{x}^{(3)}\quad\ket{\phi}_{{\cal C}(\iden)}
=\ket{\phi}_{{\cal C}(\iden)}.
\end{equation}
Applying $P_{x,s_2}^{(2)}$ to both sides of this equation gives
\begin{equation}
\sigma_{x}^{(1)}\sigma_{x}^{(3)}\quad\ket{\psi}_{{\cal C}(\iden)}
=\ket{\psi}_{{\cal C}(\iden)}.
\end{equation}

Now, since qubits 1 and 3 represent the input and output qubits
respectively, the assumption of Theorem~\ref{gatecond},
equation~\eqref{gatecheck}, is satisfied for $U=\iden$. The byproduct
operator $U_\Sigma$ is obtained from equation~\eqref{ByProp}, and we
find that the full unitary operation realized by the gate is
$\tilde U=\iden\  \sigma_x^{s_2}\ \sigma_z^{s_1}= \sigma_x^{s_2}\
\sigma_z^{s_1} \iden$. 

Also note that a wire with length one (${\cal{C}}_I(H)=1$,
${\cal{C}}_M(H)=\emptyset$, ${\cal{C}}_O(H)=2$), i.e. half of the above
elementary wire, implements a Hadamard transformation. As in this
construction the input- and output qubits lie on different
sub-lattices of ${\cal{C}}$, one on the even and one on the odd
sub-lattice, we do not use it in the universal set of
gates. Nevertheless, this realization of the Hadamard transformation
can be a useful tool in gate construction. For example, we will use it
in Section~\ref{sec:rotation-around-z} to construct the realization of the
$z$-rotations out of the realization of $x$-rotations.  

\subsubsection{Removing unnecessary measurements}
\label{sec:remov-unnec-meas}

In larger measurement patterns, whenever pairs of adjacent $\sigma_x$-
qubits in a wire are surrounded above and below by either vacant
lattice sites or
$\sigma_z$-measurements, they can be removed from the pattern without
changing the logical operation of the gate. 
This is simple to show in the case of a linear cluster. Consider six
qubits, labelled $a$ to $f$, which are part of a longer line of
qubits, prepared in a cluster state. Four of the eigenvalue equations which
define the state are 
\begin{equation}\label{remqubeq1}
\begin{split}
\sigma_z^{(a)}\sigma_x^{(b)}\sigma_z^{(c)}\ket{\psi}_{\cal
  C}&=\ket{\psi}_{\cal C},\\ 
\sigma_z^{(b)}\sigma_x^{(c)}\sigma_z^{(d)}\ket{\psi}_{\cal
  C}&=\ket{\psi}_{\cal C},\\ 
\sigma_z^{(c)}\sigma_x^{(d)}\sigma_z^{(e)}\ket{\psi}_{\cal
  C}&=\ket{\psi}_{\cal C},\\ 
\sigma_z^{(d)}\sigma_x^{(e)}\sigma_z^{(f)}\ket{\psi}_{\cal
  C}&=\ket{\psi}_{\cal C}. 
\end{split}
\end{equation}

Suppose, a measurement pattern $\cal M$ on these qubits contains measurements
of the observable $\sigma_x$ on qubits $c$ and $d$. Measurements in
the  $\sigma_x$ basis can be made before any other measurements in
$\cal M$. If these two measurements alone are carried out, the new
state fulfills the following eigenvalue equations, derived from
equation \eqref{remqubeq1} in the usual way, 
\begin{equation}
\begin{split}
\sigma_z^{(a)}\sigma_x^{(b)}\sigma_z^{(e)}\ket{\psi}_{\cal
  C}&=(-1)^{s_d}\ket{\psi}_{\cal C},\\ 
\sigma_z^{(b)}\sigma_x^{(e)}\sigma_z^{(f)}\ket{\psi}_{\cal
  C}&=(-1)^{s_c}\ket{\psi}_{\cal C}. 
\end{split}
\end{equation}
The resulting state is therefore a cluster state from which qubits $c$
and $d$ have been removed, and $b$ and $e$ play the role of adjacent
qubits. Thus, the two measurements have mapped a cluster state onto a
cluster state and thus do not contribute to the logical operation
realized by $\cal M$, which, in the case where both $s_c$ and $s_d$
equal 0, is completely equivalent to the reduced measurement pattern
$\cal M'$, from which these adjacent $\sigma_x$ measurements have been
removed.

\subsubsection{One-qubit rotation around $x$-axis}
\label{sec:xrot}

A one-qubit rotation through an angle $\alpha$ about the $x$-axis
$U_x[\alpha]=\exp[-i\alpha/2\sigma_x]$ is  realized on the same three qubit
layout as the identity gate. Labeling the qubits 1, 2 and 3 as in the
previous section,   $1 = {\cal C}_I$,  $2 = {\cal C}_M$ and
$3 = {\cal C}_O$. The measurement pattern ${\cal M}(U_x)$ consists of
a measurement, on qubit 2, of the observable represented by the vector
$\vec{r}_{xy}(\eta)=(\cos(\eta),\sin(\eta),0)$, 
\begin{equation}\label{rxydef}
    \vec{r}_{xy}(\eta)\cdot\vec{\sigma}=\cos\eta\,\sigma_x+\sin\eta\,
    \sigma_y=U_z[\eta]\sigma_xU_z[-\eta],    
\end{equation}
whose eigenstates lie in the $x$-$y$-plane of the Bloch sphere at an
angle of $\eta$ to the $x$-axis. 
 
The cluster state $\ket{\phi}_{{\cal C}(U_x)}$ is defined by
equations~\eqref{3qubeig}. After the measurement of ${\cal M}(U_x)$,
the resulting state is $\ket{\psi}_{{\cal
    C}(U_x)}=P_{xy(\eta)}^{(2)}\ket{\phi}_{{\cal C}(U_x)}$ where
$P_{xy(\eta)}^{(2)}=\frac{1+{(-1)}^{s_2} \vec{r}_{xy}(\eta)
  \cdot\vec{\sigma}}{2}$.   
To generate an eigenvalue equation whose operator commutes with
$\vec{r}_{xy}(\eta)\cdot\vec{\sigma}$ we manipulate
equation~\eqref{3qubeig3} in the following way, 
\begin{align}
&&\sigma_z^{(2)}\sigma_x^{(3)}\ket{\phi}_{{\cal
    C}(U_x)}&=\ket{\phi}_{{\cal C}(U_x)}\\ 
\text{i.e.}&& \sigma_z^{(2)}\ket{\phi}_{{\cal
    C}(U_x)}&=\sigma_x^{(3)}\ket{\phi}_{{\cal C}(U_x)}\nonumber\\ 
\text{i.e.}&& [\sigma_z^{(2)}-\sigma_x^{(3)}]\ket{\phi}_{{\cal
    C}(U_x)}&=0\nonumber\\ 
\therefore&&\exp[-\im(\theta/2)[\sigma_z^{(2)}-\sigma_x^{(3)}]\ket{\phi}_{{\cal
    C}(U_x)}&=\ket{\phi}_{{\cal C}(U_x)} 
\end{align}
\noindent
where the last equation is true for all $\theta\in[0,2\pi]$. This
takes a more useful form, if we write it in terms of one-qubit rotations, 
\begin{equation}
    U_z^{(2)}[\theta]\,U_x^{(3)}[-\theta]\ket{\phi}_{{\cal
    C}(U_x)}=\ket{\phi}_{{\cal C}(U_x)} .
\end{equation}

We use this and equation~\eqref{3qubeig2} to construct the following
eigenvalue equation for $\ket{\phi}_{{\cal C}(U_x)}$, 
\begin{equation}\label{xroteq}
    \begin{split}
        \ket{\phi}_{{\cal C}(U_x)} =&\sigma_z^{(1)}\, U_z^{(2)}[\eta]\,
        \sigma_x^{(2)}\,U_z^{(2)}[-\eta] \\& 
        \quad U_x^{(3)}[-\eta]\, \sigma_z^{(3)}\, U_x^{(3)}[\eta]\,
        \ket{\phi}_{{\cal C}(U_x)}.
    \end{split}
\end{equation}

Applying $P_{xy(\eta),2}$ to both sides, we obtain the following
eigenvalue equation for $\ket{\psi}_{{\cal C}(U_x)}$, 
\begin{equation}\label{xroteq2}
    \begin{split}
        &\sigma_z^{(1)}
        U_x^{(3)}[-\eta] \, \sigma_z^{(3)}\, U_x^{(3)}[\eta]\,
        \ket{\psi}_{{\cal C}(U_x)}=(-1)^{s_2}\ket{\psi}_{{\cal C}(U_x)}.
    \end{split}
\end{equation}
In the same way as for the identity gate we also apply the projector
to an eigenvalue equation generated from equations~\eqref{3qubeig1}
and ~\eqref{3qubeig3} to obtain 
\begin{equation}
    \begin{split}
        \ket{\psi}_{{\cal
            C}(U_x)}&=\sigma_{x}^{(1)}\sigma_{x}^{(3)}\ket{\psi}_{{\cal
            C}(U_x)} \\ 
        &=\sigma_{x}^{(1)} \, U_x^{(3)}[-\eta] \, \sigma_x^{(3)} \,
        U_x^{(3)}[\eta] \,\ket{\psi}_{{\cal C}(U_x)}
    \end{split}
\end{equation}
and thus we see that equation~\eqref{gatecheck} is satisfied for
$U=U_x(-\eta)$ and
$U_\Sigma=\sigma_z^{s_1}\sigma_x^{s_2}$. Interchanging the order of
these operators is not as trivial here as for the identity gate. When
$\sigma_z$ is propagated through $U_x(\eta)$ the sign of the angle is
reversed, so we find that the gate operation realized by this
${\cal M}(U_x)$ in the \QC is 
\begin{equation}
    U_g=U_x\left[(-1)^{s_1}(-\eta)\right]. 
\end{equation}
The sign of the rotation realized by this gate is a function of
$s_1$, the outcome of the measurement on qubit 1. This is an example
of the temporal ordering of measurements in the \QCns. In order to
realize $U_x[\alpha]$ deterministically, the angle of the
measurement, $\eta$, on qubit~2   must be  
$\eta=(-1)^{s_1}(-\alpha)$, thus this measurement can only be 
realized after the measurement of qubit~1.   

\subsubsection{Rotation around $z$-axis}
\label{sec:rotation-around-z}

The measurement pattern for a rotation around the $z$-axis
$U_z[\beta]=\exp[-\im\beta/2\sigma_z]$ is illustrated in
Fig.~\ref{Gates}. It requires 5 qubits for its realization. 
 
\begin{figure}
    \begin{center}
        \epsfig{file=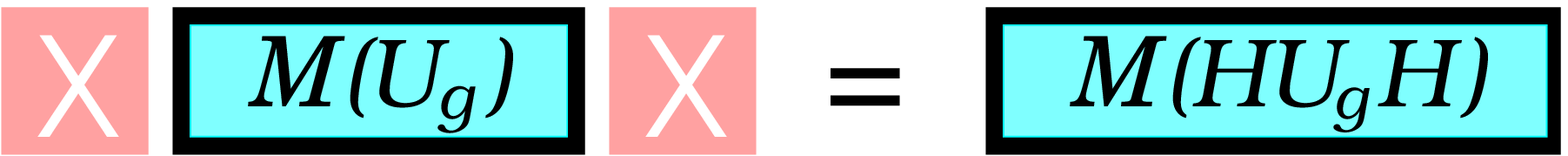,width=6cm}
        \caption{\label{fig:stretch}Useful identity for the
          realization of the rotation 
          $U_z[\alpha]$ as the sequence $H\,U_x[\alpha]\,H$.}
    \end{center}
\end{figure}

The measurement layout ${\cal M} (U_z)$ is similar to the rotation
about the $x$-axis, except for two additional $\sigma_x$ measurements
on either side of the central qubit. The simplest way to understand
this gate is regard it as the concatenation
$U_z[\alpha]=H\,U_x[\alpha] \, H$. The Hadamard transformations may be
realized as wires of length one, see Section~\ref{sec:id}. Thus, the
measurement pattern of the $z$-rotation is that of the $x$-rotation
plus one cluster qubit on either side measured in the eigenbasis of
$\sigma_x$, as displayed in Fig~\ref{fig:stretch}.

The explanation in terms of eigenvalue equations obeyed by cluster
states is as follows. Let us label the qubits 1 to 5. The cluster
state $\ket{\phi}_{{\cal 
    C}(U_z)}$ is defined by eigenvalue equations of the usual form. If
qubits 2 and 4 are measured in the $\sigma_x$ basis, the resulting
state $\ket{\phi'}_{{\cal C}(U_z)}=P_{x, s_2}^{(2)}P_{x,
  s_4}^{(4)}\ket{\phi}_{{\cal C}(U_z)}$ fulfills the following set of
eigenvalue equations 
\begin{subequations}
    \begin{align}
        \sigma_x^{(1)}\sigma_x^{(3)}\sigma_x^{(5)}\ket{\phi'}_{{\cal
            C}(U_z)}&=\ket{\phi'}_{{\cal C}(U_z)},\\ 
        \sigma_z^{(1)}\sigma_z^{(3)}\phantom{\sigma_x^{(5)}}\ket{\phi'}_{{\cal
            C}(U_z)}&=(-1)^{s_2}\ket{\phi'}_{{\cal C}(U_z)},\\ 
        \phantom{\sigma_x^{(1)}}\sigma_z^{(3)}\sigma_z^{(5)}\ket{\phi'}_{{\cal
            C}(U_z)}&=(-1)^{s_4}\ket{\phi'}_{{\cal C}(U_z)}. 
    \end{align} 
\end{subequations}
This set of equations is analogous to equations~\eqref{3qubeig},
except for the different eigenvalues and that the input and output
qubits $x$- and $z$-bases have been exchanged. From here on the analysis
of the measurement pattern runs parallel to the previous section. 

One finds ${\cal M}(U_z)$ realizes the operation $U_z(\beta)$  if
the basis of the measurement on qubit 3 is chosen to be the eigenbasis
of $\vec{r}_{xy}((-1)^{s_2}(-\beta))\cdot\vec{\sigma}$, where
$\vec{r}_{xy}(\eta)$ is defined in equation~\eqref{rxydef}. Qubit 2
must thus be measured prior to qubit 3. The byproduct operator for
this gate is $U_{\Sigma,U_z}=\sigma_x^{s_2+s_4}\sigma_z^{s_1+s_3}$.

\subsubsection{Arbitrary Rotation}
\label{sec:arbitrary-rotation}

\begin{figure}
    \begin{center}
        \epsfig{file=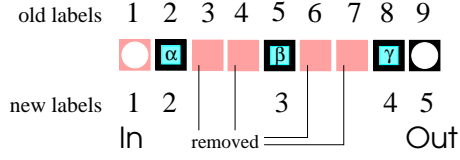,width=6cm}
        \caption{\label{rotremillu}General rotation composed of two
          $x$-rotations and a $z$-rotation in between (Euler
          representation). In the \QCns-realization pairs of adjacent cluster
          qubits measured in the $\sigma_x$-eigenbasis may be removed
          from the measurement pattern.}
    \end{center}
\end{figure}

The arbitrary Euler rotation can be realized by combining the
measurement patterns of rotations around $x$- and $z$-axes by
overlaying input and output qubits of adjacent patterns, as described
in section~\ref{connect}. This creates a measurement pattern of 7
qubits plus input and output qubits, labelled as in
Fig.~\ref{rotremillu}, with measurements of 
$\sigma_x$ on qubits 3, 4, 6 and 7, and measurements in the
$x$-$y$-plane at angles $\alpha$, $\beta$ and $\gamma$ on qubits 2, 5
and 8, respectively. 
The unitary operation realized by these connected measurement patterns is,
\begin{equation}
    \begin{split}
        U_\Sigma U_{Rot}[\xi,\eta,\zeta]=&\sigma_z^{s_7}\sigma_x^{s_8}
        U_x[(-1)^{s_7}(-\gamma)]
        \sigma_z^{s_3+s_5}\sigma_x^{s_4+s_6}\\ 
        &\quad U_z[(-1)^{s_4}(-\beta)]\sigma_z^{s_1}\sigma_x^{s_2} \\ 
        &\quad U_x[(-1)^{s_1}(-\alpha)] 
    \end{split}
\end{equation}

As we have shown above, adjacent pairs of $\sigma_x$ measurements can be
removed from the pattern without changing the operation realized by
the gate. The operation realized by this reduced measurement
pattern is obtained by setting the measurement results from the
removed qubits to zero,  $s_3,s_4,s_6,s_7=0$. After relabelling the
remaining qubits in the measurement pattern 1 to 5, we obtain 
\begin{equation} 
    \begin{split}
        U_\Sigma U_{Rot}[\xi,\eta,\zeta]=&\sigma_x^{s_4} U_x[-\gamma]
        \sigma_z^{s_3} U_z[(-\beta)]\\&\quad \sigma_z^{s_1}\sigma_x^{s_2}
        U_x[(-1)^{s_1}(-\alpha)] 
    \end{split}
\end{equation}
Propagating all byproduct operators to the left hand side we find the
unitary operation realized by the measurement pattern is 
\begin{equation}
    \begin{split}    
        U_{Rot}[\xi,\eta,\zeta]=&U_x[-(-1)^{s_1+s_3}\gamma]
        U_z[-(-1)^{s_2}\beta] \\&\quad U_x[-(-1)^{s_1}\alpha]
    \end{split} 
\end{equation}
with byproduct operator $U_\Sigma=\sigma_x^{s_2+s_4}\sigma_z^{s_1+s_3}$. 
 One finds that, to realize a specific rotation
 $U_{Rot}[\xi,\eta,\zeta] = U_x[\zeta]U_z[\eta]U_x[\xi]$, the angles
 $\alpha$, $\beta$, $\gamma$ specifying the measurement bases of the
 qubits 2,3, and 4 are again
 dependent on the measurement results of other qubits. We see that
 $\alpha=(-1)^{s_1}(-\xi)$, $\beta=(-1)^{s_2}(-\eta)$,
 $\gamma=(-1)^{s_1+s_3}(-\zeta)$. 
To realize a specific rotation deterministically, qubit 2 must
 thus be measured before qubits 3 and 4, and qubit 3 before qubit 4, in the
 bases specified in Section~\ref{unigate}. 

\subsubsection{Hadamard- and $\pi/2$-phase gate}
\label{sec:hapha}

The Hadamard- and the $\pi/2$-phase gate have the property that under
conjugation with these gates Pauli operators are mapped onto Pauli
operators,
\begin{equation}
    \label{Hadaconj}
    \begin{array}{rcl}
        H\sigma_x H^\dagger &=& \sigma_z, \\  
        H\sigma_z H^\dagger &=& \sigma_x,
    \end{array}
\end{equation}
and
\begin{equation}
    \label{Uzconj}
    \begin{array}{rcl}
        U_z[\pi/2] \sigma_x U_z[\pi/2]^\dagger &=& \sigma_y, \\
        U_z[\pi/2] \sigma_z U_z[\pi/2]^\dagger &=& \sigma_z,
    \end{array}
\end{equation}
from which the propagation relations (\ref{Hadaproprel}) follow.
 Related to this property is the fact that these two special
rotations may be realized via $\sigma_x$- and
$\sigma_y$-measurements. Such measurement bases need not be adapted to
previously obtained measurement results and therefore, while these
rotations might be realized in the same way as any other rotation,
there is a more advantageous way to do so. 

To realize either of the gates we use again a cluster state of 5 qubits
in a chain ${\cal{C}}(H)$. Let the labeling of the qubits be as in
Fig.~\ref{Gates}d 
and e, i.e. qubit 1 is the input- and qubit 5 the output qubit.

A cluster state $|\phi\rangle_{{\cal{C}}(H)}$ obeys the two eigenvalue
equations
\begin{equation}
    \begin{array}{rcl}
        |\phi\rangle_{{\cal{C}}(H)} &=& K^{(1)} K^{(3)} K^{(4)}
        |\phi\rangle_{{\cal{C}}(H)}\\
        &=& \sigma_x^{(1)} \sigma_y^{(3)} \sigma_y^{(4)} \sigma_z^{(5)}
        |\phi\rangle_{{\cal{C}}(H)}, \vspace{1.5mm} \\ 
        |\phi\rangle_{{\cal{C}}(H)} &=& K^{(2)} K^{(3)} K^{(5)}
        |\phi\rangle_{{\cal{C}}(H)} \\
        &=&  \sigma_z^{(1)} 
        \sigma_y^{(2)} \sigma_y^{(3)} \sigma_x^{(5)}
         |\phi\rangle_{{\cal{C}}(H)}. 
    \end{array}
\end{equation}
When the qubits 2, 3 and 4 of this state are measured in the
$\sigma_y$-eigenbasis and thereby the measurement outcomes $s_2,\,
s_3, \, s_4 \in \{0,1\}$ are obtained, the resulting state
$|\psi\rangle_{{\cal{C}}(H)}$ obeys the eigenvalue equations
\begin{equation}
    \label{Hadacorr}
    \begin{array}{rcl}
        \sigma_x^{(1)} \sigma_z^{(5)}
        |\phi\rangle_{{\cal{C}}(H)} &=& (-1)^{s_3+s_4}
        |\phi\rangle_{{\cal{C}}(H)}, \\  
        \sigma_z^{(1)} \sigma_x^{(5)} |\phi\rangle_{{\cal{C}}(H)} &=&
        (-1)^{s_2+s_3} |\phi\rangle_{{\cal{C}}(H)}. 
    \end{array}
\end{equation}  
From equation (\ref{Hadaconj}) we see that the correlations
(\ref{Hadacorr}) are precisely those we need to explain the
realization of the Hadamard gate. Using Theorem~\ref{gatecond} 
we find that by procedure~\ref{Hadaproc} with measurement of the
operators $\sigma_x^{(1)}$, $\sigma_y^{(2)}$, $\sigma_y^{(3)}$ and 
$\sigma_y^{(4)}$ a Hadamard gate with a byproduct operator
as given in (\ref{Hadabyprop}) is realized.

A cluster state $|\phi\rangle_{{\cal{C}}(U_z[\pi/2])}$ of a chain of 5
qubits obeys the eigenvalue equations
\begin{equation}
    \begin{array}{rcl}
        |\phi\rangle_{{\cal{C}}(U_z[\pi/2])} &=& K^{(1)} K^{(3)}
        K^{(4)} K^{(5)}
        |\phi\rangle_{{\cal{C}}(U_z[\pi/2])},\\
        &=& -\sigma_x^{(1)} \sigma_y^{(3)} \sigma_x^{(4)} \sigma_y^{(5)}
        |\phi\rangle_{{\cal{C}}(U_z[\pi/2])} \vspace{1.5mm} \\ 
        |\phi\rangle_{{\cal{C}}(U_z[\pi/2])} &=& K^{(2)} K^{(4)}
        |\phi\rangle_{{\cal{C}}(U_z[\pi/2])} \\
        &=&  \sigma_z^{(1)} 
        \sigma_x^{(2)} \sigma_x^{(4)} \sigma_z^{(5)}
         |\phi\rangle_{{\cal{C}}(U_z[\pi/2])}. 
    \end{array}
\end{equation}
When the qubits 2, and 4 of this state are measured in the
$\sigma_x$- and qubit 3 is measured in the $\sigma_y$-eigenbasis, with
the measurement outcomes $s_2,\,
s_3, \, s_4 \in \{0,1\}$ obtained, the resulting state
$|\psi\rangle_{{\cal{C}}(U_z[\pi/2])}$ obeys the eigenvalue equations
\begin{equation}
    \label{Uzcorr}
    \begin{array}{rcl}
        \sigma_x^{(1)} \sigma_y^{(5)}
        |\psi\rangle_{{\cal{C}}(U_z[\pi/2])} &=& (-1)^{s_3+s_4+1}
        |\psi\rangle_{{\cal{C}}(U_z[\pi/2])}, \\  
        \sigma_z^{(1)} \sigma_z^{(5)} |\psi\rangle_{{\cal{C}}(U_z[\pi/2])} &=&
        (-1)^{s_2+s_4} |\psi\rangle_{{\cal{C}}(U_z[\pi/2])}. 
    \end{array}
\end{equation}  
Using Theorem~\ref{gatecond} 
we find that by procedure~\ref{Hadaproc} with measurement of the
operators $\sigma_x^{(1)}$, $\sigma_x^{(2)}$, $\sigma_y^{(3)}$ and 
$\sigma_x^{(4)}$ a $\pi/2$-phase gate is realized, where the
byproduct operator is given by (\ref{Hadabyprop}).

\subsubsection{The CNOT gate}
\label{sec:cnot-gate}

A measurement pattern which realizes a CNOT gate is illustrated in
Fig.~\ref{Gates}. Labeling the qubits as in Fig.~\ref{Gates}, we
use the same analysis as above to show that this measurement pattern
does indeed realize a CNOT gate in the \QCns. 
 
Of the cluster ${\cal C}(CNOT)$ on which the gate is realized,
qubits 1 and 9 belong to ${\cal C}_I$, qubits 7 
and 15 belong to  ${\cal C}_O$ and the remaining qubits belong to
${\cal C}_M$. Let $\ket{\phi}$ be a cluster state on
${\cal C}(CNOT)$, which obeys the set of eigenvalue equations (\ref{EVeqn}).

\begin{figure}
    \begin{center}
        \epsfig{file=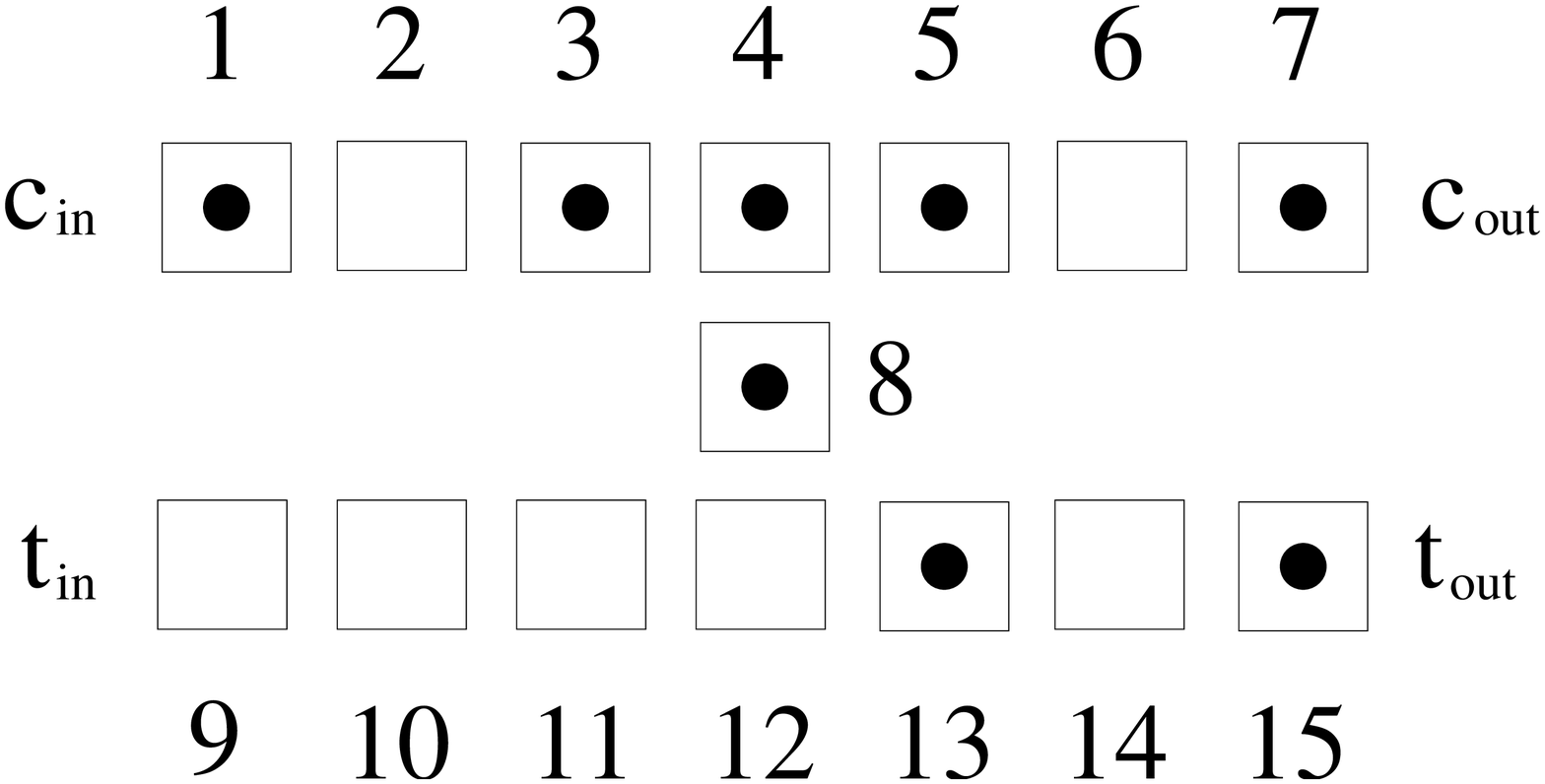, width=4.5cm}
        \caption{\label{fig:CNOTcorr1}Pattern of correlation centers
          representing the eigenvalue equation (\ref{CNOTcorr1}).}
    \end{center}
\end{figure}

From these basic eigenvalue equations there follow the equations
\begin{subequations}   
    \begin{align}
        \ket{\phi} &= K^{(1)} K^{(3)} K^{(4)}  K^{(5)} K^{(7)}
        K^{(8)} K^{(13)} K^{(15)}  \, \ket{\phi} \nonumber \\ 
        &= - \sigma_x^{(1)}  \sigma_y^{(3)}  \sigma_y^{(4)}
        \sigma_y^{(5)} \sigma_x^{(7)} \sigma_y^{(8)} \sigma_x^{(13)}
        \sigma_x^{(15)} \, \ket{\phi}, \label{CNOTcorr1}  \\ 
        \ket{\phi} &= K^{(2)} K^{(3)}  K^{(5)} K^{(6)}
        \, \ket{\phi} \nonumber \\
        &=   \sigma_z^{(1)} \sigma_y^{(2)}  \sigma_y^{(3)}  \sigma_y^{(5)}
        \sigma_y^{(6)}  \sigma_z^{(7)} \, \ket{\phi}, \label{CNOTcorr2} \\
        \ket{\phi} &= K^{(9)} K^{(11)}  K^{(13)} K^{(15)}
        \, \ket{\phi} \nonumber \\
        &=  \sigma_x^{(9)}  \sigma_x^{(11)}  \sigma_x^{(13)}
        \sigma_x^{(15)} \, \ket{\phi}, \label{CNOTcorr3} \\
        \ket{\phi} &= K^{(5)} K^{(6)}  K^{(8)} K^{(10)}
        K^{(12)} K^{(14)} \, \ket{\phi} \nonumber \\
        &=  \sigma_y^{(5)}  \sigma_y^{(6)}  \sigma_z^{(7)}
        \sigma_y^{(8)} \sigma_z^{(9)}  \sigma_x^{(10)} \sigma_y^{(12)}
        \sigma_x^{(14)} \sigma_z^{(15)} \, \ket{\phi}. \label{CNOTcorr4} 
    \end{align}
\end{subequations}  
Subsequently we will often use a graphic representation of eigenvalue
equations like (\ref{CNOTcorr1}) - (\ref{CNOTcorr4}). Each of these
equations is specified by the set of correlation centers $q$ for which
the basic correlation operators $K^{(q)}$ (\ref{BasCorr}) enter the
r.h.s. of the equation. While the information content is
the same, it is often more illustrative to display the
pattern of correlation centers than to write down the corresponding
cluster state eigenvalue equation. As an example, the pattern of
correlation centers which represents the eigenvalue equation
(\ref{CNOTcorr1}) is given in Fig.~\ref{fig:CNOTcorr1}.

If the qubits 10, 11, 13 and 14 are measured in the $\sigma_x$- and
the qubits 2, 3, 4, 5, 6, 8 and 12 are measured in the
$\sigma_y$-eigenbasis, whereby the measurement results $s_2$ - $s_6$,
$s_8$, $s_{10}$ - $s_{14}$ are obtained, then the cluster state
eigenvalue equations (\ref{CNOTcorr1}) - (\ref{CNOTcorr4}) induce the following
eigenvalue equations for the projected state $\ket{\psi}$  
\begin{subequations}
    \begin{align}
        \label{CNOTeig1}
        \sigma_x^{(1)}\,\sigma_x^{(7)}\,\sigma_x^{(15)}
        \,\ket{\psi}&=(-1)^{1+s_3+s_4+s_5+s_8+s_{13}} \ket{\psi},  \\
        \label{CNOTeig2} 
        \sigma_z^{(1)}\,\sigma_z^{(7)} \, \ket{\psi}&=(-1)^{s_2+s_3+s_5+s_6}
        \ket{\psi} \\ \label{CNOTeig3}          
        \sigma_x^{(9)}\, \sigma_x^{(15)} \,
        \ket{\psi}&=(-1)^{s_{11}+s_{13}}\ket{\psi},  \\ \label{CNOTeig4}   
        \sigma_z^{(9)}\,\sigma_z^{(7)}\, \sigma_z^{(15)} \,
        \ket{\psi}&=(-1)^{s_{5}+s_{6}+s_{8}+s_{10}+s_{12}+s_{14}}
        \ket{\psi}.
    \end{align} 
\end{subequations}
Therein, qubits 1 and 7 represent the input and output for  the control qubit
and qubits 9 and 15 represent the input and output for the target
qubit. Writing the CNOT unitary operation on control and target
qubits $CNOT(c,t)$, we find 
\begin{subequations}
    \begin{align}
        CNOT(c,t)\sigma_x^{(c)}CNOT(c,t)&=\sigma_x^{(c)}\sigma_x^{(t)},\\
        CNOT(c,t)\sigma_z^{(c)}CNOT(c,t)&=\sigma_z^{(c)},\\
        CNOT(c,t)\sigma_x^{(t)}CNOT(c,t)&=\sigma_x^{(t)},\\
        CNOT(c,t)\sigma_z^{(t)}CNOT(c,t)&=\sigma_z^{(c)}\sigma_z^{(t)}.
    \end{align}
\end{subequations}
Comparing these equations to the eigenvalue equations \eqref{CNOTeig1}
to \eqref{CNOTeig4}, one sees that $\cal M$ does indeed realize a
CNOT gate. Furthermore, after reading off the operator $U_\Sigma$
using equations \eqref{gatecheck} and (\ref{ByProp}) and propagating
the byproduct operators through to the output side of the CNOT gate, one
finds the expressions for the byproduct operators, reported in
equation~(\ref{CNOTbyprop}). 

\subsection{Upper bounds on resource consumption}
\label{Upperbounds}

Here we discuss the spatial, temporal and operational resources
required for the \QC and compare with resource requirements of a
network quantum computer.   

To run a specific quantum algorithm, the \QC requires a cluster of a certain
size. Therefore the \QCns-{\em{spatial resources}} $S$  are the number
of cluster qubits in the required cluster state $\csns$,
i.e. $S=|{\cal{C}}|$. The 
computation is driven by one-qubit measurement only. Thus, a single
one-qubit measurement is one unit of operational resources, and  
the \QCns-{\em{operational resources}} $O$ are defined as the total
number of one-qubit 
measurements involved. The operational resources $O$ are always 
smaller or equal to the spatial resources $S$,
\begin{equation}
    \label{OS}
    O \leq S,
\end{equation} 
since each cluster qubit is measured at most once.
As for the temporal resources, the \QCns-{\em{logical depth}} $T$ is the
minimum number of measurement rounds 
to which the measurements can be parallelized.


Let us briefly recall the definition of these resources in the network
model. The temporal resources are specified by the network logical
depth $T_\text{qln}$, which is the minimal number of steps to which quantum
gates and readout measurements can be 
parallelized. The spatial resources $S_\text{qln}$ count the number of logical
qubits on which an algorithm runs. Finally, the operational resources
$O_\text{qln}$ are the number of elementary operations required to carry out an
algorithm, i.e. the number of gates and measurements.  

The construction kit for the simulation of quantum
logic networks on the \QC shall contain a universal set of
gates, in our case the CNOT gate between arbitrary qubits and the one
qubit rotations. Already  the next-neighbor CNOT with general
rotations is universal since a 
general CNOT can be assembled of a next-neighbor CNOT and swap gates
which can themselves be composed of next-neighbor CNOTs. However, in
the following we would like to use for the general CNOT the
less cumbersome construction described in
Section~\ref{sec:non-neigh-cnot}.  For this gate, the distance between logical
qubits, i.e. between parallel qubit wires, is 4. The virtue of this
gate is that it can always
be realized on a vertical slice of width 6 on the cluster, no matter
how far control and target qubit are separated. A slice of width 6
means that the distance between an input qubit of the gate and the
corresponding  input of the consecutive gate is 6 lattice
spacings. This general CNOT 
gate determines the spatial dimensions of a unit cell in the measurement
patterns. The size of this unit cell is $4 \times 6$. The other
elementary gates, the next-neighbor CNOT and the rotations are
smaller than a unit cell and therefore have to be stretched. This is easily
accomplished. The next-neighbor CNOT as displayed in Fig.~\ref{Gates}a
has a size of $2\times6$ and is extended to size $4\times 6$ by
inserting two adjacent cluster qubits into the vertical bridge connecting the 
horizontal qubit lines. The general rotation as in Fig.~\ref{Gates}b
has width 4 and is stretched to width 6 by inserting two cluster
qubits just before the output. 

Concerning the temporal resources we first observe that we can realize the
gates in the same temporal order as in the network model. To realize
a general CNOT on the \QC takes one step of measurements, to realize
a general rotation takes at most three. For the network model we do
not assume that a general rotation has to be Euler-decomposed. Rather
we assume that in the network model a rotation can be realized in a
single step. Thus the temporal resources of the \QC and in the network
model are related via
\begin{equation}
    \label{Tres}
    T \leq \,3 T_\text{qln}.
\end{equation} 

As for the spatial resources, let us consider a rectangular cluster of
height $h$ and width $w$ on which the qubit wires are oriented
horizontally, with the network register state propagating from left to
right. As the logical qubits have distance 4, the height of the
cluster has to be $h=4 S_\text{qln}-3$ where $S_\text{qln}$ is equal
to the number $n$ 
of logical qubits. Further, the number of gates in the circuit is at
most $S_\text{qln} T_\text{qln}$ because, in the network model, in
each step at most 
$S_\text{qln}$ gates can be realized. On each vertical slice of width 6 on
the cluster there fits at least one gate such that  --taking
into account an extra slice of width 1 for the readout cluster
qubits-- for the width holds $w \leq 6 S_\text{qln}
T_\text{qln}+1$. With $S=h\,w$ one finds that
\begin{equation}
    \label{Sres}
    S \leq 24 \,{S_\text{qln}}^2 T_\text{qln}. 
\end{equation}

In a similar way, a bound involving the network operational resources
can be obtained. The spatial overhead $S$ and the operational overhead
$O$ per elementary network operation is $\leq 24 S_\text{qln}$ if this
operation is a unitary gate from the universal set described before,
and is equal to one if this operation is a readout measurement. Thus,
we also have 
\begin{equation}
    \label{OSres}
    \begin{array}{rcl}
        S &\leq& 24 \,O_\text{qln} S_\text{qln},\\
        O &\leq& 24 \,O_\text{qln} S_\text{qln}.
    \end{array}
\end{equation}       

The purpose of this section was to demonstrate that the scaling of
spatial and temporal resources is at worst polynomial 
as compared to the network model. In \cite{QCmodel} it has been shown,
as stated in Section~\ref{IP}, that the required classical processing
increases the computation time only marginally (logarithmically in the
number $n$ of logical qubits) and
thus there is no exponential overhead in either classical or quantum
resources. 

The upper bounds in (\ref{Tres}), (\ref{Sres}) and  (\ref{OSres}) should not be
taken for estimates. For algorithms of 
practical interest the required resources usually scale much more
favorably and there do not even have to be overheads at all. This is
illustrated for the temporal complexity of Clifford circuits in
Section~\ref{Onestep} and in the examples of Section~\ref{examples}. A
spatial overhead always exists. However, this is compensated by the
fact that the operational effort to create a cluster state is independent
of the cluster size.

\subsection{Quantum circuits in the Clifford group can
  be realized in a single step}
\label{Onestep}

The measurement bases to realize the Hadamard- and
the $\pi/2$-phase gate need not be adapted since only operators
$\sigma_x$ and $\sigma_y$ are measured. The same holds for the
realization of the CNOT gate, see Fig.~\ref{Gates}. Thus, all the
Hadamard-, $\pi/2$-phase- and CNOT-gates of a quantum circuit can be
realized simultaneously in the first measurement round, 
regardless of their location in the network. In particular, quantum
circuits which consist only of such gates, i.e. circuits in the
Clifford group, can be realized in a single time step. As an example,
many circuits for coding and decoding are in the Clifford group.  

The fact that quantum circuits in the Clifford group can be realized
in a single time step has previously not been known for networks. The
best upper bound on the logical depth that was known previously scales
logarithmically with the number of logical qubits \cite{M&N}.

Note that, as stated by the Gottesman-Knill-Theorem \cite{GKth}, there
is no need for fast Clifford circuits if the quantum output is
measured in a Pauli basis because these circuits can be simulated
efficiently classically. However, the purpose of this section is to
point out that the whole Clifford part of {\em{any}} quantum circuit can be
performed in a single time step. We will discuss this point further in Section~\ref{graphs}. 

Here we find a first aspect of \QCns-computation which is not
adequately described within the network model, and with this
observation we conclude the discussion of 
the \QC as a simulator of quantum logic networks.

\section{Computational model underlying the \QC}
\label{model}

\subsection{Processing of information}
\label{IP}

In the network model of quantum computation one usually regards a
quantum register as the carrier of  
information. The quantum register is prepared in some input state and
processed to some output state by applying a suitable unitary
transformation composed of quantum gates. Finally, the output state of
the quantum register is measured by which the classical readout is
obtained. 

For the \QC the notions of ``quantum
input'' and ``quantum output'' have no genuine meaning if we
restrict ourselves to the situation where the 
input state is known. As stated before, Shor's factoring
algorithm \cite{fac} and Grover's search algorithm
\cite{searoot} are both examples of such a situation. 
In these cases the final result of any computation --including
quantum computations-- is a
classical number. In a \QCns-computation this number is extracted from the
outcomes of all the one-qubit measurements. The entire computation
amounts to just measurements of the cluster qubits in a certain order
and basis.  

We have divided the set ${\cal{C}}$ of cluster qubits into subsets
$I$, $M$ and $O$ to describe the \QC in terms of
the network model. Such a terminology is not required for the \QC a
priori. It is true that when a quantum logic network is
realized on a cluster state there is a subset of cluster qubits
which play the role of the output register. However, these qubits are not the
final ones to be measured, but among the first (!).  
The measurement outcomes from all the cluster qubits contribute to the
result of the computation. The qubits 
of $O \subset {\cal{C}}$ simulate the output state of the quantum
register and thus contribute obviously to the computational result. The
cluster qubits in the set  
$I \subset {\cal{C}}$ simulate the fiducial input state of the quantum register
and their measurement contributes via the
accumulated byproduct operator on $O$. Finally, the qubits in the section $M
\subset {\cal{C}}$ of the cluster whose measurements simulate the quantum
gates also contribute via the byproduct operator. 

Naturally there arises the question whether there is any difference in
the way how
measurements of cluster qubits in $I$, $O$ or $M$ contribute to the
final result of the computation. As shown in \cite{QCmodel}, it turns out
that there is none. This is why we can abandon the notions of quantum
input, quantum output and quantum register altogether from the
description of the \QCns. 
\begin{figure}
    \begin{center}
    \epsfig{file=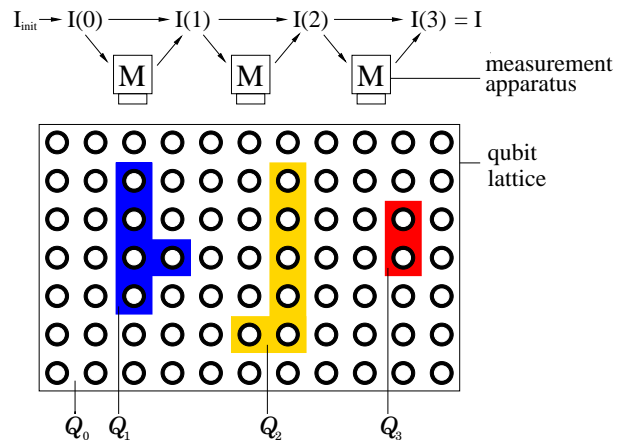,width=8cm}
    \parbox{\linewidth}{\caption{\label{Schema}General scheme of
        the quantum computer via one-qubit measurements. The sets
    $Q_t$ of lattice qubits are measured one after the other. The
    results of earlier measurements determine the measurement bases of
    later ones. All classical information from the measurement results
    needed to steer the \QC is contained in the information flow vector
    ${\bf{I}}(t)$. After the last measurement round
    $t_{\mbox{\footnotesize{max}}}$,
    ${\bf{I}}(t_{\mbox{\footnotesize{max}}})$ contains the result of
    the computation.}}  
    \end{center}
\end{figure}
Furthermore, quantum gates are not constitutive elements of the \QCns;
these are instead one-qubit measurements performed in a
certain temporal order and in a spatial pattern of adaptive
measurement bases. In fact, the most efficient temporal order of the
measurements does not 
follow from the temporal order of the simulated gates in the network
model.

The general view of a \QCns-computation is as follows. The cluster
${\cal{C}}$ is divided into disjoint subsets $Q_t \subset {\cal{C}}$
with $0 \leq t \leq t_{\mbox{\footnotesize{max}}}$, i.e.
$\bigcup_{t=0}^{t_{\mbox{\tiny{max}}}}Q_t = {\cal{C}}$ and $Q_s \cap
Q_t = \emptyset$ for all $s\not=t$. The cluster qubits within each set
$Q_t$ can be measured simultaneously and the sets are measured one
after another. The set $Q_0$ consists of all those qubits for which no
measurement bases have to be adjusted, i.e. those of which the
operator $\sigma_x$, $\sigma_y$ or $\sigma_z$ is measured. In the
subsequent measurement rounds only operators of 
the form $\cos \varphi \, \sigma_x \pm \sin \varphi \, \sigma_y$ are
measured where $|\varphi|<\pi/2$, $\varphi \neq 0$. The
measurement bases are adaptive in these rounds, i.e. they are adapted
to measurement results obtained in previous rounds. The measurement
outcomes from the qubits in $Q_0$ determine the measurement bases for
the qubits in $Q_1$, which are measured in the second round, those from
$Q_0$ and $Q_1$ together determine the bases for the measurements of the
qubits in $Q_2$ which are measured in the third round, and so
on. Finally, the result of the computation is 
calculated from the measurement outcomes obtained in all the measurement rounds.     

Now there arises the question of how complex the required classical
processing is. In principle it could be that all the obtained
measurement results had to be stored separately and the functions to
compute the measurement bases were so complicated that one would gain
no advantage over the classical algorithm for the considered problem.
This is not at all the case. If the network algorithm runs on
$n$ qubits then the classical data that the \QC has to keep track of
is entirely contained in a
$2n$-component binary valued vector, which we have called the
information flow vector 
${\bf{I}}(t)$ \cite{QCmodel}. The update of ${\bf{I}}(t)$ is a 
classical computation that is needed to adapt the  
measurement bases of cluster qubits according to previous measurement outcomes.
These updates and the final identification of the computational result
from ${\bf{I}}(t_\text{max})$ are all
elementary.

Concerning the resources for the classical processing of the
measurement outcomes in a \QCns-computation, we point out that this
processing increases the total time of 
computation only marginally \cite{QCmodel}.     

In summary, the formal description of the \QC is based on primitive
quantities of which the most important ones are the sets $Q_t \subset
{\cal{C}}$ of cluster 
qubits defining the temporal ordering of measurements on the cluster
state, and the binary valued information flow vector ${\bf{I}}(t)$
which is the carrier of the algorithmic information. The reader who is
interested in how this computational model arises and in its detailed
description is referred to \cite{QCmodel}.  

\subsection{Algorithms and graphs}
\label{graphs}

In this section we relate \QCns-algorithms to graphs. We do this by
considering non-universal graph states suited for the 
specific algorithm in question. For the \QCns, the Clifford part of
each algorithm can be removed. A mathematical graph comprises all the
information that needs to be kept the Clifford part.

While the network formulation of a quantum algorithm is given as a
sequence of quantum gates applied to a fiducial input state, the
\QCns-version of a quantum algorithm is specified by a measurement
pattern on the universal cluster state plus the structure
\cite{QCmodel} for the processing of the measurement outcomes.

The measurement pattern is, in the simplest case, just a copy of the
network layout to the substrate cluster state, imprinted by the
measurements. As such it contains information about the precise
location of the gate simulations and about the way the ``wires''
connecting the gates are bent around. These are all details of the
realization of an algorithm but do not belong to the description of
the algorithm itself. Thus, the measurement pattern introduces a large
amount of redundancy into the description of a \QCns-algorithm. This
redundancy may be reduced to a large extent by allowing for non-universal,
algorithm-specific quantum resources.

Clearly, at this point one has to specify how special the
algorithm-specific resource is allowed to be. Obviously it would make
no sense to take the quantum output of the entire network as the
required quantum resource and to regard the subsequent readout
measurements as the algorithm. Here, we allow for any {\em{graph state}}
\cite{Schlingel1}, (\ref{EVeqnG}) as the quantum resource. Graph
states are both easy to create and to describe. Every algorithm may be
run with a graph state as the quantum resource since the cluster state
is a particular graph state.

To allow for an algorithm-specific graph state as the quantum
resource of a \QCns-computation reduces the redundancy of both the
description and the realization of a quantum algorithm. This can
easily be seen
from the material presented in Section~\ref{redrem}. All the cluster
qubits $q \in {\cal{C}} \backslash {\cal{C}}_N$ can be get rid of
either by measuring them in the $\sigma_z$-eigenbasis or equivalently by not
placing them initially into their positions at all. The remaining state on the
sub-cluster ${\cal{C}}_N$ is again a cluster state. Hence it is also a 
graph state. It is less redundant and no longer universal. 

But we can go further. Not only the qubits measured in the
$\sigma_z$-eigenbasis may be removed from the cluster but instead all
those qubits of which one of the Pauli operators $\sigma_x$,
$\sigma_y$ or $\sigma_z$ is measured, i.e. all the qubits which form
the set $Q_0$. The state of the unmeasured qubits that emerges after
the measurement of the cluster qubits in $Q_0$ is again (local
equivalent to) a graph state.

This may be seen as follows. First note that the
operators $ \sigma_x^{(a)} \bigotimes_{b \in V} \left( \sigma_z^{(b)} 
    \right)^{\Gamma_{ab}}$ which appear in (\ref{EVeqnG}) form a
stabilizer of the state  $|\phi \{\kappa\} \rangle_G$. The generator
of the stabilizer contains $|{\cal{C}}|$ elements for a state of
$|{\cal{C}}|$ qubits. After all the qubits $q \in Q_0$ have been
measured, the resulting state $|\Psi\rangle_{{\cal{C}}\backslash
  Q_0}$ of the $|{\cal{C}}\backslash Q_0|$
unmeasured qubits is again described by a stabilizer of the form 
\begin{equation}
    \label{stabil}
    \begin{array}{r}
    \ds{\bigotimes_{i=1}^{|{\cal{C}}\backslash Q_0|}
    \left(\sigma_x^{(i)}\right)^{X_{a,i}}
    \left(\sigma_z^{(i)}\right)^{Z_{a,i}} \,
    |\Psi\rangle_{{\cal{C}}\backslash Q_0}} = \ds{\pm
    |\Psi\rangle_{{\cal{C}}\backslash Q_0}} \\
    \ds{\forall a=1\,..\,{|{\cal{C}}\backslash Q_0}|}, 
    \end{array}
\end{equation}           
with two $|{\cal{C}}\backslash Q_0|\times |{\cal{C}}\backslash
Q_0|$-matrixes $X$ and $Z$, for which 
$X_{a,i},Z_{a,i} \in \{0,1\}$. The $|{\cal{C}}\backslash Q_0| \times
2|{\cal{C}}\backslash Q_0|$-compound matrix $(X|Z)$ is 
called the generator matrix of the stabilizer for 
$|\Psi\rangle_{{\cal{C}}\backslash Q_0}$. The state 
$|\Psi\rangle_{{\cal{C}}\backslash Q_0}$ is uniquely
determined by the generator 
of its stabilizer.

The state $|\Psi\rangle_{{\cal{C}}\backslash Q_0}$ can thus be regarded as a
$[|{\cal{C}}\backslash Q_0|, 0, d]$-stabilizer code, with the distance
$d$ not specified. Whether a code with only one code word that
encodes 0 qubits should  be regarded as a code in the sense of
coding shall not concern us here. 
For the present purpose, it is important to note that the state 
$|\Psi\rangle_{{\cal{C}}\backslash Q_0}$ fulfills the assumptions of
Theorem 1 in \cite{Grassl}. The cited theorem states that any
stabilizer code over the alphabet $A=\mathbb{F}_{p^m}$ is [local
unitary] equivalent to a graph code. If we specialize to our case,
$A=\mathbb{F}_{2^2}$, we find that the state
$|\Psi\rangle_{{\cal{C}}\backslash Q_0}$ specified in (\ref{stabil}) is local
unitary equivalent to a graph state
$|\phi\{\kappa\}\rangle_{G({\cal{C}}\backslash Q_0,
E_{{{\cal{C}}\backslash Q_0}})}$ (\ref{EVeqnG}).  

That is, the state $|\Psi\rangle_{{\cal{C}}\backslash Q_0}$ needed for the
computation can be obtained from a graph state
$|\phi\{\kappa\}\rangle_{G({\cal{C}}\backslash Q_0,
E_{{{\cal{C}}\backslash Q_0}})}$ via local unitary
transformations. Subsequently in the process of computation, the qubits of
$|\Psi\rangle_{{\cal{C}}\backslash Q_0}$ are measured in their given
temporal order and in the appropriate adapted bases. An alternative
way to proceed is to use the graph state
$|\phi\{\kappa\}\rangle_{G({\cal{C}}\backslash Q_0, 
E_{{{\cal{C}}\backslash Q_0}})}$
directly, only modifying the measurement bases instead of performing
the local rotations prior to the measurements. Thus, in a
\QCns-computation with a special graph state as the quantum resource
and the first measurement round omitted, the way of processing the
classical information is the same as in a \QCns-computation with a
universal resource and the first measurement round performed.

The graphs associated with states (\ref{stabil}) are in general not
unique \cite{Grassl}. A constructive way to obtain graphs on
${{\cal{C}}\backslash Q_0}$ from $G({\cal{C}},E_{{\cal{C}}})$ and the
measurement 
bases of the qubits in $Q_0$ has been described in \cite{Schlingel2}.

Now note that the measurement of the qubits in $Q_0$ realize the
Clifford part of a quantum circuit. The fact that we can reduce the quantum
resource by these qubits means that {\em{we can remove from each quantum
algorithm its Clifford part}}. This represents, in a way, an extension
to the Knill-Gottesman-Theorem \cite{GKth}, stating that
a quantum 
computation that consist only of quantum input state preparation in
the computational basis, unitary gates in the Clifford group,
measurement of observables in the Pauli group, and gates in the
Clifford group conditioned on the outcomes of such measurements, may
be simulated efficiently classically and thus 
requires no quantum resources at all.  

With only a single non-Clifford operation in the circuit, such as a
one-qubit rotation about most axes and angles, the efficient classical
formalism upon which the Gottesman-Knill theorem rests can no longer
be applied. The \QCns-construction, on the other hand, is not affected
by this. Each quantum network algorithm 
in question may be reduced by its Clifford part. Only the non-Clifford gates 
require quantum resources. The price is that
the universal quantum resource, the cluster state, is changed into a
non-universal, algorithm-specific resource --a graph state
(\ref{EVeqnG})-- on fewer qubits. 
The Clifford part of the network algorithm specifies the corresponding graph.

In conclusion, instead of describing a quantum algorithm as a network of
gates applied to some fiducial input state, a quantum algorithm may
(arguably more effectively) be characterized by a graph specifying the
quantum resource and the structure \cite{QCmodel} for the processing
of the measurement outcomes. 

\section{Examples of practical interest}
\label{examples}

\subsection{Multi-qubit swap gate}
\label{sec:multiswap}

\begin{figure}
    \begin{center}
        \begin{tabular}{ll}
        a) \\
        & \epsfig{file=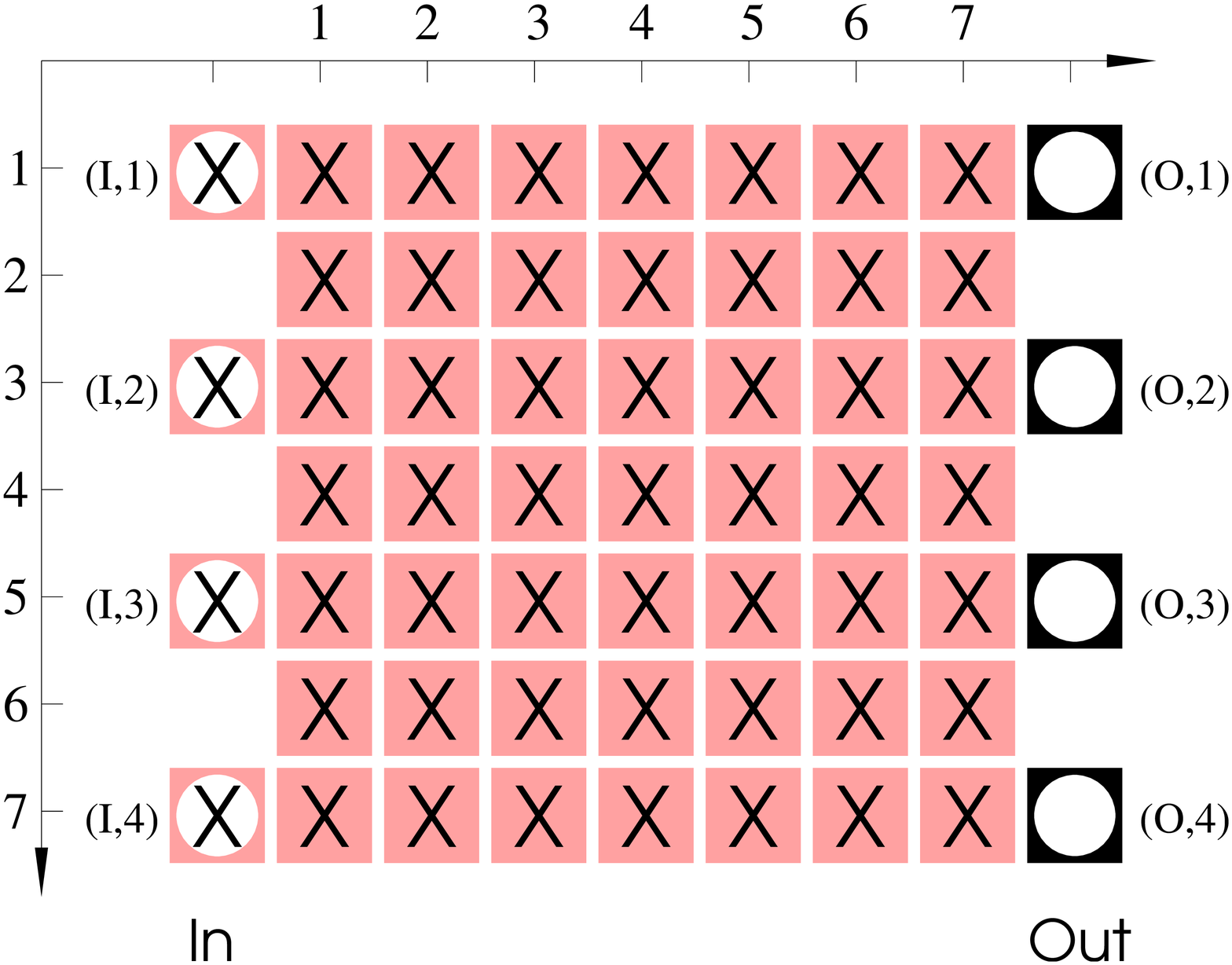, width=5.5cm} \\
        b) \\
        & \epsfig{file=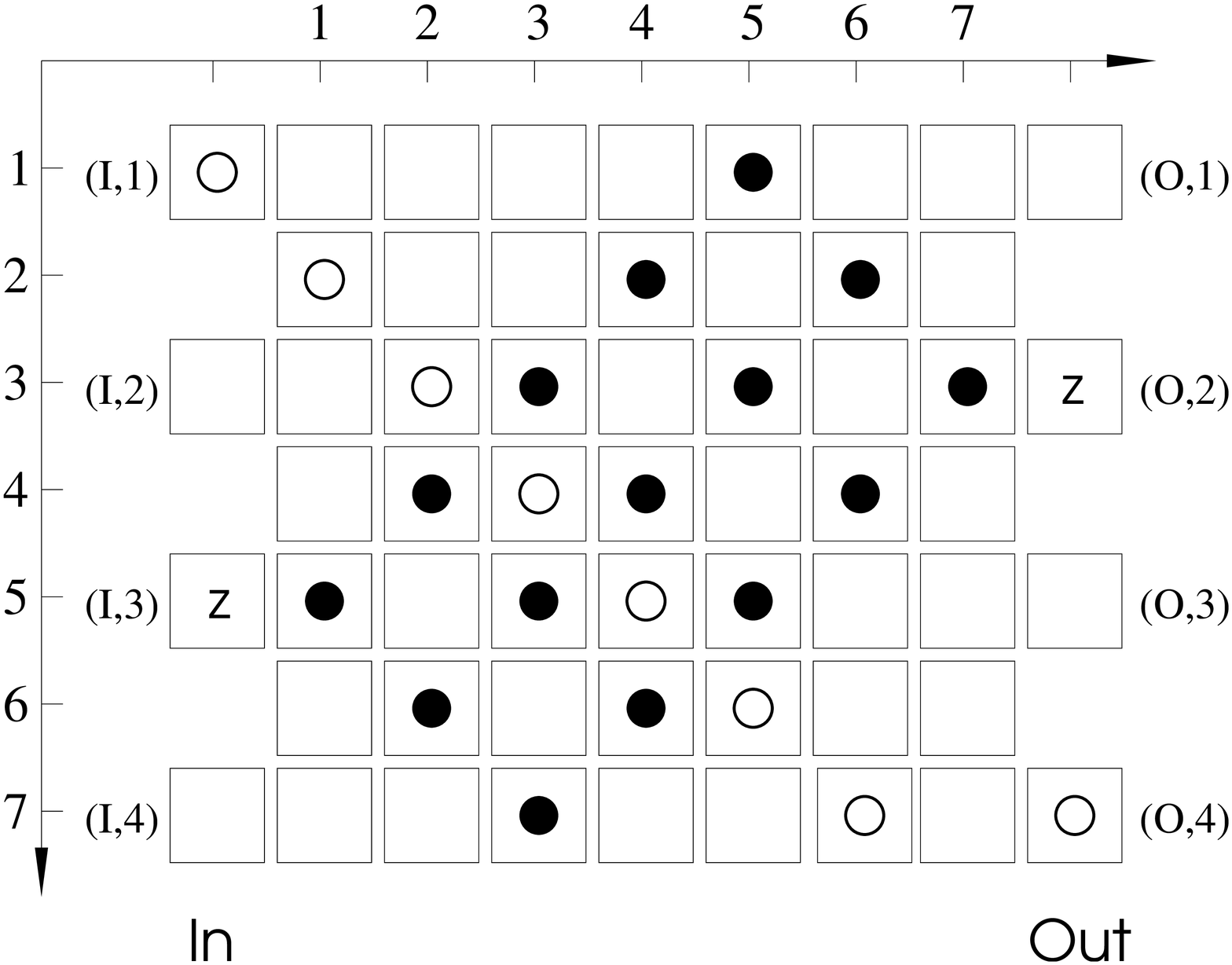, width=5.5cm}
        \end{tabular}
        \caption{\label{fig:swap}The multi-qubit swap gate. a)
        measurement pattern to realize the swap gate. b) Correlation
        centers for two correlations of the projected state
        $|\psi\rangle_{{\cal{C}}(\text{swap})}$ as inherited from
        correlations of $|\phi\rangle_{{\cal{C}}(\text{swap})}$. The
        correlation $\sigma_x^{(I,1)}\sigma_x^{(O,4)}$ of
        $|\psi\rangle_{{\cal{C}}(\text{swap})}$ stems from the product 
        correlation for $|\phi\rangle_{{\cal{C}}(\text{swap})}$ with
        the centers $a$ of basic correlation operators $K^{(a)}$
        denoted by ``$\circ$''. The centers of the initial correlation,
        which after the measurements induces the correlation
        $\sigma_z^{(I,3)}\sigma_z^{(O,2)}$ of  
        $|\psi\rangle_{{\cal{C}}(\text{swap})}$, are denoted by
        ``$\bullet$''.}
  \end{center}
\end{figure}

A multi-qubit swap gate is an $n$-qubit generalization of the
two-qubit swap gate. It reverses the order of the $n$-qubits,
interchanging qubit $i$ with $n+1-i$, $i=1,2,..,N$. This can be
realized in a simple way
on the \QC, as shown in Fig.~\ref{fig:swap}a. The measurement pattern
$\cal M$ on ${\cal{C}}_M$ consists of a square of 
$\sigma_x$ measurements, with sides of $2n-1$ cluster qubits. The
input qubits are, simultaneously with the qubits in ${\cal{C}}_M$,
also measured in the $\sigma_x$-eigenbasis. 

It can be verified using the methods introduced above that realizing $\cal
M$ leads to correlations between the $i$th input qubit and the
$n+1-i$th output qubit. Here, we discuss the four-qubit swap as a
particular example. 

After the
$\sigma_x$-measurements of the qubits in ${\cal{C}}_M$ we obtain for
the projected state $|\psi\rangle_{{\cal{C}}(\text{swap})}$ the
eigenvalue equations 
\begin{equation}
    \label{swapEveqs}
    \begin{array}{rcl}
        \ds{\sigma_x^{(I,1)}
        \sigma_x^{(O,4)}\,|\psi\rangle_{{\cal{C}}(\text{swap})}} &=& 
        \ds{{(-1)}^{\lambda_{x,1}} |\psi\rangle_{{\cal{C}}(\text{swap})}},\\
        \ds{\sigma_x^{(I,2)}
        \sigma_x^{(O,3)}\,|\psi\rangle_{{\cal{C}}(\text{swap})}} &=& 
        \ds{{(-1)}^{\lambda_{x,2}} |\psi\rangle_{{\cal{C}}(\text{swap})}},\\
        \ds{\sigma_x^{(I,3)}
        \sigma_x^{(O,2)}\,|\psi\rangle_{{\cal{C}}(\text{swap})}} &=& 
        \ds{{(-1)}^{\lambda_{x,3}} |\psi\rangle_{{\cal{C}}(\text{swap})}},\\
        \ds{\sigma_x^{(I,4)}
        \sigma_x^{(O,1)}\,|\psi\rangle_{{\cal{C}}(\text{swap})}} &=& 
        \ds{{(-1)}^{\lambda_{x,4}}
        |\psi\rangle_{{\cal{C}}(\text{swap})}}, \vspace{3mm}\\
      
        \ds{\sigma_z^{(I,1)}
        \sigma_z^{(O,4)}\,|\psi\rangle_{{\cal{C}}(\text{swap})}} &=& 
        \ds{{(-1)}^{\lambda_{z,1}} |\psi\rangle_{{\cal{C}}(\text{swap})}},\\
        \ds{\sigma_z^{(I,2)}
        \sigma_z^{(O,3)}\,|\psi\rangle_{{\cal{C}}(\text{swap})}} &=& 
        \ds{{(-1)}^{\lambda_{z,2}} |\psi\rangle_{{\cal{C}}(\text{swap})}},\\
        \ds{\sigma_z^{(I,3)}
        \sigma_z^{(O,2)}\,|\psi\rangle_{{\cal{C}}(\text{swap})}} &=& 
        \ds{{(-1)}^{\lambda_{z,3}} |\psi\rangle_{{\cal{C}}(\text{swap})}},\\
        \ds{\sigma_z^{(I,4)}
        \sigma_z^{(O,1)}\,|\psi\rangle_{{\cal{C}}(\text{swap})}} &=& 
        \ds{{(-1)}^{\lambda_{z,4}}
        |\psi\rangle_{{\cal{C}}(\text{swap})}}.
    \end{array}
\end{equation}
Therein, the parameters $\lambda_{k,x}, \lambda_{k,z} \in \{0,1\}$
depend linearly on the measurement outcomes $\{s_{(i,j)}\}$. Therein,
$i$ is the value of the $x$- and $j$ the value of the $y$-coordinate
of the respective qubit site.
For example, $\lambda_{x,1}=s_{(1,2)}+s_{(2,3)}+s_{(3,4)}+
s_{(4,5)}+s_{(5,6)}+s_{(6,7)}\;\text{mod}\,\,2$. 

The eigenvalue equations (\ref{swapEveqs}) can be derived from
corresponding  eigenvalue equations for the cluster state
$|\phi\rangle_{{\cal{C}}(\text{swap})}$ on the cluster
${\cal{C}}(\text{swap})$. The required initial correlations are
products of the basic 
correlation operators (\ref{BasCorr}). The way to obtain the equations
(\ref{swapEveqs}) is rather straightforward and therefore we omit 
the detailed derivations. In Fig.~\ref{fig:swap}b two examples for the
composition of product correlation operators from basic correlation
operators $K^{(a)}$ are illustrated. The first line of (\ref{swapEveqs}),
\begin{equation}
    \ds{\sigma_x^{(I,1)}
        \sigma_x^{(O,4)}\,|\psi\rangle_{{\cal{C}}(\text{swap})}} = 
        \ds{{(-1)}^{\lambda_{x,1}}
        |\psi\rangle_{{\cal{C}}(\text{swap})}}, \nonumber
\end{equation}  
for example, is derived from the eigenvalue equation
\begin{equation}
     |\phi\rangle_{{\cal{C}}(\text{swap})} = K^{({\cal{C}}_{x,1})}
      |\phi\rangle_{{\cal{C}}(\text{swap})},
\end{equation}
with  
\begin{equation}
    K^{({\cal{C}}_{x,1})}= \prod_{a \in {\cal{C}}_{x,1}} K^{(a)},
\end{equation}
and ${\cal{C}}_{x,1}= \{ (I,1), (1,2), (2,3), (3,4), (4,5),
(5,6),$ $(6,7),$ $(O,4) \}$. Multiplying it out we find that all
operators $\sigma_z$ cancel and that
\begin{equation}
    K^{({\cal{C}}_{x,1})}= \prod_{a \in {\cal{C}}_{x,1}} \sigma_x^{(a)}.
\end{equation}
It is now easy to see that after the $\sigma_x$ measurements of the
qubits in ${\cal{C}}_M$ there remains a strict $\sigma_x^{(I,1)}
\sigma_x^{(O,4)}$-correlation for the state
$|\psi\rangle_{{\cal{C}}(\text{swap})}$. A similar construction can be
given to obtain the $\sigma_z^{(I,1)} \sigma_z^{(O,4)}$-correlation.

With the eigenvalue equations (\ref{swapEveqs}) the assumptions of
theorem~\ref{gatecond} are fulfilled and thus via the described
measurement pattern a unitary operation $U=SWAP$ is realized modulo
a byproduct operator as specified in (\ref{ByProp}). To exchange the
order of the swap-gate $U_{\text{swap}}$ and the byproduct operator
  $U_\Sigma$ the byproduct operator is conjugated under
  $U_{\text{swap}}$, as usual for gates in the Clifford group. 

\subsection{Simulating multi-qubit Hamiltonians}
\label{sec:HamilSim}

Here we display a gate which simulates the unitary evolution with
$U=\exp(-i H_4 t)$ of the quantum 
input for the multi-particle Hamiltonian
\begin{equation}
    \label{simHamil}
    H_4=g\, \sigma_z^{(1)} \sigma_z^{(2)} \sigma_z^{(3)} \sigma_z^{(4)}
\end{equation}
and {\em{arbitrary}} times $t$. In addition, the gate performs a swap 
which can be corrected for by a subsequent swap gate as described in
Section~\ref{sec:multiswap}.  

\begin{figure}
    \begin{center}
        \begin{tabular}{ll}
        a) \\
        & \epsfig{file=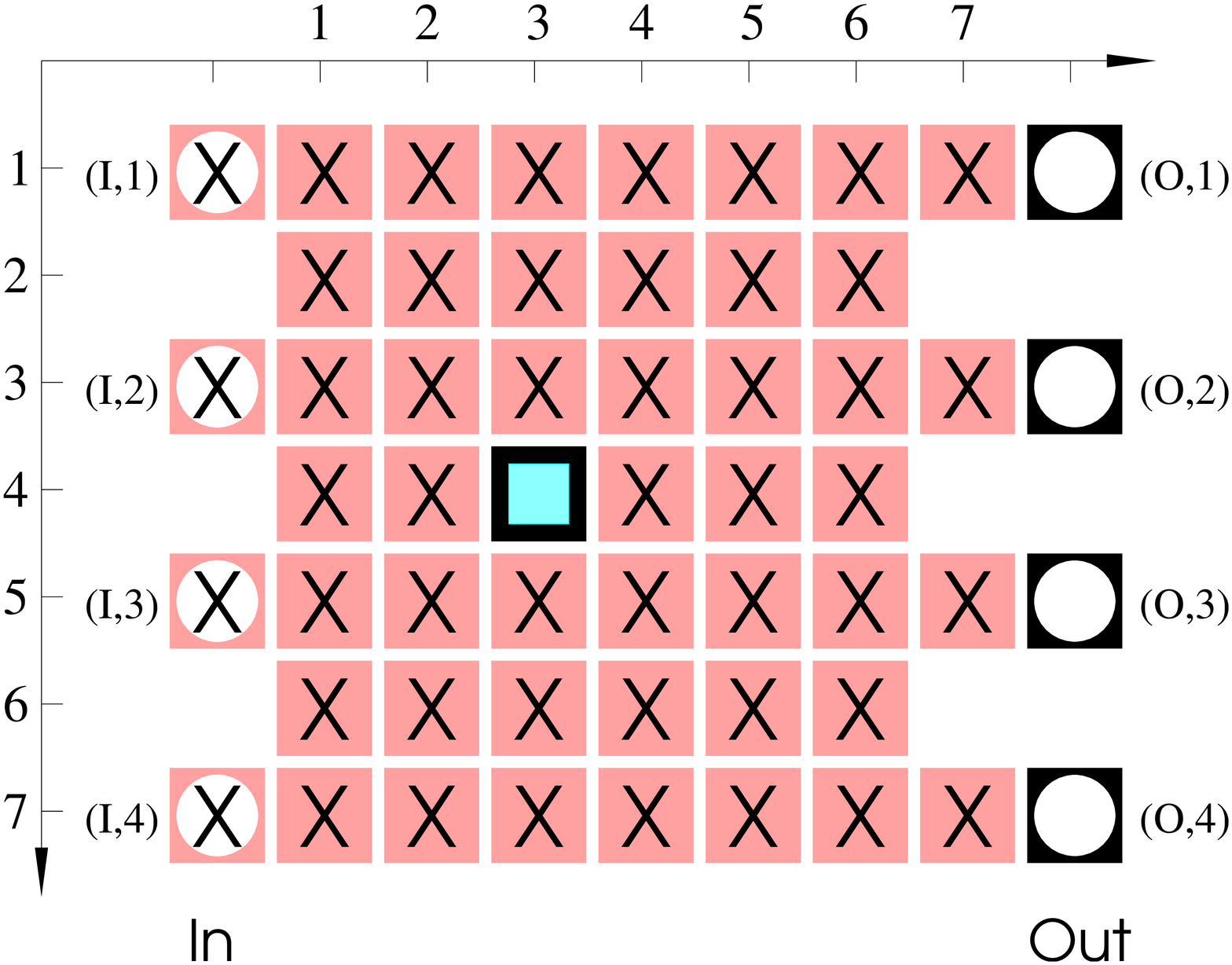, width=5.5cm} \\
        b) \\
        & \epsfig{file=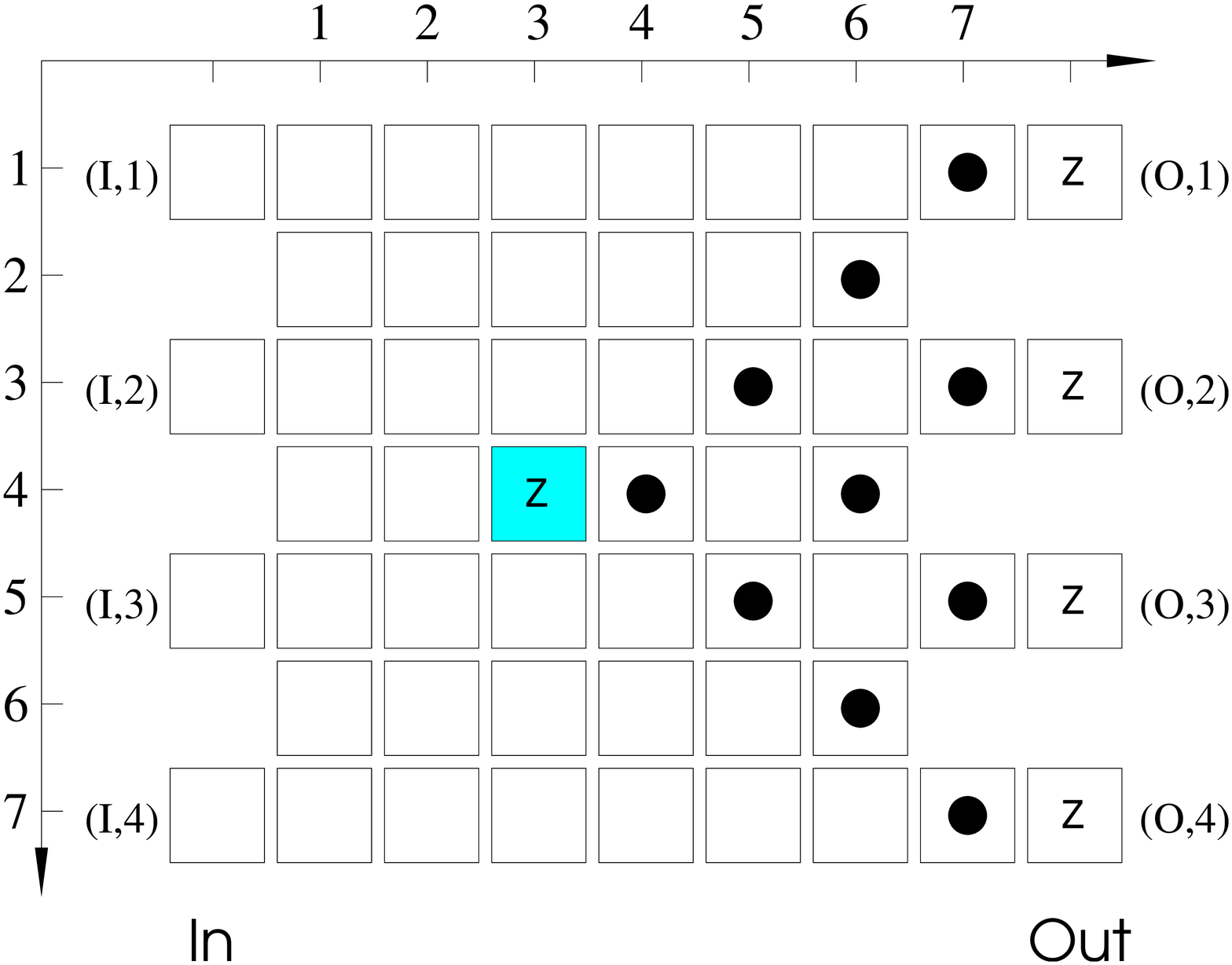, width=5.5cm}
        \end{tabular}
        \caption{\label{fig:hamil}Simulation of the Hamiltonian $H_4$
        as specified in eq. (\ref{simHamil}). a)
        measurement pattern. b) Correlation centers for additional
        correlation. Shaded squares (in b)) represent cluster qubits
        measured in adaptive bases.} 
    \end{center}
\end{figure}

The procedure to realize the measurement pattern ${\cal{M}}$ for
Hamiltonian simulation, as shown in Fig.~\ref{fig:hamil}, 
requires two rounds of measurements. In the first round all the
$\sigma_x$-measurements are performed. In the second measurement
round, of the qubit $(3,4)$ the operator 
\begin{equation}
    \label{mQ1}
    \vec{r}_{(3,4)}\cdot \vec{\sigma} = U_z[(-1)^{\lambda_M} 2 \varphi]
    \, \sigma_x \, U_z^\dagger[(-1)^{\lambda_M} 2\varphi]
\end{equation} 
is measured, where $U_z[\alpha] = \exp(-i \alpha \sigma_z/2)$. Therein, the
angle $\varphi$ is given by 
\begin{equation}
    \label{gt}
    \varphi=gt,
\end{equation} 
and $\lambda_M \in \{0,1\}$, which depends linearly on outcomes of measurements
in the first round, will be specified below. 
 
To understand the functioning of the Hamiltonian simulator let us
first discuss the state $|\psi^\prime\rangle$ on the cluster
${\cal{C}}(\text{sim})$ after the first round of measurements. By
arguments analogous to those used in Section~\ref{sec:multiswap}, the
state $|\psi^\prime\rangle$ obeys the following eigenvalue equations:
\begin{equation}
    \label{simEveqs1}
    \begin{array}{rcl}
        \ds{\sigma_x^{(3,4)} \sigma_x^{(I,1)}
        \sigma_x^{(O,4)}\,|\psi^\prime\rangle}&=& 
        \ds{{(-1)}^{\lambda_{x,1}} |\psi^\prime\rangle,}\\
        \ds{\sigma_x^{(3,4)} \sigma_x^{(I,2)}
        \sigma_x^{(O,3)}\,|\psi^\prime\rangle}&=& 
        \ds{{(-1)}^{\lambda_{x,2}} |\psi^\prime\rangle,}\\
        \ds{\sigma_x^{(3,4)} \sigma_x^{(I,3)}
        \sigma_x^{(O,2)}\,|\psi^\prime\rangle}&=& 
        \ds{{(-1)}^{\lambda_{x,3}} |\psi^\prime\rangle,}\\
        \ds{\sigma_x^{(3,4)} \sigma_x^{(I,4)}
        \sigma_x^{(O,1)}\,|\psi^\prime\rangle}&=& 
        \ds{{(-1)}^{\lambda_{x,4}}
        |\psi^\prime\rangle,} \vspace{3mm}\\
      
        \ds{\sigma_z^{(I,1)}
        \sigma_z^{(O,4)}\,|\psi^\prime\rangle} &=& 
        \ds{{(-1)}^{\lambda_{z,1}} |\psi^\prime\rangle},\\
        \ds{\sigma_z^{(I,2)}
        \sigma_z^{(O,3)}\,|\psi^\prime\rangle} &=& 
        \ds{{(-1)}^{\lambda_{z,2}} |\psi^\prime\rangle,}\\
        \ds{\sigma_z^{(I,3)}
        \sigma_z^{(O,2)}\,|\psi^\prime\rangle} &=& 
        \ds{{(-1)}^{\lambda_{z,3}} |\psi^\prime\rangle,}\\
        \ds{\sigma_z^{(I,4)}
        \sigma_z^{(O,1)}\,|\psi^\prime\rangle} &=& 
        \ds{{(-1)}^{\lambda_{z,4}}
        |\psi^\prime\rangle.}
    \end{array}
\end{equation}
Further, the state $|\psi^\prime\rangle$ obeys the eigenvalue equation
\begin{equation}
    \label{simEveqs2}
    \sigma_z^{(3,4)}\,
        \sigma_z^{(O,1)} \sigma_z^{(O,2)} \sigma_z^{(O,3)}
        \sigma_z^{(O,4)} \,|\psi^\prime\rangle = 
        {(-1)}^{\lambda} |\psi^\prime\rangle,
\end{equation}  
with $\lambda \in \{0,1\}$ linear in the measurement outcomes of the
first round. Equation (\ref{simEveqs2}) can be
easily verified with the pattern of correlation centers displayed in
Fig.~\ref{fig:hamil}b. From (\ref{simEveqs2}) it follows that
\begin{equation}
    \label{simEveqs2b}
    \exp\left(i \theta \,\sigma_z^{(3,4)}\right) U_4[
     (-1)^\lambda \theta]
     \,|\psi^\prime\rangle = |\psi^\prime\rangle
\end{equation}
for arbitrary angles $\theta$, with 
\begin{equation}
    \label{U4}
    U_4[\alpha]= \exp\left(-i \alpha\, 
        \sigma_z^{(O,1)} \sigma_z^{(O,2)} \sigma_z^{(O,3)}
        \sigma_z^{(O,4)}\right).
\end{equation}
Equation (\ref{simEveqs2b}) is now inserted in both the l.h.s. and r.h.s. of
the equations (\ref{simEveqs1}). For example, with the first equation
from (\ref{simEveqs1}) one obtains  
\begin{equation}
    \label{simEveqs2c}
    \begin{array}{rcl}
        \ds{(-1)^{\lambda_{x,1}}\,|\psi^\prime\rangle} &\!=\!&   
        \ds{{\left( U_z[2\theta] \sigma_x U_z^\dagger[2\theta] \right)}^{(3,4)}
        \sigma_x^{(I,1)}} \vspace{1.5mm} \\ 
        && \ds{\left(U_4[-(-1)^\lambda \theta] \sigma_x^{[4]}
        U_4^\dagger [-(-1)^\lambda \theta]\right)^{(O)} \!
        |\psi^\prime\rangle.} 
    \end{array} 
\end{equation}
In the second measurement round the qubit $(3,4)$ is the only one left
to be measured. As can be seen from (\ref{simEveqs2c}), if of the
operator $U_z[2\theta] \sigma_x U_z^\dagger[2\theta]$  of qubit (3,4)
is measured then the state $|\psi\rangle$, into which the cluster qubits are
projected after the second measurement round, obeys the eigenvalue
equation
\begin{equation}
    \label{simEveqs3a}
    \begin{array}{l}
        \ds{(-1)^{\lambda_{x,1}+s_{(3,4)}}\,|\psi\rangle} =   
        \vspace{1.5mm} \\ 
        \hspace*{5mm} \ds{\sigma_x^{(I,1)} \left(U_4[-(-1)^\lambda
        \theta] \sigma_x^{[4]} 
        U_4^\dagger [-(-1)^\lambda \theta]\right)^{(O)} |\psi\rangle.}
    \end{array}
\end{equation}
If we carry out this procedure for all equations in (\ref{simEveqs1})
we find that the state $|\psi\rangle$ that emerges after the second
measurement round obeys the eigenvalue equations
\begin{equation}
    \label{simEveqs3}
    \begin{array}{rcl}
        \ds{\sigma_x^{(I,i)} \left( U_4 U_{\text{swap}} \sigma_x^{[i]}
        {U_{\text{swap}}}^{\!\!\dagger} {U_4}^{\!\!\dagger}
        \right)^{(O)}|\psi\rangle} &\!\!=\!\!& 
        \ds{{(-1)}^{\lambda_{x,i}+s_{(3,4)}}\,|\psi\rangle,}
        \vspace{1mm} \\
        \ds{\sigma_z^{(I,i)} \left( U_4 U_{\text{swap}} \sigma_z^{[i]}
        {U_{\text{swap}}}^{\!\!\dagger} {U_4}^{\!\!\dagger}
        \right)^{(O)}|\psi\rangle}  &\!\!=\!\!& 
        \ds{{(-1)}^{\lambda_{z,i}}\,|\psi\rangle,} 
    \end{array}
\end{equation}
for $i=1\,..\,4$ and with $U_4$ written in short for
$U_4[-(-1)^\lambda \theta]$. 

With the set of equations (\ref{simEveqs3}) the assumptions
(\ref{gatecheck}) of theorem~\ref{gatecond} are
fulfilled. With theorem~\ref{gatecond} it follows that the measurement
pattern displayed in Fig.~\ref{fig:hamil} realizes a unitary
transformation
\begin{equation}
    \label{Usim1}
    U_\text{sim} = U_4[-(-1)^\lambda \theta] U_{\text{swap}} U_\Sigma,
\end{equation} 
where the byproduct operator is given by
\begin{equation}
    \label{ByprU4}
    U_\Sigma= \bigotimes_{i=1}^4
    \left(\sigma_z^{[i]}\right)^{s_{(I,i)}+\lambda_{x,i}+s_{(3,4)}}
    \left(\sigma_x^{[i]}\right)^{\lambda_{z,i}}.
\end{equation}
Finally, the order of the operators has to be exchanged. Note that
$U_{\text{swap}}$ and $U_4$ commute. From (\ref{Usim1}) one finds
\begin{equation}
    U_\text{sim} = U_\Sigma^\prime U_{\text{swap}}
    U_4[-(-1)^{\lambda+\sum_{i=1}^4 \lambda_{z,i}} \,\theta],
\end{equation}
with
\begin{equation}
    U_\Sigma^\prime = U_{\text{swap}} \, U_\Sigma \,
    {U_{\text{swap}}}^\dagger.
\end{equation}
Thus, in order to realize $U_4[\varphi]$ with $\varphi$ specified in
(\ref{gt}) we must choose 
\begin{equation}
    \theta=(-1)^{1+\lambda+\sum_{i=1}^4 \lambda_{z,i}} \varphi.
\end{equation}
That is, in the second measurement round we measure on the qubit
$(3,4)$ the operator given in (\ref{mQ1}), where
\begin{equation}
    \lambda_M = \left(1+ \lambda+\sum_{i=1}^4 \lambda_{z,i}\right)
    \;\text{mod}\, 2.
\end{equation}
The $\{\lambda_{x,i}\}$, $\{\lambda_{z,i}\}$ and $\lambda$ depend
linearly on the measurement outcomes $\{s_{(i.j)} \}$ obtained in the
first measurement round.

The sub-circuit we have described in this section simulates the
unitary evolution according to a particular four-particle Hamiltonian
in a two-step process of measurements. The time for which the
simulated Hamiltonian acts is encoded in the basis of the measurement
in the second round. 

The generalization of the simulation of the 4-particle Hamiltonian
$H_4$, shown in Fig.~\ref{fig:hamil}, to an arbitrary number
$n$ of qubits, i.e. the simulation of the Hamiltonian
$H_n=\bigotimes_{i=1}^n \sigma_z^{[i]}$, is straightforward. 
   
\subsection{CNOT between non-neighbouring qubits}
\label{sec:non-neigh-cnot}

The CNOT gate described in Section~\ref{sec:cnot-gate} operates on two
logical qubits whose input qubits are adjacent to each
other on the cluster. However, for universal quantum computation, one
must be able to realize a CNOT gate between any two logical
qubits. While this could be achieved using a combination of the CNOT
gate, introduced above, and the swap gate, the width of the
measurement pattern needed to realize this would grow linearly with
the separation of the two logical qubits. There
is, however, an alternative measurement pattern, which, at the cost of doubling
the spacing between the input qubits on the cluster, has a fixed width. 
\begin{figure}
  \epsfig{file=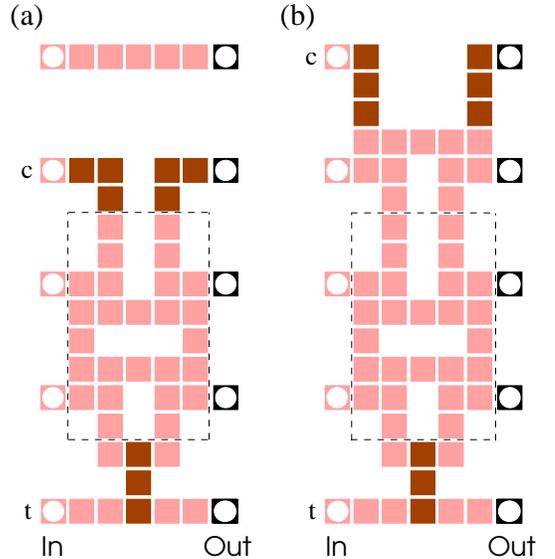, width=7cm}
  \caption{\label{fig:nonneighcnot}  Measurement pattern for a
    CNOT gate between two logical qubits whose input and output qubits
    are not neighbors. Squares in light gray denote cluster qubits
    measured in the eigenbasis of $\sigma_x$, in dark gray of $\sigma_y$.
    Pattern (a) is for the case where the two  qubits are  separated by an
    odd number of logical qubits. Pattern (b) is for an even
    numbered separation. The patterns can be adapted to any separation by
    repeating the  
    section enclosed by the dashed line. The width of the pattern remains
    the same for all separations.}
\end{figure}
The measurement pattern is illustrated in
Fig.~\ref{fig:nonneighcnot} for qubits separated by an odd and even
number of logical qubits, respectively.

\begin{figure}
\epsfig{file=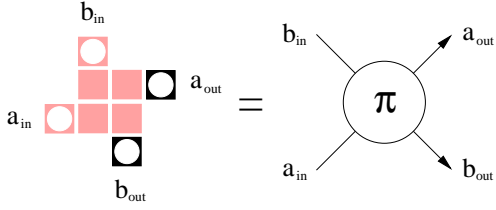, width=6.5cm}
\caption{\label{fig:CNOTsubcircuit}This measurement pattern is one of
  the key components of the measurement pattern in
  Fig.~\ref{fig:nonneighcnot}. It performs a conditional
  $\pi$-phase-gate and a swap-gate.} 
\end{figure}

This layout can be understood within the quantum logic network model. The
``wires'' for the logical qubits in between the target- and the
control qubit are crossed using the measurement sub-pattern
illustrated in Fig.~\ref{fig:CNOTsubcircuit}a. However, as well as swapping
the qubits, this pattern also realizes the a controlled $\pi$-phase gate,
also known as a  controlled $\sigma_z$ gate, illustrated in
Fig.~\ref{fig:CNOTsubcircuit}b. 

\begin{figure}
  \epsfig{file=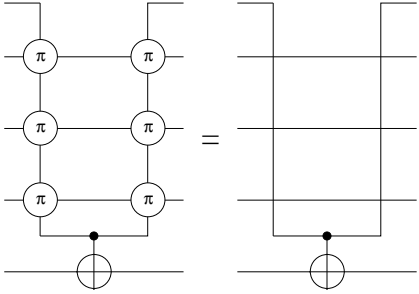, width=5.5cm}
\caption{\label{fig:cnotcircuit}  The measurement pattern in
  Fig.~\ref{fig:nonneighcnot} realizes the quantum logic circuit
  on the left hand side of this figure. This network is
  equivalent to the one on the right hand side, where the only
  gate realized is the CNOT between the two desired non-adjacent
  qubits. } 
\end{figure}

The quantum logic circuit realized by the whole measurement
pattern, illustrated on the left-hand side of
Fig.~\ref{fig:cnotcircuit} uses these sub-patterns to swap the
positions of adjacent qubits. This brings non-neighboring qubits
together so that a CNOT operation may be performed on them.  

The networks on the left and on the right of
Fig.~\ref{fig:cnotcircuit} act identically, and thus the measurement
pattern displayed in Fig.~\ref{fig:nonneighcnot} realizes a distant
CNOT-gate.

\subsection{Controlled Phase Gate}
\label{sec:quantum-phase-gate}

Here, we give an example of another two-qubit gate which can be
realized  without 
decomposing it into CNOTs and  rotations, the controlled phase gate
$U_\text{CPG}(\theta)$. This gate realizes the unitary operation
\begin{equation}
\label{CPGdef}
U_{\text{CPG}}[\theta]= \iden^{(ab)}+ (e^{i\theta}-1)\,
|11\rangle_{ab}\langle 11 |, 
\end{equation}
applied to the two qubits $a$ and $b$.

We can write this in terms of the following one- and two-qubit rotations,
\begin{equation}
\label{CPGdecomp}
U_{\text{CPG}}[\theta]=
 e^{i\frac{\theta}{4}} U_{zz}^{(ab)}[-\theta/{2}]
 U_{z}^{(a)}[\theta/{2}] U_{z}^{(b)}[\theta/{2}], 
\end{equation}
where the two-qubit rotation is 
\begin{equation}
    \label{tworot}
    U_{zz}^{(ab)}[\theta]=\exp\left(-\im \theta/2 \sigma_z^{(a)}
    \sigma_z^{(b)}\right). 
\end{equation} 
 
\begin{figure}
    \begin{center}
        \epsfig{file=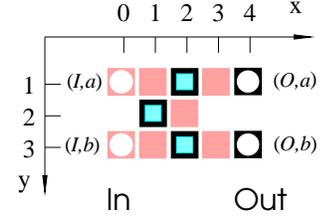, width=4cm}
        \caption{\label{fig:CPG}Controlled phase gate with additional swap.}
    \end{center}
\end{figure}

This representation is particularly convenient for finding the
measurement pattern that realizes the gate, since rotations
$U_z[\theta/2]$ and $U_{zz}[-\theta/2]$ are realized on the \QC in a
simple natural way. The measurement pattern is illustrated in
Fig.~\ref{fig:CPG}, in which the labelling of the qubits is also
defined.  

We follow the same method as above, beginning with the eigenvalue
equations of the cluster state $\ket{\phi}_{\cal C}$ on the qubits
shown. The $\sigma_x$-measurements can be considered first, using the
methods already illustrated in this paper. The resultant state of the
remaining qubits $\ket{\psi^\prime}$, after this sub-set of the
measurements has been carried out, is defined by the following set of
eigenvalue equations. 
\begin{subequations}
\label{cpgeq}
\begin{align}
\sigma_x^{(a,I)}\sigma_x^{(1,2)}\sigma_x^{(2,3)}\sigma_x^{(b,O)}
\ket{\psi^\prime}&=\ket{\psi^\prime},\label{cpgeq1}\\ 
\sigma_x^{(b,I)}\sigma_x^{(1,2)}
\sigma_x^{(2,1)}\sigma_x^{(a,O)} \ket{\psi^\prime}&=\ket{\psi^\prime},\\ 
\sigma_z^{(a,I)}\sigma_z^{(b,O)}
\ket{\psi^\prime}&=(-1)^{s_{(1,1)}+s_{(2,2)}+s_{(3,3)}} \ket{\psi^\prime},\\ 
 \sigma_z^{(b,I)}\sigma_z^{(a,O)}
\ket{\psi^\prime}&=(-1)^{s_{(1,3)}+s_{(2,2)}+s_{(3,1)}} \ket{\psi^\prime},
\end{align}
\end{subequations}
and
\begin{subequations}
\label{cpgeqsB}
\begin{align}
\sigma_z^{(2,1)}\sigma_z^{(a,O)} \ket{\psi^\prime}&=(-1)^{s_{(3,1)}}
\ket{\psi^\prime}, \label{cpgeq4}\\ 
\sigma_z^{(2,3)}\sigma_z^{(b,O)}
\ket{\psi^\prime}&=(-1)^{s_{(3,3)}}\ket{\psi^\prime}, \label{cpgeq5}\\ 
\sigma_z^{(1,2)}\sigma_z^{(a,O)}  \sigma_z^{(b,O)}
\ket{\psi^\prime}&=(-1)^{s_{(3,1)}+s_{(2,2)}+s_{(3,3)}}
\ket{\psi^\prime}. \label{cpgeq6}  
\end{align}
\end{subequations}

As in section~\ref{sec:xrot}, eigenvalue equations are now generated
which commute with the remaining measurements in $\cal M$, namely the
measurements of $\sigma_{xy}^{(i)}(\alpha_i)$ on qubits
$i\in\{(2,1),(1,2),(2,3)\}$. First, we manipulate the equations
(\ref{cpgeqsB})such that, for example, the eigenvalue equation
\eqref{cpgeq6} attains the form 
\begin{equation}
    \label{cpgeq7}
    \begin{array}{l}
        U_z^{(1,2)}[\xi] \,U_{zz}^{\left( (O,a), (O,b)
          \right)}\left[-(-1)^{s_{(3,1)}+s_{(2,2)}+s_{(3,3)}} 
            \xi\right] \, \ket{\psi^\prime}\vspace{1mm} \\ \;\;\;=
          \ket{\psi^\prime}. 
    \end{array}
\end{equation}  
Similar equations containing one-qubit rotations on qubits $(2,1)$ and
$(O,a)$, and $(2,3)$ and $(O,b)$ are derived from the other equations
of (\ref{cpgeqsB}) in the same way. These
equations are inserted into both sides of the eigenvalue equations
(\ref{cpgeq}) so
that, using the method introduced above, we obtain a set of four 
eigenvalue equations for $|\psi^\prime\rangle$ which induce a set of four
eigenvalue equations for the state $|\psi\rangle$ that one obtains after
the remaining measurements have been carried out.  

Specifically, in the second measurement round the qubits
$(1,2)$, $(2,1)$ and $(2,3)$ are measured. Of these qubits one
measures the observables   
\begin{equation}
    \label{CPGround2}
    \vec{r}_a \cdot \vec{\sigma}^{(a)} = \left(U_z[\alpha_a] \sigma_x
    U_z[\alpha_a]^\dagger \right)^{(a)},
\end{equation}
for $a \in \{ (1,2), (2,1), (2,3) \}$ and the $\{\alpha_a\}$ specified
below.

The induced eigenvalue equations for the state $|\psi\rangle$ are of
the form of equation~(\ref{gatecheck}), and the unitary operation
realized by 
the gate can be read off from them using theorem~\ref{gatecond}.
The full unitary operation realized by the measurement pattern is 
\begin{widetext}
\begin{equation}
    \label{CPGop1}
    \begin{array}{rcl}
        U' U_\Sigma' &=& \ds{U_{zz}^{(a,b)}\bigl[-(-1)^{s_{(3,1)}+s_{(2,2)}
              +s_{{(3,3)}}} \alpha_{(1,2)} \bigr]
        U_z^{(a)}\bigl[-(-1)^{s_{(3,1)}} \alpha_{(2,1)} \bigr] \,
        U_z^{(b)}\bigl[-(-1)^{s_{{(3,3)}}} \alpha_{(2,3)} \bigr] 
        U_{\text{swap}}^{(a,b)}} \vspace{1mm} \\ & & 
        \left(\sigma_x^{(a)} \right)^{s_{(1,1)}+s_{(2,2)}+s_{(3,3)}}
        \left(\sigma_x^{(b)} \right)^{s_{(1,3)}+s_{{(2,2)}}+s_{(3,1)}}  
        \left(\sigma_z^{(a)}\right)^{s_{(I,a)}+s_{(1,2)}+s_{(2,3)}} 
        \left(\sigma_z^{(b)}\right)^{s_{(I,b)}+s_{(2,1)}+s_{(1,2)}}  
    \end{array}
\end{equation}
such that after the order of the gate and the byproduct operator is
reversed, $ U' U_\Sigma'= U_\Sigma U$, one obtains
\begin{equation}
    \label{CPGop2}
    \begin{array}{rcl}
        U_\Sigma U &=&
        \left( \sigma_x^{(a)} \right)^{s_{(1,3)}+s_{{(2,2)}}+s_{(3,1)}}
        \left( \sigma_x^{(b)} \right)^{s_{(1,1)}+s_{{(2,2)}}+s_{{(3,3)}}}  
        \left( \sigma_z^{(a)} \right)^{s_{(2,1)}+s_{(1,2)}+s_{(I,b)}}
        \left( \sigma_z^{(b)} \right)^{s_{(I,a)}+s_{(1,2)}+s_{(2,3)}}
        \vspace{1mm} \\   
        &&\ds{U_{zz}^{(a,b)}\bigl[-(-1)^{s_{(1,1)}+s_{(2,2)}+s_{(1,3)}}
        \alpha_{(1,2)} \bigr]
        U_z^{(a)}\bigl[-(-1)^{s_{(2,2)}+s_{(1,3)}} \alpha_{(2,1)} \bigr]
        U_z^{(b)}\bigl[-(-1)^{s_{{(1,1)}}+s_{(2,2)}} \alpha_{(2,3)}\bigr] 
        U_{\text{swap}}^{(a,b)}.}
    \end{array}
\end{equation}
\end{widetext}

Using (\ref{CPGop2}) one finds the following result: To realize the
controlled phase 
gate (\ref{CPGdef}) together with a swap-gate, the observables
(\ref{CPGround2}) measured in the second round have to be chosen with
the angles $\alpha_{(2,1)} = {(-1)}^{1+s_{(2,2)}+s_{(1,3)}}
  \,\theta/2$, $\alpha_{(1,2)} = (-1)^{s_{(1,1)}+s_{(2,2)}+s_{(1,3)}}
  \theta/2$ and  $\alpha_{(2,3)} = (-1)^{s_{(1,1)}+s_{(2,2)}+1} \, \theta/2$.
This realizes the gate $U_\Sigma U_{CPG}[\theta]$, where
the byproduct operator $U_\Sigma$ generated by the measurements may be
read off from equation (\ref{CPGop2}).

\subsection{Quantum Fourier transformation}
\label{sec:fourier}

\begin{figure}
    \begin{center}
        \begin{tabular}{lc}
        a) \\ & \epsfig{file=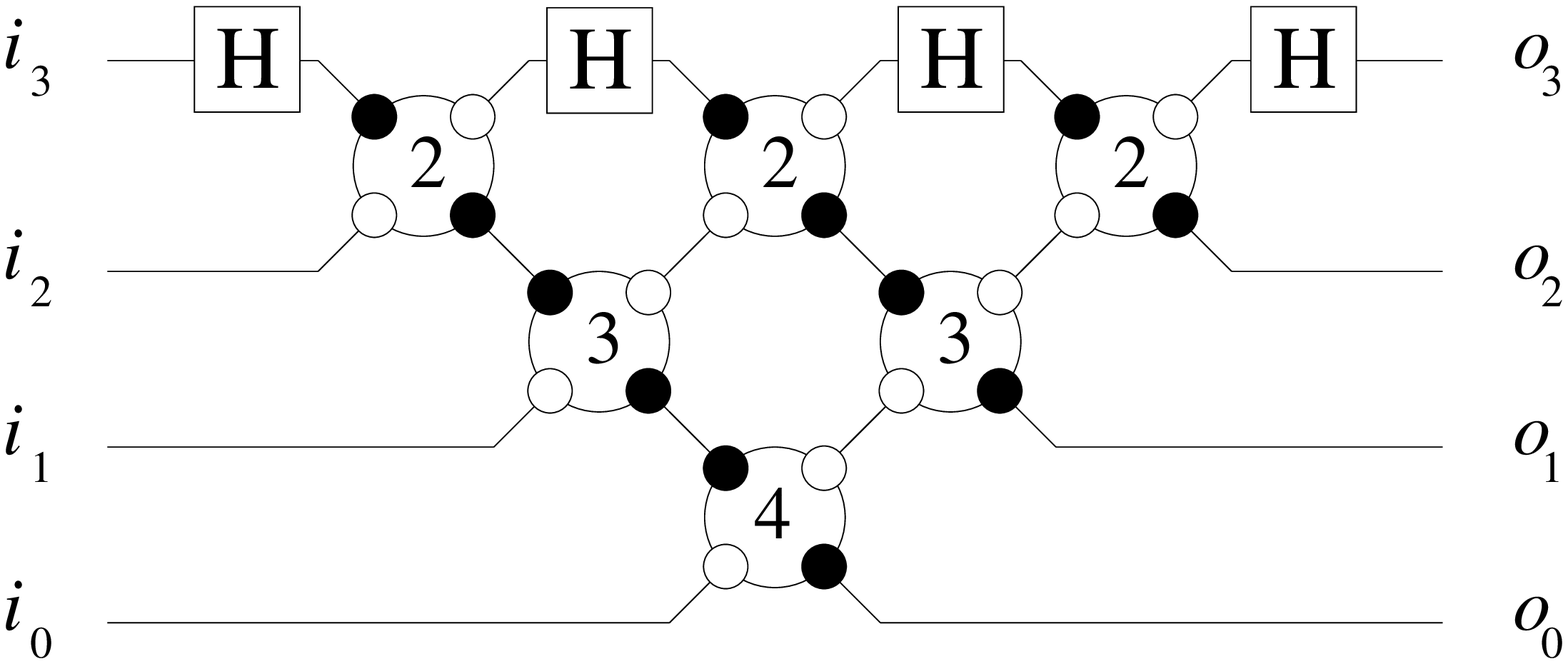, width=6cm} \vspace{3mm} \\
        b) \\ &\epsfig{file=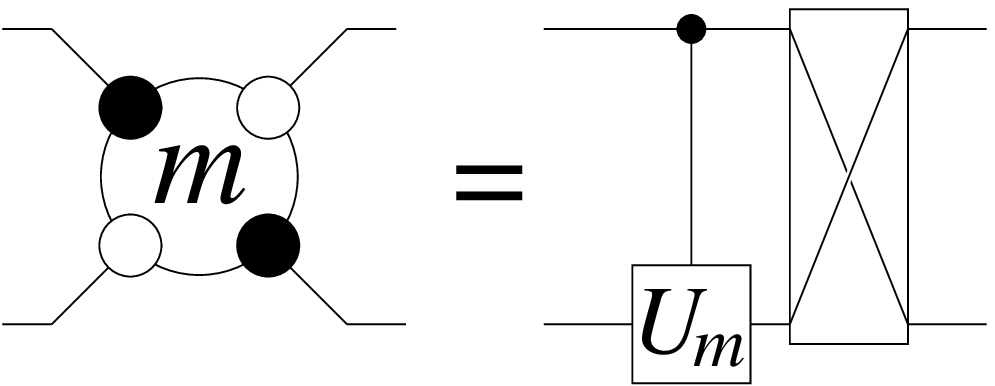, width=3.5cm}
        \end{tabular}
        \caption{\label{fig:FourierNet}Quantum Fourier
        Transformation. a) Network for quantum Fourier
          transformation on four qubits, taken from
          \cite{QFT}. b) Component of the network shown in a) which
          performs a conditional phase- and a swap-gate. Specifically,
          the gate shown is $U_{CPG}[2\pi/2^m]$,
          i.e. $U_m=|0\rangle\langle0|+e^{i2\pi/2^m} |1\rangle\langle
          1|$.}  
    \end{center}
\end{figure}

To realize the quantum Fourier transform we simulate the quantum
logic network given in Fig.~\ref{fig:FourierNet}a. The arrangement of
the gates in this network is taken from \cite{QFT}. Note that in
\cite{QFT} it was demonstrated that the setup to perform a quantum
Fourier transformation simplifies considerably in a situation where
the output state is measured right after the transformation. Here,
however, the quantum Fourier transformation may constitute part 
of a larger quantum circuit and we do not measure its output state. 

As can be seen from Fig.~\ref{fig:FourierNet}, the quantum Fourier
transform consists of Hadamard gates and combined gates which perform
a conditional phase shift and a swap. These gates have been discussed
in Sections~\ref{unigate} and \ref{sec:quantum-phase-gate}. All that
remains to do is put the measurement patterns simulating these gates
together, using the network-like composition principle described in
Section~\ref{connect}. 

\begin{figure}
    \begin{center}
        \epsfig{file=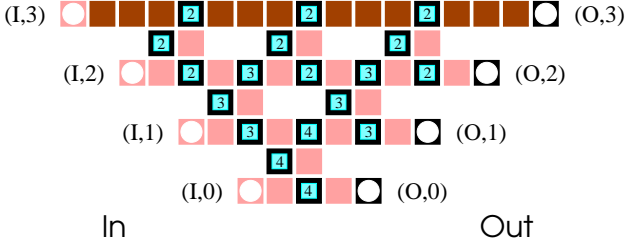, width=8.3cm}
        \caption{\label{Fourier}\QCns-realization of a quantum
          Fourier transformation on four qubits. The cluster qubits
          displayed as framed squares are measured in adapted
          bases. For the labels see text.} 
    \end{center}
 \end{figure}

In this way we obtain a measurement pattern in which there are
adjacent cluster qubits in ``wires'' that are measured in the
$\sigma_x$-eigenbasis. As described in
Section~\ref{sec:remov-unnec-meas}, such pairs of cluster qubits may
be removed from the measurement pattern. Note, that by removing
adjacent pairs of $\sigma_x$-measured cluster qubits  we have moved the
$\sigma_y$-measurements of the Hadamard transformations ``into'' the
subsequent conditional phase gates, i.e. we removed a cluster qubit
which was not from a wire. It can 
be easily verified that this is an allowed extension of the method
described in Section~\ref{sec:remov-unnec-meas}. Finally, one obtains the
\QCns-circuit displayed in Fig.~\ref{Fourier}.  

In this circuit, as in all the others, the adaptive measurements
are of observables  
\begin{equation}
    \label{QFTround2}
    U_z[\pm \eta] \sigma_x U_z[\pm \eta]^\dagger,
\end{equation}
with $\eta = \pi/4$ for cluster qubits marked with ``2'' in Fig.~\ref{Fourier},
$\eta  = \pi/8$ for qubits marked with ``3'' and $\eta= \pi/16$ for
the qubits marked with ``4''.
The sign factors of the angles in (\ref{QFTround2}) depend on the
results of previous measurements.

The \QCns-circuit, shown in Fig.~\ref{Fourier} for the
case of four qubits,  is straightforwardly generalized to an
arbitrary number $n$ of logical qubits. The temporal spatial and
operational resources $T, S$ and $O$ are, to leading order
\begin{equation}
        T = n, \;\; S,O = 2n^2.
\end{equation}
The corresponding network
resources are $T_\text{qln} =2n$, $S_\text{qln}=n$ and
$O_\text{qln}=n^2/2$. Thus, the 
scaling of the \QC spatial resources is worse than in the network
model, but the temporal and operational resources scale in the same 
way as the corresponding resources for the network. The
\QCns-simulation of the network displayed in Fig.~\ref{fig:FourierNet}
requires half as many time steps and four times as many operations,
albeit only one-qubit operations.  

\subsection{Multi-qubit controlled gates}
\label{sec:toffoli-gate}

In this section we describe the realization of the Toffoli phase
gate and the three-qubit controlled gate $CARRY$ which we will both
need for the construction of the \QCns-adder circuit described in
Section~\ref{sec:adder}. 

The Toffoli phase gate is a three-qubit generalization of the
two-qubit controlled phase gate.  If all three qubits are in the state
$\ket{1}$, the state gains a phase of $\exp(\im \phi)$, while all other
logical basis states remain unchanged by the gate,
\begin{equation}
    \label{Toffdef}
    U_{\text{Toffoli}}^{(c_1,c_2,t)}[\phi]=\iden^{(c_1,c_2,t)}+
    (e^{i\phi}-1)\,|111\rangle_{c_1,c_2,t}\langle111|.  
\end{equation}
Like the controlled phase gate it can be represented as a product of
multi-qubit rotations, 
\begin{equation}
\label{ToffSeq}
\begin{split}
U_{\text{Toffoli}}^{(c_1,c_2,t)}[\phi]=\,&
U_{zzz}^{(c_1,c_2,t)}\!\left[\frac{\phi}{4}\right] U_{zz}^{(c_1,c_2)}
\!\left[-\frac{\phi}{4}\right] U_{zz}^{(c_1,t)}
\!\left[-\frac{\phi}{4}\right]\\   
&U_{zz}^{(c_2,t)}\!\left[-\frac{\phi}{4}\right]
U_{z}^{(c_1)}\!\left[\frac{\phi}{4}  
\right]U_{z}^{(c_2)}\!\left[\frac{\phi}{4}\right] U_{z}^{(t)}
\!\left[\frac{\phi}{4}\right]. 
\end{split}
\end{equation}
where we have dropped the global phase, and
$U_{zzz}^{(c_1,c_2,t)}[\alpha]=\exp\left(-\im \alpha/2 
\sigma_z^{(c_1)}\sigma_z^{(c_2)}\sigma_z^{(t)}\right)$ is a three qubit
generalized rotation. The two-qubit rotations $U_{zz}$ are as
defined in (\ref{tworot}). 

\begin{figure}
    \epsfig{file=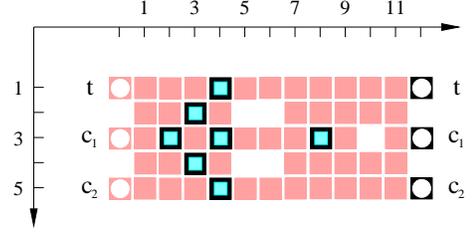,width=6cm}
    \caption{\label{tpgfig}A measurement layout to realize a
      Toffoli phase gate with 
      phase $\phi$. The qubits marked by black boxes are
      simultaneously measured 
      in adapted bases depending on previous measurement outcomes.} 
\end{figure}

\begin{figure}
    \epsfig{file=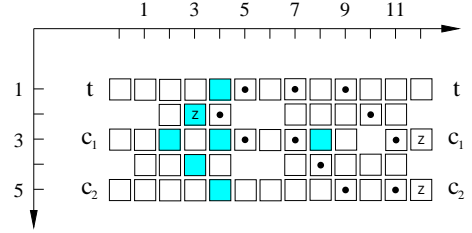,width=6cm}
    \caption{\label{fig:Toffcorrs}Cluster state quantum correlations
     for the realization of $U_{zz}^{(c_1,c_2)}[\phi/4]$, used  in
     the Toffoli phase gate.} 
\end{figure}

The way to convert the sequence (\ref{ToffSeq}) of generalized
rotations into a measurement pattern is as in the examples before. 
The measurement layout for the Toffoli phase gate is illustrated in
Fig.~\ref{tpgfig}. Each of the generalized rotations that make up
the gate is directly associated with of one of the measurements made
in the eigenbasis of $U_z[\pm \phi/4] \sigma_x U_z[\pm
\phi/4]^\dagger$. An initial cluster-state 
correlations which is used for the realization of a
generalized rotation is shown in Fig.~\ref{fig:Toffcorrs}: the
rotation $U_{zz}^{(c_1,c_2)}[\phi/4]$ is realized via the 
measurement of the cluster qubit at the lattice site $(3,1)$ in the
appropriate basis.    

The sign factors of the angles that specify the measurement bases depend on
the outcome of $\sigma_x$-measurements only. Thus, after all $\sigma_x$-
measurements have been performed, the measurement bases for
the remaining qubits can be
deduced and the Toffoli phase-gate is realized in a single further
time-step. The measurement pattern realizes the generalized rotations
directly and 
is not derived from a quantum logic network. 

\medskip

Now we describe the realization of a four-qubit gate
$CARRY$, which has one target and three control qubits. It performs a
phase-flip $\sigma_z$ on the target if at least two of the control
qubits are in state $|1\rangle$ and otherwise does nothing, i.e.
\begin{equation}
    \label{Ucarry0}
    U_{CARRY} = \exp\left(-i \pi \!\!\!\!  \sum_{i=000_d\, |\,
    w(i)\geq2}^{111_d} \!\!\!\!\! |i\rangle_{c_1 c_2 c_3} \langle i |  \otimes
    |1\rangle_t \langle 1| \right),
\end{equation}
Expanding the projectors on the control qubits into products of Pauli
operators one obtains
\begin{widetext}
\begin{equation}
\label{Ucarry}
\begin{array}{rcl}
        U_{CARRY} &=&
        \ds{e^{-i\frac{\pi}{4}} \underbrace{\exp\left(-i\frac{\pi}{8}
        \sigma_z^{(t)} \sigma_z^{(c_3)} \right)}_{U_i}
        \underbrace{\exp\left(-i\frac{\pi}{8}  
            \sigma_z^{(t)}\sigma_z^{(c_2)}\right)}_{U_h}
        \underbrace{\exp\left(i \frac{\pi}{8} \sigma_z^{(c_3)}
        \right)}_{U_g} \underbrace{\exp\left(i \frac{\pi}{8}
        \sigma_z^{(c_2)}\right)}_{U_f} \underbrace{\exp\left(i
        \frac{\pi}{8} \sigma_z^{(c_1)} \right)}_{U_e}} \\
        && \ds{\underbrace{\exp\left(i\frac{\pi}{4}\sigma_z^{(t)}
        \right)}_{U_d} \underbrace{\exp\left(-i\frac{\pi}{8}
        \sigma_z^{(t)}\sigma_z^{(c_1)} \right)}_{U_c} \underbrace{\exp 
        \left(-i \frac{\pi}{8} \sigma_z^{(c_1)}
          \sigma_z^{(c_2)}\sigma_z^{(c_3)}\right)}_{U_b}
        \underbrace{\exp\left( i \frac{\pi}{8}  \sigma_z^{(t)}    
        \sigma_z^{(c_1)}\sigma_z^{(c_2)}\sigma_z^{(c_3)}\right)}_{U_a}.}
\end{array}   
\end{equation}
\end{widetext}
The global phase is henceforth discarded.

\begin{figure}
    \begin{center}
        \epsfig{file=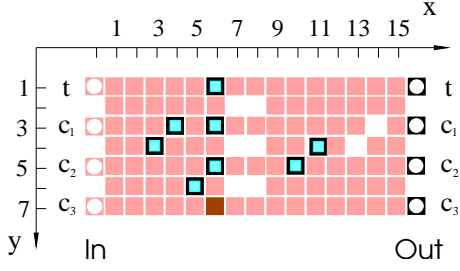,width=6cm}
        \caption{\label{Addelm}The three qubit controlled gate. Qubits
        displayed as squares in light gray are measured in the
        $\sigma_x$-eigenbasis, the qubit displayed in dark gray is
        measured in the $\sigma_y$-eigenbasis, and the measurement
        bases of the qubits displayed as framed squares are adaptive.}
    \end{center}
\end{figure}

The unitary transformation is now subdivided into two parts,
\begin{equation}
    U_{CARRY}= U_{h,i} \, U_{a-g},
\end{equation} 
with $U_{a-g}=U_gU_f U_e U_d U_c U_b U_a$ and $U_{h,i} =U_i
U_h$. Correspondingly, the 
cluster on which $U_{CARRY}$ is realized is divided into two
sub-clusters. On the first sub-cluster the transformations $U_a$ to
$U_g$ are realized, on the second sub-cluster $U_{h,i}$. The
measurement pattern to 
realize $U_{CARRY}$ is displayed in Fig.~\ref{Addelm}. The first
sub-cluster stretches from $x=0$ to $x=8$, with the input at $x=0$
and the intermediate output 
at $x=8$. The qubits with $8\leq x \leq 16$ belong to the second
sub-cluster.

\begin{figure}
    \begin{center}
        \begin{tabular}{ll}
            a) & b)\\    
            \epsfig{file=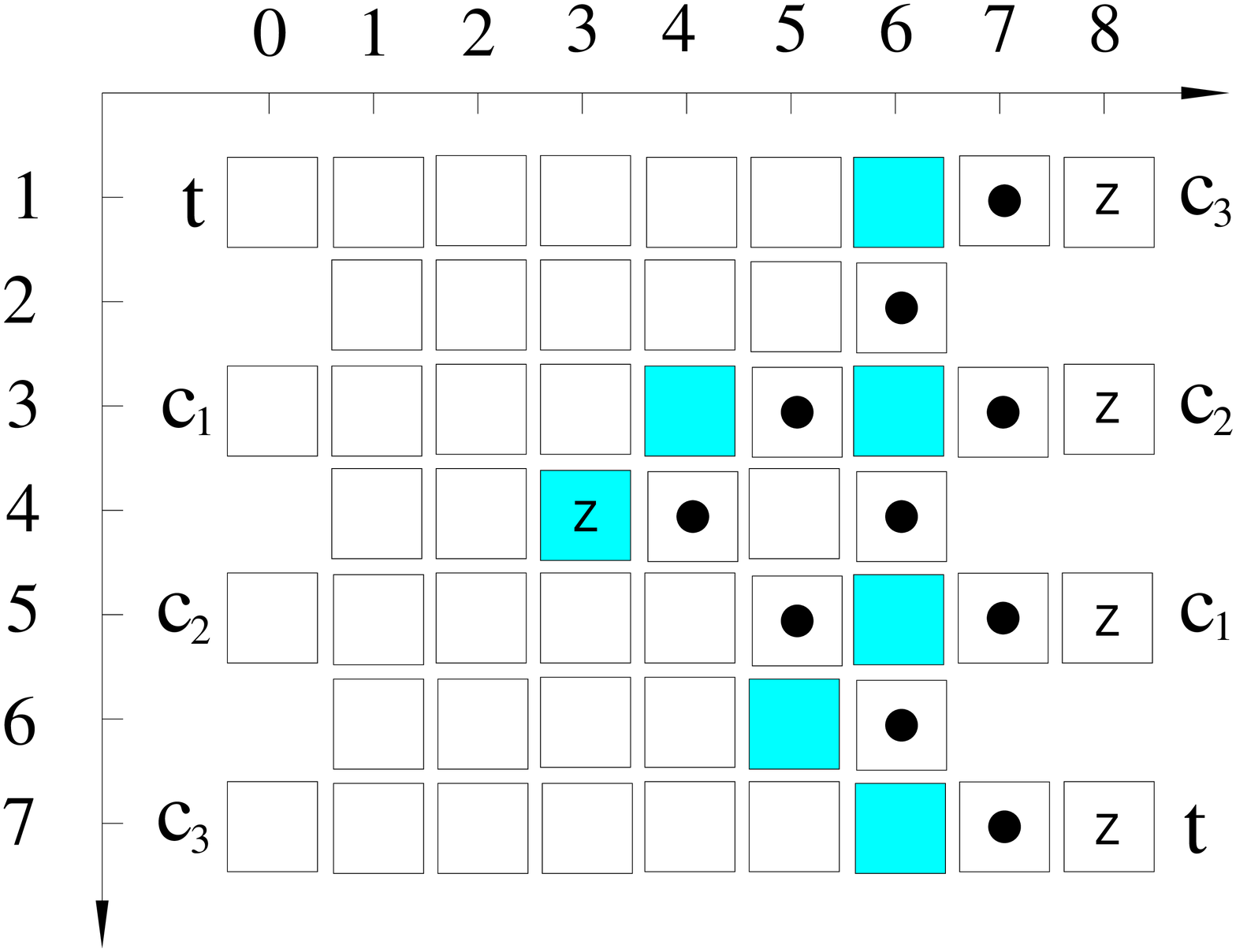,width=4.2cm} &
            \epsfig{file=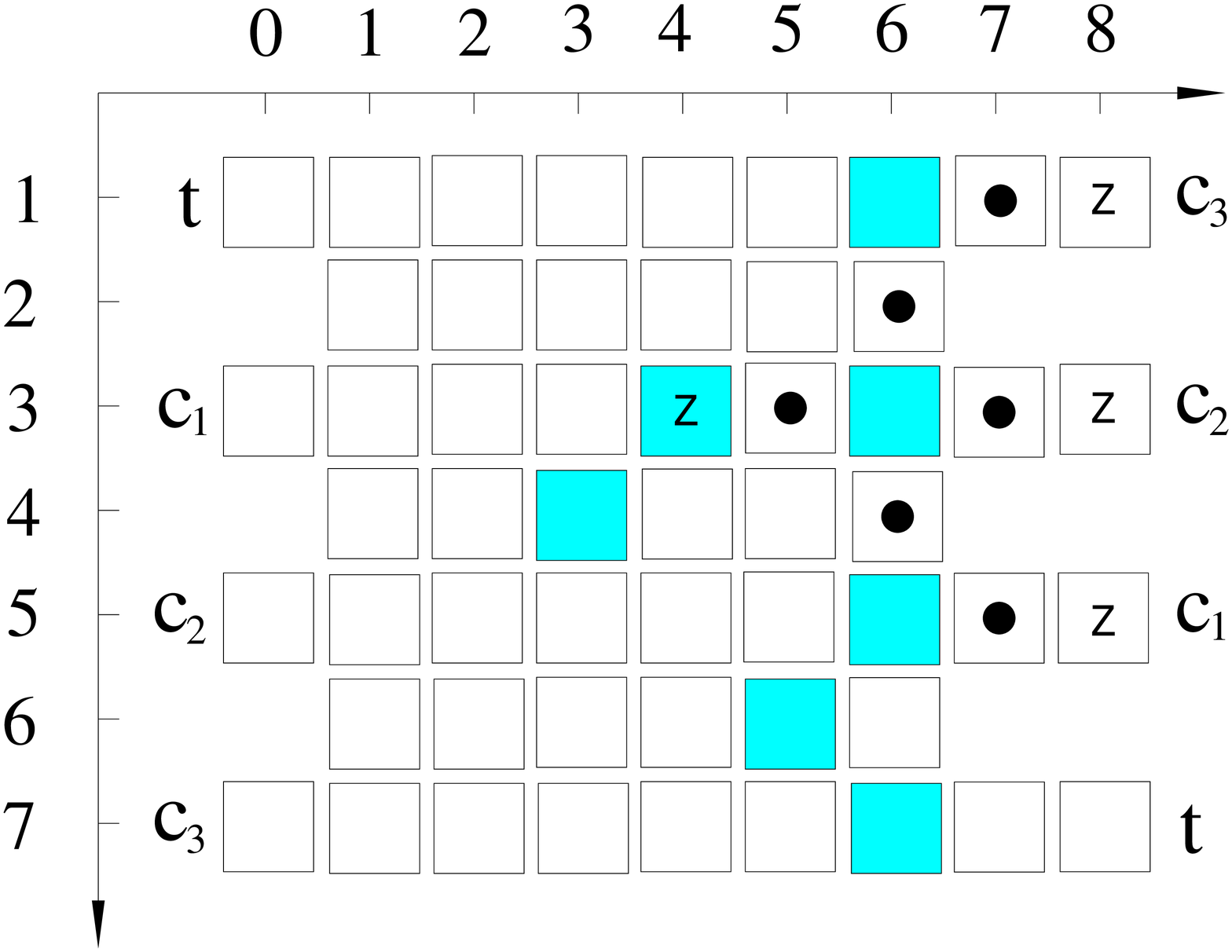,width=4.2cm} \vspace{3mm}\\
            c) & d)\\    
            \epsfig{file=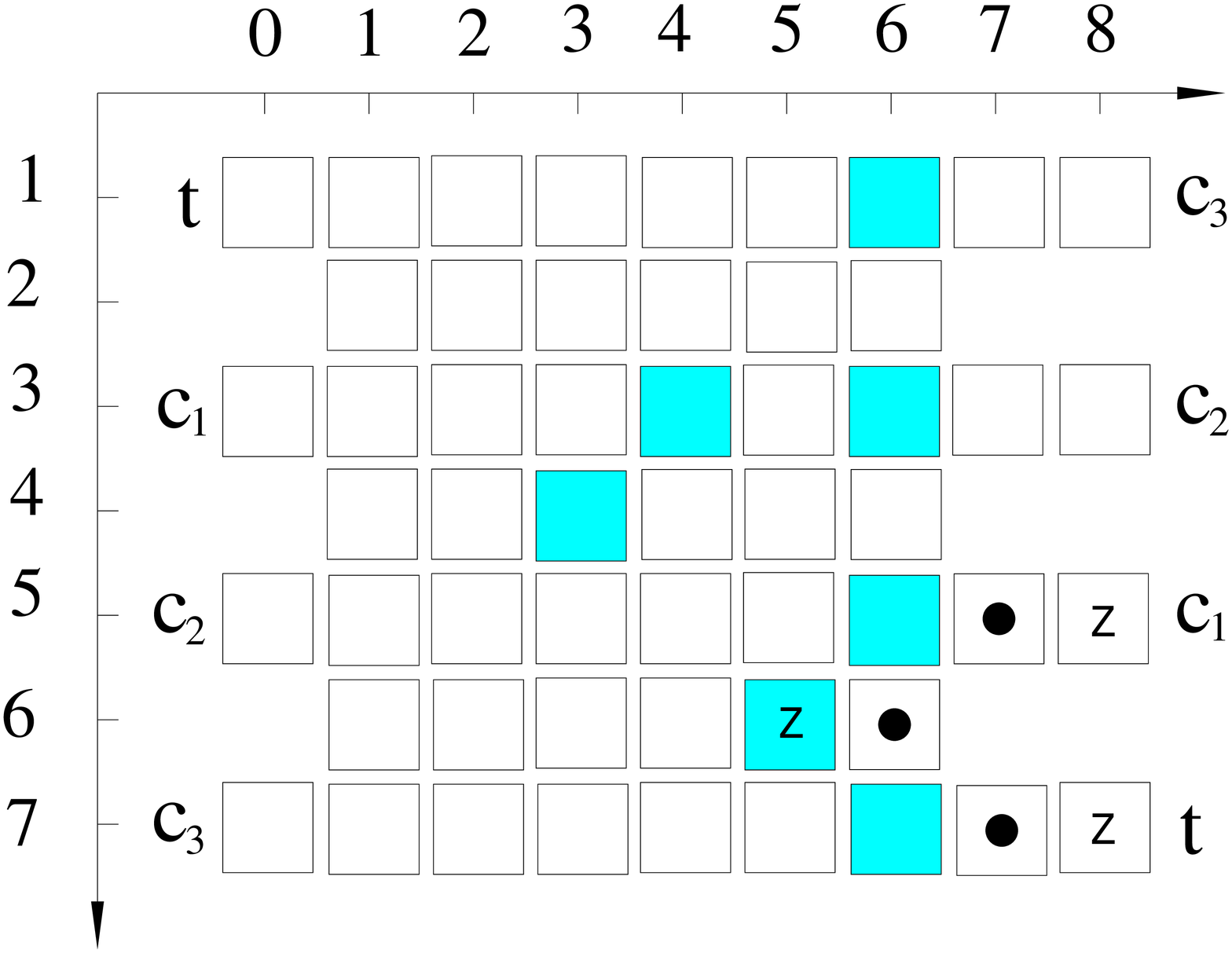,width=4.2cm} &
            \epsfig{file=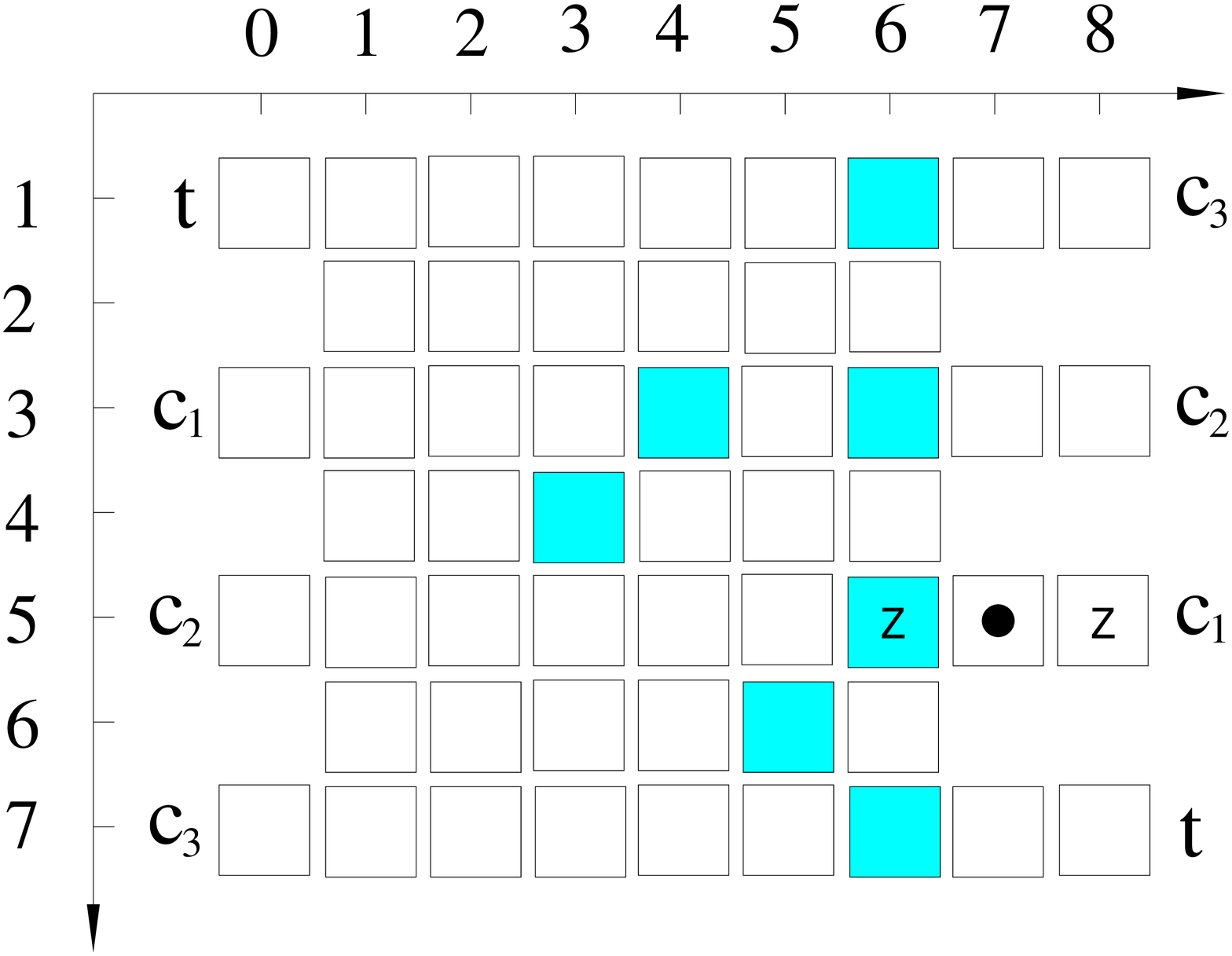,width=4.2cm} 
        \end{tabular} 
        \caption{\label{fig:CorrExpl2}Quantum correlations of the
          initial cluster state $|\phi\rangle_{{\cal{C}}(U_{a-g})}$ on
          the cluster ${\cal{C}}(U_{a-g})$. These correlations induce via the
          $\sigma_x$-measurements the quantum correlations for the
          state $|\psi^\prime\rangle$ which act only on the output qubits
          and one cluster qubit in ${\cal{C}}_M(U_{a-g})$. The
          pattern of correlation centers in a) displays
          the correlation required to realize $U_a$; b),c) and d)
          display the correlations for $U_b$, $U_c$ and $U_e$,
          respectively. The correlations used for the 
          realization of $U_d$, $U_f$ and $U_g$ are  
          not shown. They are analogous to the
          one in d) used for the realization of $U_e$.} 
    \end{center}
\end{figure}

Let us now explain the sub-gate $U_{a-g}$. The conversion of the
sequence (\ref{Ucarry}) of generalized rotations is as in the
previous examples. For each generalized rotation there is one cluster
qubit in ${\cal{C}}_M(U_{a-g})$ whose measurement basis
specifies the respective rotation angle. 
Specifically, the measurement of the cluster qubit $(3,4)$
sets the rotation angle of $U_a$, the measurement of qubit $(4,3)$
sets the angle for $U_b$, $(5,6)$ sets $U_c$, $(6,7)$ sets
$U_d$, $(6,5)$ sets $U_e$, $(6,3)$ sets $U_f$ and 
qubit $(6,1)$ sets $U_g$. The quantum correlations of the initial
cluster state which induce via the measurements of the cluster qubits in
${\cal{C}}_M(U_{a-g})$ the quantum correlations associated with
the generalized rotations are
displayed in Fig.~\ref{fig:CorrExpl2}. 

The realization of the gate requires two
measurement rounds. In the first round the standard measurements of
$\sigma_x$ and $\sigma_y$ are performed. 
Note that the rotation angle of $U_d$ is twice as big as for the other
rotations. To realize $U_d$ of the cluster qubit $(6,7)$ the observable
\begin{equation}
    \label{MB2}
    U_z\left[\pm \frac{\pi}{4} \right]
        \, \sigma_x U_z\left[\mp  \frac{\pi}{4} \right]= \pm \sigma_y
\end{equation}
is measured. Thus, the realization of $U_d$ belongs to the first
round of measurements. Strictly speaking, this
measurement round does not belong to the gate but to the
circuit as a whole since all standard measurements are performed
simultaneously.

In the second measurement round, of the remaining qubits in
${\cal{C}}_M(U_{a-g})$ one measures the observables
\begin{equation}
    \label{MB1}
    U_z\left[\pm \frac{\pi}{8} \right] \, \sigma_x U_z\left[\mp
    \frac{\pi}{8} \right].
\end{equation}
The procedure to infer the sign factors in (\ref{MB1})
and (\ref{MB2}) is explained in Section~\ref{QcorrQcomp}.

The reason why the measurements in the tilted bases may all be performed
simultaneously in the second round can be seen as follows. Be
$Q_\nearrow$ the set of qubits measured in tilted bases. The
contribution $U_{\Sigma,Q_\nearrow}$ of the cluster qubits measured in
tilted bases to the byproduct operator
$U_\Sigma$ in (\ref{ByProp}) contains only a $z$-part but no
$x$-part. That is, it has the form 
\begin{equation}
    \label{By2}
    U_{\Sigma,Q_\nearrow}=\bigotimes_{i \in I
  \subset\{ t, c_1, c_2,c_3\}} \sigma_z^{[i]}.
\end{equation}
In (\ref{acts}) the
byproduct operator appears ``on the wrong side'' of $U_{a-g}$ as does
the contribution $U_{\Sigma,Q_\nearrow}$. When the order of the gate
and the byproduct operator is 
exchanged, the byproduct operator may modify the gate. While this is,
not surprisingly, indeed the case for the whole $U_\Sigma$, it is not
so for the contribution $U_{\Sigma,Q_\nearrow}$ coming from the
measurements in the tilted bases. Because $U_{\Sigma,Q_\nearrow}$ has only
a $z$-part it commutes with $U_{a-g}$. Therefore, the results of
measurements in a tilted basis do not mutually affect the choice of
their measurement bases.

The fact that hat the byproduct operator $U_{\Sigma,Q_\nearrow}$ is
indeed of form 
(\ref{By2}) we do not show here explicitly. For the byproduct operator
created in the measurement of qubit $(3,4)$ realizing the
transformation $U_a$ it may be verified from equation (\ref{ByprU4}) in
Section~\ref{sec:HamilSim}. 

\begin{figure}
    \begin{center}
        \begin{tabular}{ll}
        a) & b)\\    
        \epsfig{file=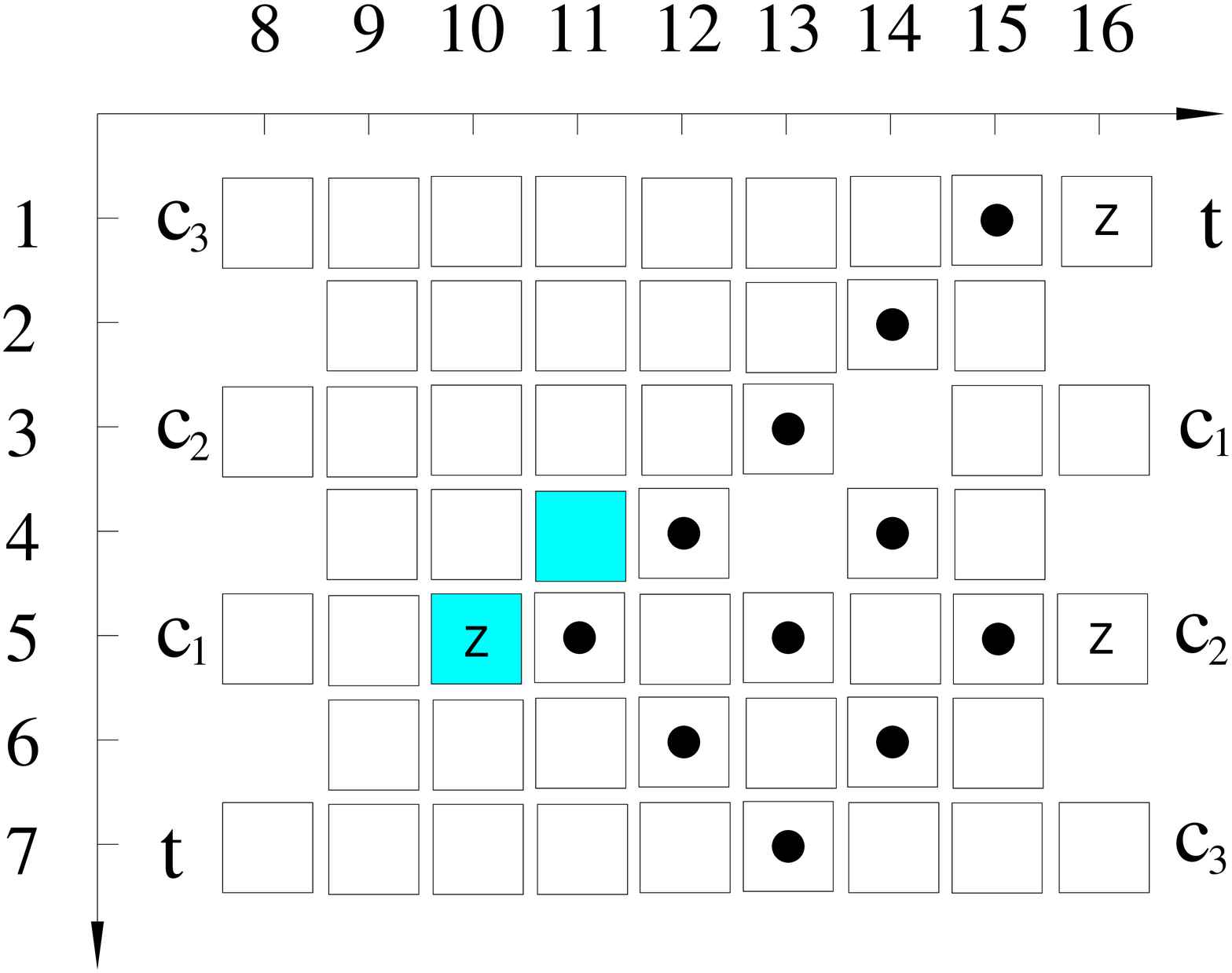,width=4.2cm} &
        \epsfig{file=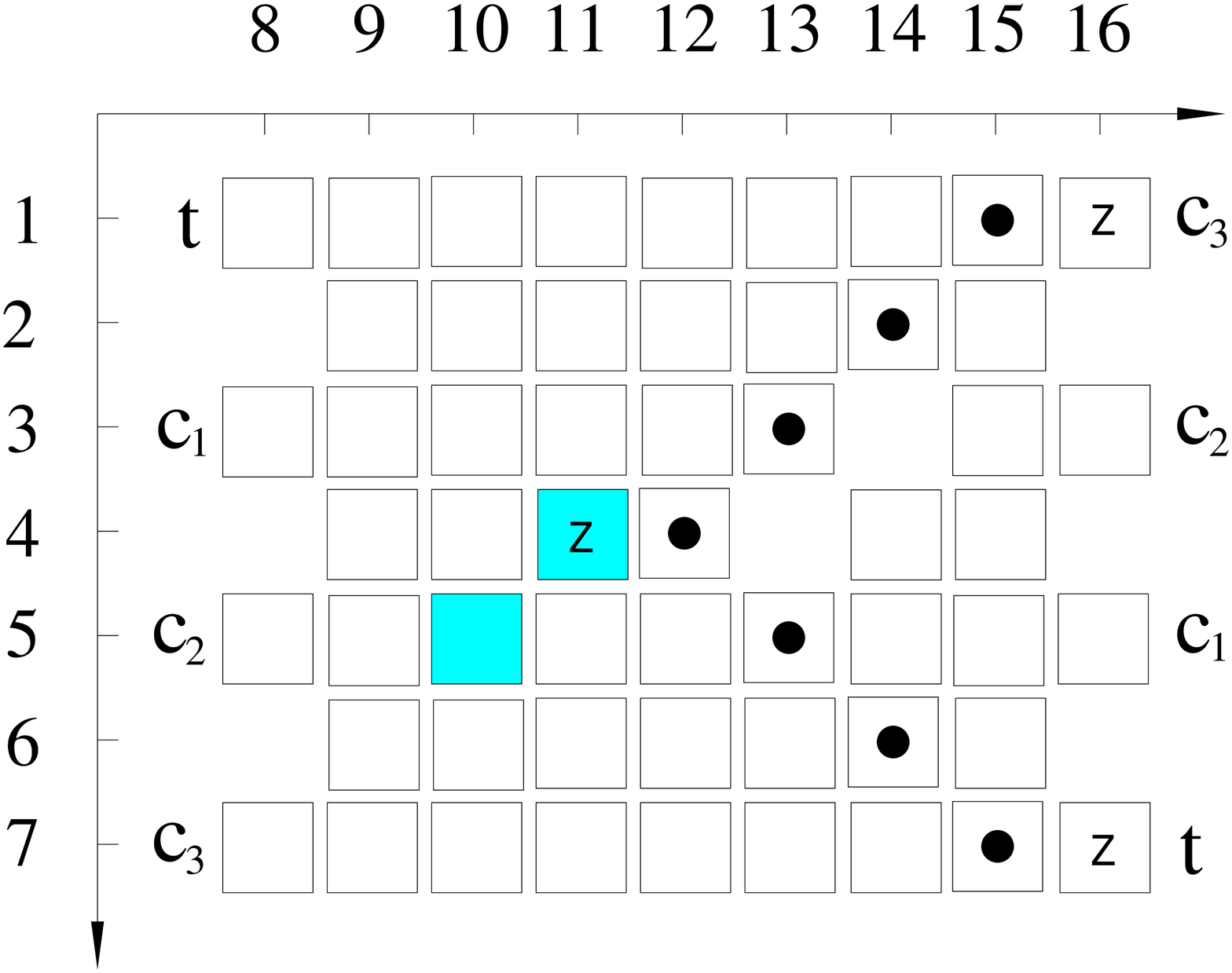,width=4.2cm} 
        \end{tabular} 
        \caption{\label{fig:CorrExpl3}Quantum correlations of the
        initial cluster states on ${\cal{C}}(U_h)$ and
        ${\cal{C}}(U_i)$. These correlations induce, via the 
        $\sigma_x$-measurements, the quantum correlations for the
        states $|\psi^\prime\rangle_{{\cal{C}}(U_h)}$  and
        $|\psi^\prime\rangle_{{\cal{C}}(U_i)}$ that involve only the
        respective output qubits 
        and one qubit in the gate body. The
        pattern of correlation centers in a) displays
        the correlation required to realize $U_h$ and b) the
        correlation for $U_i$.} 
    \end{center}
\end{figure}

The explanation of the second sub-gate, $U_{h,i}$,
is analogous. Fig.~\ref{fig:CorrExpl3} displays the
quantum correlations of the initial cluster state which, via the
measurements in ${\cal{C}}_M(U_{h,i})$, induce the required quantum
correlations associated with $U_h$ and $U_i$. 

Two further points we would like to address in this section. The first
is to note that the whole gate $U_{CARRY}$ can be performed on the 
\QC in two measurement rounds. The first measurement
round is that of the
$\sigma_x$-, $\sigma_y$- and $\sigma_z$-measurements which, strictly
speaking, does not belong to the gate but to the circuit as a
whole. The second measurement round is that of the simultaneous
measurements in tilted measurement bases. 

We have already seen that the measurements that realize the unitary
transformations $U_a, \, ..\, ,U_g$ may be realized simultaneously, and
this argument may be extended to the entire gate $U_{CARRY}$. All the
byproduct operators created with the measurements in tilted bases have
only a $z$- but no $x$-part. Therefore they all commute with
$U_{CARRY}$. Thus, to choose the right measurement bases neither of
the measurements in a tilted basis that realizes one of the rotations
$U_a, \, ..\, ,U_i$ needs to wait for another measurement in a tilted basis. 

\begin{figure}
    \begin{center}
        \epsfig{file=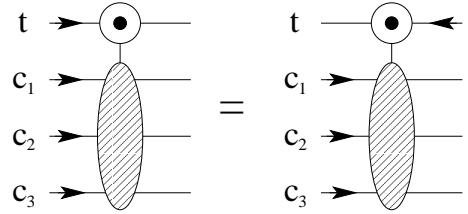,width=6cm}
        \caption{\label{fig:backforth}In the three-qubit controlled
          gate $CARRY$, the target qubit may travel either back or forth.}
    \end{center}
\end{figure}

Second, note that the for $U_{CARRY}$ the target-input and the
target-output can be interchanged, see Fig.~\ref{fig:backforth}. This
holds because the 
(conditional) phase-flip on the target qubit is its own inverse.
Thus, the target qubit may travel through the gate backwards. This
property also holds for the Toffoli phase gate. We will make use of it in
the construction of the quantum adder in the next section. 
 
\subsection{Circuit for addition}
\label{sec:adder}

\begin{figure}
    \begin{center}
        \epsfig{file=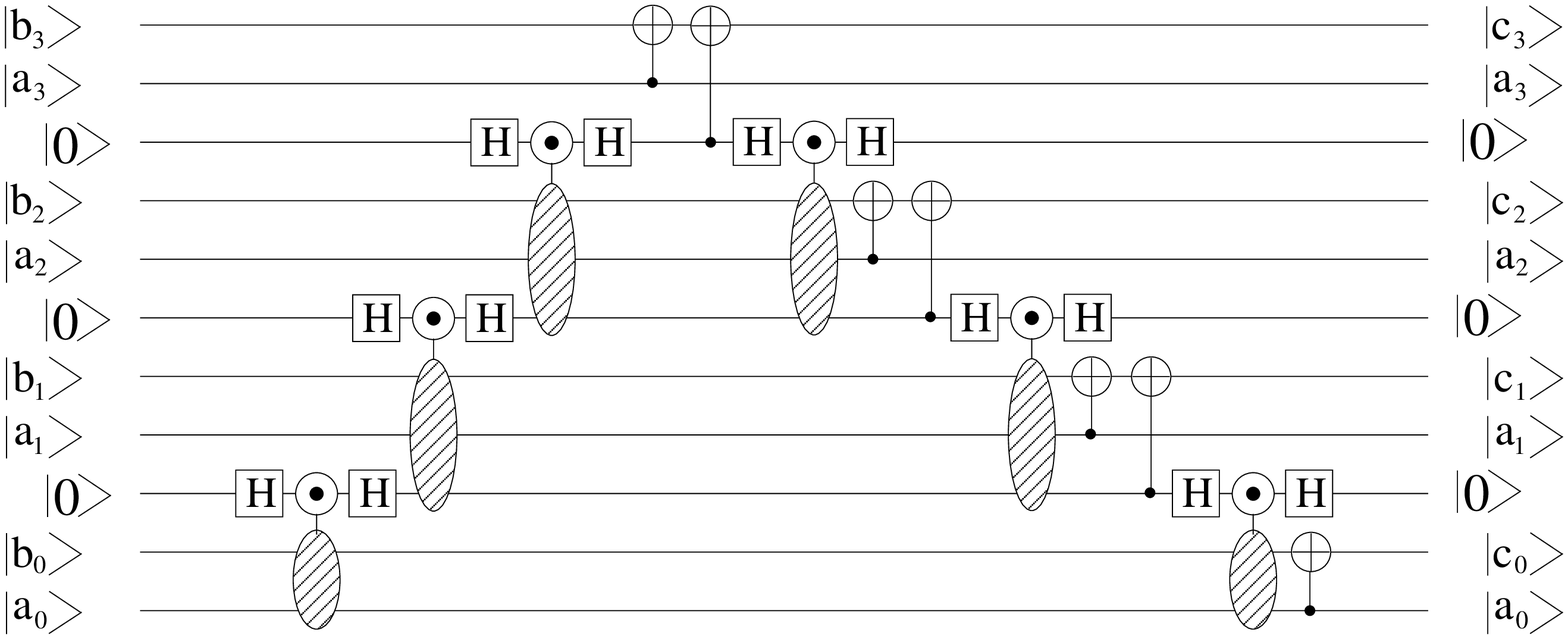,width=8cm}
        \caption{\label{fig:addernetw}Quantum logic network for
          4-qubit adder, $c=a+b\;\text{mod}\,2^4$. The adder 
          network is taken from \cite{AdderNW}. The two-qubit
          controlled gate in this network is the Tofolli phase gate as
          discussed in Section~\ref{sec:toffoli-gate}. A
          straightforward simulation of this network on the \QC
          would result in a quadratic scaling of spatial resources.
          However, the more compact realization discussed below
          requires only a linear overhead.}  
    \end{center}
\end{figure}

The \QCns-version of the quantum adder corresponds to
the quantum logic network as given in
\cite{AdderNW}, see Fig.~\ref{fig:addernetw}. In this paper we use the
three-qubit controlled phase gate 
$CARRY$ together with a prior and subsequent Hadamard gate on the
target qubit while in \cite{AdderNW} the equivalent three-qubit
controlled spin-flip gate is used directly. 

\begin{figure}
    \begin{center}
        \epsfig{file=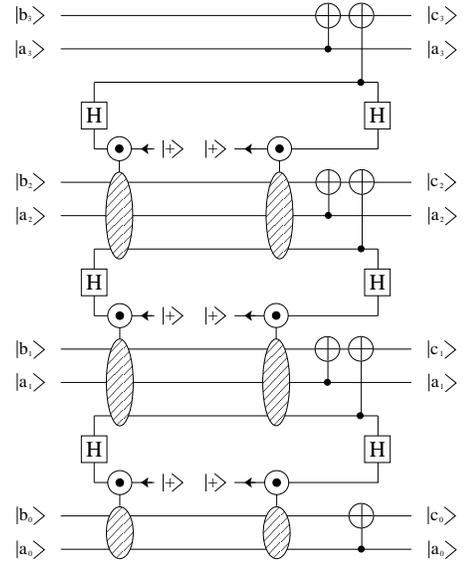,width=6cm}
        \caption{\label{fig:bended_adder}Quantum logic network for
          4-qubit adder, bent.}  
    \end{center}
\end{figure}

At first sight it appears as if the horizontal dimension of the
cluster to realize the adder circuit would grow linearly with the
number of logical qubits $n$. This is, however, not the case. The
\QCns-circuit may be formed in such a way that the horizontal size of
the required cluster is constant such
that the cluster size increases only linearly with the number $n$ of
logical qubits.  To see what the \QCns-realization of the quantum
adder will look like, the network displayed in
Fig.~\ref{fig:addernetw} may be bent in a way displayed in
Fig.~\ref{fig:bended_adder}. 

To ``bend a network'' is a rather informal notion. We therefore now specify
what we mean by this. If a quantum circuit is displayed as a quantum
logic network, the vertical axis usually denotes some spatial
dimension, i.e. the location of the qubit carriers, and the
horizontal axis corresponds to the sequence of steps of a quantum
computation, i.e. a logical time. As the basic blocks of quantum
computation in the network model, the universal gates, are unitary
transformations generated by suitably chosen Hamiltonians the logical
time becomes associated with physical time. This is, however,
a peculiarity of the network model. If on the \QC a quantum logic
network is simulated, the temporal axis is converted into an additional
spatial axis. The temporal axis in a \QCns-computation emerges
anew. It has no counterpart in the network model.
If we modify a quantum logic network in such a way that 
qubits travel from right to left, as done in
Fig.~\ref{fig:bended_adder}, it does not mean that we propose to use  
particles that travel backwards in time because we do not need to
respect the temporal axis implied by the network model. If one wants a
semi-network picture that accounts for this, one may imagine the
logical qubits as traveling through pipes on a two-dimensional surface. 
  
\begin{figure}
    \begin{center}
        \epsfig{file=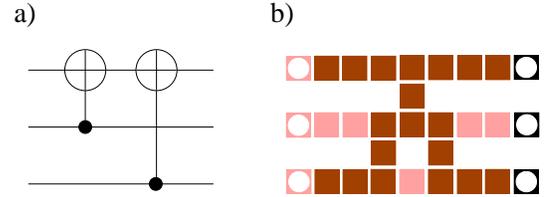,width=7cm}
        \caption{\label{fig:twoadder}Combination of two CNOT-gates (a) and
          its \QCns-realization (b).} 
    \end{center}
\end{figure}

The reason why we may let the auxiliary qubits travel
``backwards'' is the identity displayed in
Fig.~\ref{fig:backforth}. This arrangement 
of gates makes the circuit more compact.
To complete the description of components from which the \QCns-version of
the quantum adder is built, a compact measurement pattern for the two
combined CNOT-gates is displayed in Fig.~\ref{fig:twoadder}.

\begin{figure*}
    \begin{center}
        \epsfig{file=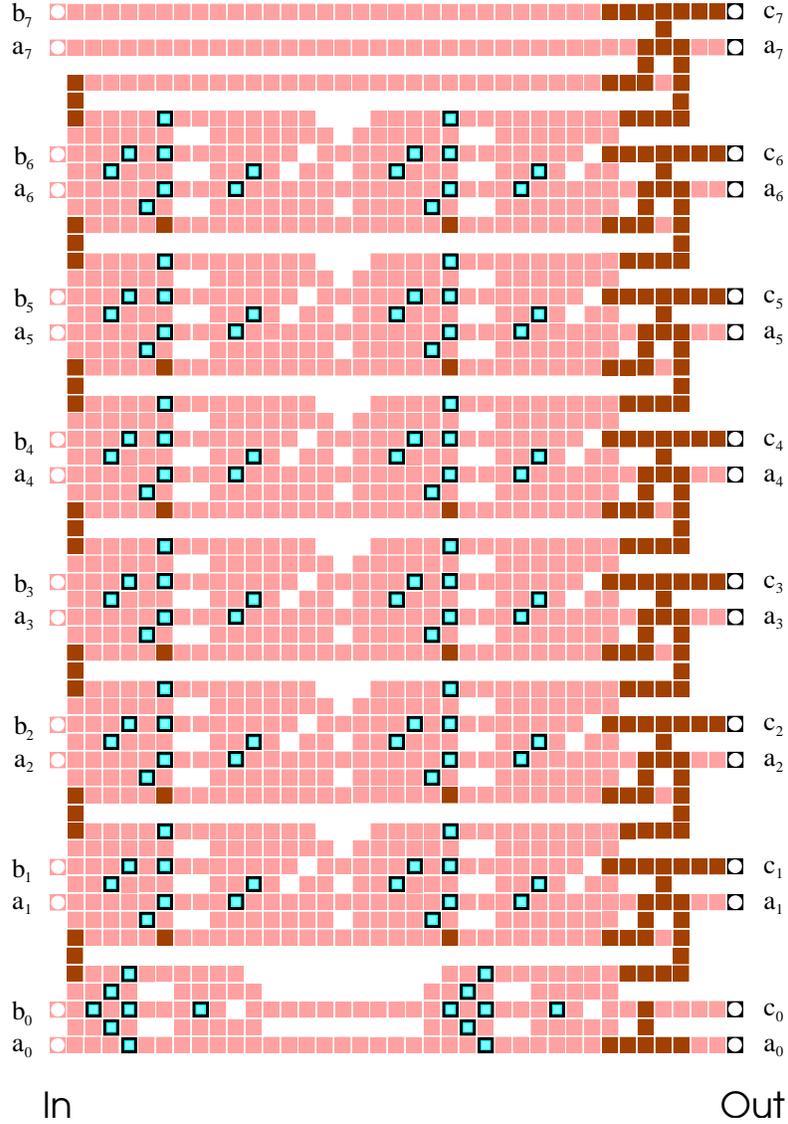,width=10.3cm}
        \parbox{0.85\textwidth}{\caption{\label{fig:Adder}Quantum
            adding circuit for two 
          8-qubit states. As in all figures displaying \QCns-circuits,
          squares in light and dark gray represent cluster qubits measured
          in the $\sigma_x$- and $\sigma_y$-eigenbasis,
          respectively. The measurement bases of qubits displayed as
          framed squares are adaptive.}}
    \end{center}
\end{figure*}

The realization of the quantum adder in the network layout of
Fig.~\ref{fig:bended_adder} directly leads to the \QCns-circuit for the
quantum adder displayed in Fig.~\ref{fig:Adder}. Please note that the
displayed \QCns-adder is for eight qubits while the networks in
Figs.~\ref{fig:addernetw} and \ref{fig:bended_adder} are only for four
qubits.

For the quantum adder circuit in Fig.~\ref{fig:Adder} we
have made two further minor simplifications. The first concerns  the ancilla
preparation. To prepare an ancilla qubit on the cluster in the state
$|+\rangle$  means to measure the respective cluster qubit in the 
$\sigma_x$-eigenbasis (the randomness of the measurement outcome does
not jeopardize the deterministic character of the circuit). As
can be seen from the Toffoli gate and the three-qubit controlled gate
displayed in Figs.~\ref{tpgfig} and \ref{Addelm}, the ancilla qubits
are located on cluster qubits which have only one next neighbor. As
can be verified from the eigenvalue equations (\ref{EVeqn}), to
measure a qubit of a cluster state which only has one next neighbor in
the eigenbasis of $\sigma_x$ also has the effect of projecting this
neighboring cluster qubit into an eigenstate of $\sigma_z$. Such cluster
qubits may be removed from the cluster as explained in
Section~\ref{redrem}. With these neighboring qubits removed the
cluster qubits on which the initial ancilla qubits were located become
disconnected from the remaining cluster and may thus 
be removed as well. With the same argument, the cluster qubits
carrying the ancillas in their output state, and their next neighbors
may also be removed.

Second, between the \QCns-realization of the CARRY-gates on the
left and the subsequent blocks of CNOT-gates we have removed pairs of
adjacent cluster qubits that would be measured in the eigenbasis of
$\sigma_x$. Why 
this can be done has been explained for adjacent qubits in wires in
Section~\ref{sec:remov-unnec-meas}. Here the situation is a little
more involved since, like in case of the circuit for Fourier
transformation displayed in Section~\ref{sec:fourier}, one 
of the removed qubits in each pair has more than two neighbors. But the
method still works as can be easily verified. 

Let us now briefly discuss the resources required for the
\QCns-realization of an $n$-qubit adder. As can be seen
directly from the circuit displayed in Fig.~\ref{fig:Adder} and the
underlying network shown in Fig.~\ref{fig:bended_adder} with its
repeating sub-structure, the adder requires a cluster of height $8n-5$
and of constant width $38$. Thus the spatial and operational resources
are, to leading order,
\begin{equation}
    S=O=304\,n.
\end{equation} 
Concerning the temporal resources note that each pair of
three-qubit controlled phase gates using the same control qubits and
the pair of Toffoli phase gates may be completed at one time instant
but that one pair of gates is completed after another. The reason
why the measurements in the tilted bases that complete each pair
of gates may be performed simultaneously is the same as the one given
previously for the measurements in tilted bases of a
single three-qubit controlled gate. The propagation of byproduct
operators is most easily 
followed in the network of Fig.~\ref{fig:addernetw}. The temporal
complexity $T$ of an $n$-qubit \QCns-adder is
\begin{equation}
    T=n,
\end{equation}
plus one step of $\sigma_x$-,$\sigma_y$- and $\sigma_z$-measurements
for the entire circuit. 

The corresponding network resources are to leading order $S_\text{qln}=3n$ and
$O_\text{qln}=T_\text{qln}=8n$. For the counting of the operational and
temporal network resources  we have assumed that the three-qubit
controlled spin-flip 
gate used in the addition circuit is composed of two Toffoli gates and
one CNOT-gate as described in \cite{AdderNW}, and that the CNOT- and
the Toffoli-gate are regarded as elementary.

Thus we find for both the network and the \QCns-realization of the
quantum adder that the spatial, temporal and operational resources
scale linearly with $n$. Therefore, the resource overheads in one
realization as compared to the other one are only constant. For the \QC
this is much better than what is indicated by the bounds (\ref{Tres}),
(\ref{Sres}) and (\ref{OSres}), in particular for the spatial and
operational resources. Equation (\ref{Sres}) yields an upper bound on
$S$ which is $\sim n^3$ and (\ref{OSres}) gives bounds on $O$ and $S$ which
are $\sim n^2$. Thus, the quantum adder is an example for which these
bounds are very loose. In general they should not be mistaken as estimates.

If the pre-factors are compared, one finds that for the realization of
a quantum adder the \QC requires about 100 times more 
spatial and 38 times more operational resources, while it is 8 times
faster. However, since we compare different objects these ratios do
not mean much apart from the fact that they 
are constant. It may be argued that in case of the \QC  spatial
resources are not as precious as they usually are, for to create cluster
states one needs a system with non-selective uniform interaction only
while for quantum logic networks one generally requires a system with
selective interactions among the qubits. Concerning the
operational and temporal 
resources, the \QC only uses one-qubit measurements while the
corresponding network uses two- and three-qubit gates as elementary
operations.

\subsection{Remarks}

We would like to add two remarks, one with regard to the elementary
constituents of the \QCns, and one with regard to their composition
principle. 

For the particular set of gate simulations used in the \QC
universality proof in Section~\ref{universal}, the CNOT-gate and arbitrary
one-qubit rotations, there is only a single instance of where one of these
gates has been used as part of a more complicated gate in all examples
of this section. Namely, the next-neighbor 
CNOT-gate has been used as part of the long-distance CNOT described in
Section~\ref{sec:non-neigh-cnot}. Of universal gate simulations one might
expect that any circuit is composed of them rather that they occur
almost not at all. One could say, though, that the used set of gates
is not a good choice for the universal set. In fact, in realizations
of network quantum computers it is often the physics  
of the specific implementation that determines which gates are
elementary. For the \QC this is not so. The \QC may simulate, for
example, general one-qubit rotations and Toffoli gates alike. Any gate
simulation may be called ``elementary'' with the same 
right as any other, but they cannot be all elementary.  
The elementary constituents of the \QC are not gate simulations.
 
As a consequence, the composition
principle for these elements will be different from gate
composition. At first sight, if we go through 
the examples of this section, we find that this is not yet reflected in
the larger and more complicated constructions.
For the quantum Fourier transform and the addition
circuit we have, though playing with some tricks,
ultimately imitated network composition. 

However, in the smaller gates and sub-circuits such as the controlled phase
gate, the Toffoli phase gate and the gate $CARRY$ we find something
that might give rise to a new and more appropriate composition principle. 
First, for the \QC
it is not the one-qubit and two-qubit operations that are particularly
simple. In the Hamiltonian simulation circuit of
Section~\ref{sec:HamilSim} we found that it is easy 
to realize generalized rotations $\exp(i\varphi \,\sigma^{(J)})$ where
$\sigma^{(J)}$ is a composite Pauli operator, $\sigma^{(J)} =
\bigotimes_{a \in J} \sigma_{k_a}^{(a)}$, $k_a=x,y,z$.  
Furthermore, in the subsequent examples of the multi-qubit gates in 
Sections~\ref{sec:quantum-phase-gate} and \ref{sec:toffoli-gate} we
have decomposed the gates into such generalized rotations rather
than into known standard gates on fewer qubits.

Any unitary transformation may be decomposed into a unitary transformation
in the Clifford group followed by generalized rotations. So, is this a
new composition principle? With our present state of knowledge, the
answer must be ``Not yet.''. First, though any transformation may be
rewritten in this form, it is presently not clear how to design
quantum algorithms with these elements directly. Second, the
construction uses the very concept of applying unitary transformations
to the state of a quantum register. However, as we have explained in
\cite{QCmodel} and also briefly sketched in Section~\ref{IP}, the \QC
has no quantum register. So, 
the generalized rotations and their concatenation at least have to be
reformulated to fit the description of the \QCns. In particular, they
have to be made compatible with the graph states identified in
Section~\ref{graphs} as characteristic quantum resource to represent
algorithms. Nevertheless, it
appears that the generalized rotations should be reflected in what may
emerge as elementary constituents and composition principle for the \QCns.

\section{Computation with limited spatial resources and in the
  presence of decoherence}
\label{wenigPlatz}

In this section we describe how to perform \QCns-computation on finite
and possibly small clusters. 
If the cluster that may be provided by a specific device
is too small for a certain measurement pattern it does not mean that
the respective \QCns-algorithm cannot be run on
this device. Instead, the \QCns-computation may be split into several
parts such that each of those parts fits on the cluster.   

To see this consider Scheme~\ref{gatescheme} for the realization of
gates.  Scheme~\ref{gatescheme} is applicable to any gate or
sub-circuit. It is thus possible to divide the circuit into
sub-circuits each of 
which fits onto the cluster. The adapted scheme is a process of
repetitive re-entangling steps alternating with rounds of
measurements.

Specifically, one starts with
the realization of the first sub-circuit acting on the fiducial input
state located on $I_1 \subset {\cal{C}}$. The fiducial input is, while
being processed,  teleported to 
some subset $O_1$ of the cluster ${\cal{C}}$. The set $O_1$ of qubits
forms the intermediate output of the first sub-circuit. These qubits
remain unmeasured while all the other qubits are measured to realize
the first sub-circuit. Now the realization of the second sub-circuit
begins. Its input state has 
already been prepared, $I_2=O_1$. The cluster qubits $a \in
{\cal{C}}\backslash O_1$ which have been measured in the
realization of the first sub-circuit are now prepared individually
in the state $|+\rangle_a$. This completes step 1 of
Scheme~\ref{gatescheme} to realize the second sub-circuit.  Step 2
is to entangle the whole cluster via 
the Ising interaction. In the third step  all cluster qubits except
those of the intermediate output $O_2$ are measured whereby the
realization of the second sub-circuit is completed. The
intermediate output is now located on $O_2$. For the realization of
the subsequent sub-circuits one proceeds accordingly.

An advantage of this modified procedure is that one gets by with
smaller clusters. A disadvantage is that the Clifford part of the
circuit may no longer be performed in a single time step. 

Perhaps the most important advantage of the above construction
is that in this way a basic requirement to make the \QC fault-tolerant
can be fulfilled. Namely, decoherence can be controlled. If a single
large cluster is used the computation might reach certain cluster
qubits only after a long time such that the cluster would have already
decohered significantly and it is not clear how error-correction
could help in such a situation. This might, for any error
rate, limit the duration 
of a computation. On the contrary, if the computation is split then
the size of the sub-circuits may be adjusted such that each of them
can be performed within a fixed time $T$ and in this way, each cluster qubit
is, before being measured, exposed to a bounded amount of decoherence
specified by $T$. Thus, ``fresh'' qubits for
computation are always provided.

\section{Conclusion}

In this paper we have given a detailed account of the one-way quantum
computer. We have shown that the \QC can be regarded as a simulator of
quantum logic networks. This way, we clarified the relation of the \QC to the
network model of quantum computation and gave the universality proof.

We have based our description on the correlations exhibited by cluster
states, and states that can be created from them under one-qubit
measurements. For this purpose, theorem~1 of
Section~\ref{QcorrQcomp} is an important tool. It relates unitary
transformations to quantum correlations
exhibited by certain pure states.

In Section~\ref{examples} we have presented a number of example
circuits such as the circuit for quantum Fourier transformation and
for addition. In this way, hopefully, we also have acquainted the reader with
a number of construction principles for \QCns-circuits. Note that
the simulations of the universal gates required in the
universality proof are hardly used. Instead, more compact
measurement patterns have been found. 

The main purpose of this paper has been to provide a comprehensive description
of the \QC from the network perspective. Beyond that, we have pointed
out the non-network aspects of the \QCns, such as the 
different nature of information processing \cite{QCmodel, QCmodelsum},
and the connection to mathematical graphs.

\section*{Acknowledgements}

This work has been supported by the Deutsche Forschungsgemeinschaft
(DFG) and in part by IST-1999-13021. We would like to thank
D. Schlingemann, M. Grassl, M. Hein, H. Aschauer, B. Neuburger and
H. Wagner for valuable discussions.  

\appendix

\section{Cluster decomposition}
\label{decomp}

In Section~\ref{correl} we have associated the cluster state $\cskN$ on a
cluster ${\cal{C}}_N$ with a graph $G({\cal{C}}_N,E_{{\cal{C}}_N})$ where the
set $E_{{\cal{C}}_N}$ of edges is defined in the same way as in
(\ref{clusteredges}) for $E_{\cal{C}}$. To decompose 
the cluster means in more precise terms to decompose the associated
graph $G({\cal{C}}_N,E_{{\cal{C}}_N})$ into subgraphs, that is we decompose
both the set of vertices, 
${\cal{C}}_N$, and the set of edges, $E_{{\cal{C}}_N}$. As in
(\ref{vertexunion}), the set of vertices
is decomposed into the subsets ${\cal{C}}(g_i)$, the sub-clusters
corresponding to the gates $g_i$,
\begin{equation}
    {\cal{C}}_N= \bigcup_{i=1}^{|{\cal{N}}|} {\cal{C}}(g_i). \nonumber
\end{equation} 
Herein, the sets ${\cal{C}}(g_i)$ are overlapping. The output vertices
of some sub-cluster ${\cal{C}}(g_k)$, corresponding to the output
cluster qubits of the gate $g_k$, form --if
they are not the output vertices of the whole graph-- (some of) the input
vertices of other sub-clusters $\big\{{\cal{C}}(g_l)\big\}$. We define
the sets $I$ and $O$ of input and output vertices of ${\cal{C}}_N$, and
the set of overlapping vertices $V_{I/O}$ as follows:
\begin{equation}
    \label{IOVio}
    \begin{array}{rcl}
        I &=& \ds{\big\{a \in {\cal{C}}_N:\, \exists i\,|\, a\in
        {\cal{C}}_I(g_i) \, \wedge \, \neg \exists j\,|\, a\in
        {\cal{C}}_O(g_j) \big\}} \vspace{2mm}\\
        O &=& \ds{\big\{a \in {\cal{C}}_N:\, \neg \exists i\,|\, a\in
        {\cal{C}}_I(g_i) \, \wedge \, \exists j\,|\, a\in
        {\cal{C}}_O(g_j) \big\}} \vspace{2mm}\\
        V_{I/O} &=& \big\{a \in {\cal{C}}_N:\, \exists i\,|\, a\in
        {\cal{C}}_I(g_i) \, \wedge \, \exists j\,|\, a\in
        {\cal{C}}_O(g_j) \big\}. \vspace{2mm} 
    \end{array}
\end{equation} 
In the same way as we decompose the set of vertices, ${\cal{C}}_N$,
into subsets ${\cal{C}}(g_i)$, according to (\ref{edgeunion}), we
decompose the set of edges, $E_{{\cal{C}}_N}$, into subsets $E(g_i)$,
\begin{equation}
    E_{{\cal{C}}_N}= \bigcup_{i=1}^{|{\cal{N}}|} E(g_i). \nonumber
\end{equation}

\begin{figure}
    \begin{center}
        \epsfig{file=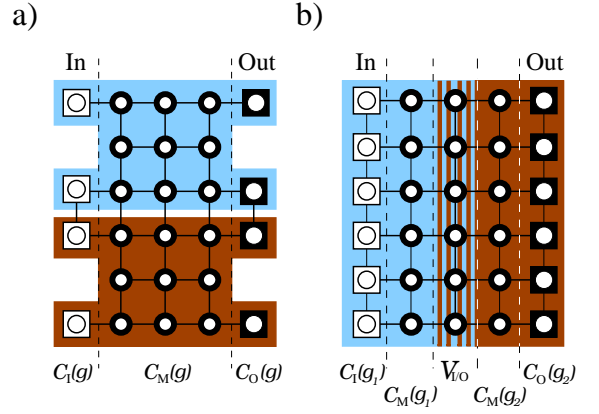, width=7.5cm}
        \caption{\label{NoGos}Two forbidden divisions of a cluster
          into sub-clusters. The functioning of the 4-qubit gate in a) cannot
          be understood from the functioning of the two 2-qubit gates
          on the sub-clusters. The displayed subdivision is excluded
          by constraint (\ref{subgraphcond}). The situation in b) can
          be treated with a careful assignment of the edges to the
          subgraphs. In this paper, however, it is excluded for simplicity by
          the assignment (\ref{inducedSG}) together with the
          constraint (\ref{doublecountavoid}).}
    \end{center}
\end{figure}

Now, for the decomposition to be useful, the subsets ${\cal{C}}(g_i)$
and $E(g_i)$ must fulfill a number of constraints. The first
of them is that each pair $({\cal{C}}(g_i),E(g_i))$ is again
a graph, $G({\cal{C}}(g_i),E(g_i))$. As in
(\ref{subgraphcond}), this requires in
particular, that the endpoints of all the edges in $E(g_i)$
are in ${\cal{C}}(g_i)$,
\begin{equation}
    \forall \, a\in {\cal{C}}_N\, | \, ( \exists e \in
    E(g_i) \,\, \text{s.th.}\,
    a \in e): \,\, a \in {\cal{C}}(g_i). \nonumber
\end{equation}
This already excludes a cluster ${\cal{C}}_N$ as displayed in
Fig.~\ref{NoGos}a. There, the cluster ${\cal{C}}_N$ is decomposed into
sub-clusters ${\cal{C}}(g_1)$ and ${\cal{C}}(g_2)$. But there are
edges, namely those which connect ${\cal{C}}(g_2)$ and
${\cal{C}}(g_1)$, which can neither be included in $G({\cal{C}}(g_1),
E(g_1))$ nor $G({\cal{C}}(g_2),E(g_2))$. Therefore, condition
(\ref{edgeunion}) cannot be satisfied and consequently the decomposition
of ${\cal{C}}_N$ into ${\cal{C}}(g_1)$ and ${\cal{C}}(g_2)$ is not
allowed. 

It is necessary to exclude a decomposition as in Fig~\ref{NoGos}a as the
circuit displayed there cannot be understood from its components on
the sub-clusters ${\cal{C}}(g_1)$ and ${\cal{C}}(g_2)$. The reason
for this is that the cluster states on the sub-clusters
${\cal{C}}(g_1)$ and ${\cal{C}}(g_2)$ are mutually entangled. This is
caused by 
precisely those interactions $\Sab$ that correspond to the edges in
$E_{{\cal{C}}_N}$ which could not be included  in either of the
subgraphs. 

The central condition (\ref{edgesdisjoint}) is that, in contrast to the
sets of vertices  ${\cal{C}}(g_i)$, the sets of edges $E(g_i)$ do not overlap,
\begin{equation}
    \forall i,j=1\,..\,|{\cal{N}}|, i\neq j:\; E(g_i)\cap E(g_j)=\emptyset. \nonumber
\end{equation}

We need a way to assign the edges to the vertices in the
sub-clusters. In this paper, for simplicity we adopt the convention
that sets of edges $E(g_i)$ are chosen such that the subgraphs
$G({\cal{C}}(g_i),E(g_i))$ are {\em{induced subgraphs}} in
$G({\cal{C}}_N,E_{{\cal{C}}_N})$, i.e.
\begin{equation}
    \label{inducedSG}
    G({\cal{C}}(g_i),E(g_i))=G[{\cal{C}}(g_i)].
\end{equation} 
As we have overlapping vertices, this simple assignment may run into
conflict with condition (\ref{edgesdisjoint}). To avoid this, we require that
\begin{equation}
    \begin{array}{rcl}
        \label{doublecountavoid}
        G[{\cal{C}}_I(g_i)]&=&G({\cal{C}}_I(g_i)),\emptyset),\;\text{and}\\
        G[{\cal{C}}_O(g_i)]&=&G({\cal{C}}_O(g_i)),\emptyset),
        \;\forall\,\, i=1\,..\,|{\cal{N}}|.
    \end{array}
\end{equation}
That is, the vertices within the sets ${\cal{C}}_I(g_i)$,
${\cal{C}}_O(g_i)$ are not connected by edges. 

The assignment convention (\ref{inducedSG}) together with the
condition (\ref{doublecountavoid}) exclude decompositions of a cluster
like the one displayed in Fig~\ref{NoGos}b. In principle, however,
such a decomposition 
is possible. In such a case the assignment of the edges to
the sub-clusters is not as simple as (\ref{inducedSG}). Instead, each
edge had to be assigned to only one sub-cluster while respecting condition
(\ref{subgraphcond}). In order to keep the
notation as simple as possible, in this paper we do not consider
decompositions of the sort displayed in Fig~\ref{NoGos}b.

The constraints displayed so far applied to the subgraphs. We have two
further constraints on the graph $G({\cal{C}}_N, 
E_{{\cal{C}}_N})$, with regard to its decomposition. Or, to put it in
different terms, the remaining constraints are for the proper
composition of the 
graph out of the subgraphs. They correspond to the usual gate
composition rules. We require that each input and
output vertex is the input or output vertex of exactly 
one gate,
\begin{equation}
    \label{oneIO}
    \begin{array}{rl}
        \forall a \in I \cup V_{I/O}:& \neg \exists \,i,j\,|\, i \neq
        j,\, a \in {\cal{C}}_I(g_i) \wedge a \in {\cal{C}}_I(g_j)
        \vspace{2mm}\\ 
        \forall a \in O \cup V_{I/O}:& \neg \exists \, i,j\,|\, i \neq
        j,\, a \in {\cal{C}}_O(g_i) \wedge a \in {\cal{C}}_O(g_j). \vspace{2mm}
    \end{array}
\end{equation}
Further, the graph
$G({\cal{C}}_N,E_{{\cal{C}}_N})$ with the vertices 
of $I \cup O \cup V_{I/O}$ and associated edges removed disintegrates
into mutually disconnected induced subgraphs corresponding to the gate bodies,
\begin{equation} 
    \label{disconnected}
    G({\cal{C}}_N,E_{{\cal{C}}_N})\backslash (I \cup O \cup V_{I/O}) =
    \bigcup_{i=1}^{|{\cal{N}}|} G[{\cal{C}}_M(g_i)]. 
\end{equation}

To summarize, the two central conditions (\ref{edgeunion}) and
(\ref{edgesdisjoint}) for the decomposition of the edges
$E_{{\cal{C}}_N}$ of the initial graph and the constraint
(\ref{subgraphcond}) are fulfilled if the subgraphs
$G({\cal{C}}(g_i),E(g_i))$ are chosen in accordance with the assignment
(\ref{inducedSG}) and the constraint (\ref{doublecountavoid}). All the 
examples for \QCns-gate simulations displayed in this paper are of this type.


\end{document}

%% file: cuts.tex
\setlength{\unitlength}{0.6cm}
\definecolor{lgray}{gray}{0.65}
\definecolor{dgray}{gray}{0.35}
\begin{picture}(12,6)
  \setlength{\unitlength}{0.6cm}
  \color{lgray}
  \multiput(0.8,3)(0,0.016){64}{\line(1,0){1.4}}
  \multiput(5.5,3)(0,0.016){64}{\line(1,0){1.4}}
  \color{white}
  \put(1.95,3.5){\circle*{0.32}}
  \put(6.67,3.5){\circle*{0.32}}
  \color{black}
  \put(1.95,3.5){\circle*{0.2}}
  \put(6.67,3.5){\circle*{0.2}}
  
  \setlength{\linewidth}{0.7mm}
  \put(0,5.5){\line(1,0){12}}
  \put(0,4.5){\line(1,0){12}}
  \put(0,3.5){\line(1,0){12}}
  \put(0,2.5){\line(1,0){12}}
  \put(0,1.5){\line(1,0){12}}
  
  \put(3.65,4.5){\line(0,-1){1.37}}
  \put(8.4,2.5){\line(0,1){1.37}}
  \put(6,4.5){\line(0,1){1.37}}
  \put(6,2.5){\line(0,-1){1.37}}
  \put(3.65,3.5){\circle{0.77}}
  \put(8.4,3.5){\circle{0.77}}
  \put(6,5.5){\circle{0.77}}
  \put(6,1.5){\circle{0.77}}
  \put(3.65,4.5){\circle*{0.2}}
  \put(8.4,2.5){\circle*{0.2}}
  \put(6,4.5){\circle*{0.2}}
  \put(6,2.5){\circle*{0.2}}

  \put(0,5.5){\circle*{0.35}}
  \put(0,4.5){\circle*{0.35}}
  \put(0,3.5){\circle*{0.35}}
  \put(0,2.5){\circle*{0.35}}
  \put(0,1.5){\circle*{0.35}}

  \put(12,5.5){\circle*{0.35}}
  \put(12,4.5){\circle*{0.35}}
  \put(12,3.5){\circle*{0.35}}
  \put(12,2.5){\circle*{0.35}}
  \put(12,1.5){\circle*{0.35}}
  \color{white}
  \put(0,5.5){\circle*{0.25}}
  \put(0,4.5){\circle*{0.25}}
  \put(0,3.5){\circle*{0.25}}
  \put(0,2.5){\circle*{0.25}}
  \put(0,1.5){\circle*{0.25}}

  \put(12,5.5){\circle*{0.25}}
  \put(12,4.5){\circle*{0.25}}
  \put(12,3.5){\circle*{0.25}}
  \put(12,2.5){\circle*{0.25}}
  \put(12,1.5){\circle*{0.25}}
  \multiput(5.6,3.1)(0,0.016){50}{\line(1,0){0.8}}
  \multiput(10.4,2.1)(0,0.016){50}{\line(1,0){0.8}}
  \multiput(10.4,3.1)(0,0.016){50}{\line(1,0){0.8}}
  \multiput(10.4,4.1)(0,0.016){50}{\line(1,0){0.8}}
  \multiput(0.9,3.1)(0,0.016){50}{\line(1,0){0.8}}
  \multiput(0.9,4.1)(0,0.016){50}{\line(1,0){0.8}}
  \color{black}
  
  \put(5.6,3.1){\line(1,0){0.81}}
  \put(5.6,3.9){\line(1,0){0.81}}
  \put(5.6,3.1){\line(0,1){0.8}}
  \put(6.4,3.1){\line(0,1){0.8}}

  \put(10.4,2.1){\line(1,0){0.81}}
  \put(10.4,2.9){\line(1,0){0.81}}
  \put(10.4,2.1){\line(0,1){0.8}}
  \put(11.2,2.1){\line(0,1){0.8}}

  \put(10.4,3.1){\line(1,0){0.81}}
  \put(10.4,3.9){\line(1,0){0.81}}
  \put(10.4,3.1){\line(0,1){0.8}}
  \put(11.2,3.1){\line(0,1){0.8}}

  \put(10.4,4.1){\line(1,0){0.81}}
  \put(10.4,4.9){\line(1,0){0.81}}
  \put(10.4,4.1){\line(0,1){0.8}}
  \put(11.2,4.1){\line(0,1){0.8}}

  \put(0.9,3.1){\line(1,0){0.81}}
  \put(0.9,3.9){\line(1,0){0.81}}
  \put(0.9,3.1){\line(0,1){0.8}}
  \put(1.7,3.1){\line(0,1){0.8}}

  \put(0.9,4.1){\line(1,0){0.81}}
  \put(0.9,4.9){\line(1,0){0.81}}
  \put(0.9,4.1){\line(0,1){0.8}}
  \put(1.7,4.1){\line(0,1){0.8}}

  \put(2.25,3.0){$a$}
  \put(7,3.0){$b$}
  \put(10.55,2.4){\normalsize{$U_z$}}
  \put(5.75,3.4){\normalsize{$U_z$}}
  \put(10.55,3.4){\normalsize{$U_x$}}
  \put(10.55,4.4){\normalsize{$U_x$}}
  \put(1,3.4){\normalsize{$U_x$}}
  \put(1,4.4){\normalsize{$U_x$}}
  \put(0.3,0.4){${\cal{O}}_i$}
  \put(4.6,0.4){${\cal{O}}_j$}
  \put(9.3,0.4){$\not\!\!{\cal{O}}\,\,$}
  \put(11.6,0.4){$\Omega$}
  \put(-0.3,5.9){IN}
  \put(11.5,5.9){OUT}

  \color{dgray}
  \put(0.5,1.0){\line(0,1){5.15}}
  
  \put(4.8,4.2){\line(0,1){2}}
  \put(4.8,1.0){\line(0,1){1.8}}
  \put(7.6,2.8){\line(0,1){1.4}}
  \put(4.8,2.8){\line(1,0){2.8}}
  \put(4.8,4.2){\line(1,0){2.8}}

  \put(11.7,1.0){\line(0,1){4.8}}

  \multiput(9.5,4.9)(0,-0.6){7}{\line(0,-1){0.32}}
  \multiput(5.2,4.9)(0.665,0){7}{\line(1,0){0.3}}
  \multiput(5.2,4.9)(0,0.6){2}{\line(0,1){0.25}}
  \put(5.2,6.05){\line(0,1){0.1}}
\end{picture}